\pdfoutput=1

\newcommand*{\ATLASLATEXPATH}{}

\documentclass[cernpreprint,texlive=2014,txfonts,USenglish]{\ATLASLATEXPATH atlasdoc}

\usepackage{pdflscape}

\usepackage[subfigure]{\ATLASLATEXPATH atlaspackage}

\usepackage{\ATLASLATEXPATH atlasbiblatex}

\usepackage{\ATLASLATEXPATH atlascontribute}

\usepackage{\ATLASLATEXPATH atlasphysics}

\graphicspath{{logos/}{figures/}}

\usepackage{siunitx}

\usepackage{zz-defs}
\newcommand{\intlumi}{36.1}

\newcommand{\extrapolatedxsectshort}{\ensuremath{17.3\pm 0.9}}
\newcommand{\extrapolatedxsectstatuncert}{\ensuremath{\pm 0.6}}
\newcommand{\extrapolatedxsectsystuncert}{\ensuremath{\pm 0.5}}
\newcommand{\extrapolatedxsectlumiuncert}{\ensuremath{\pm 0.6}}
\AtlasTitle{$\ZZllll$ cross-section measurements and search for anomalous triple gauge couplings in 13~\TeV{} $\Pproton\Pproton$ collisions with the ATLAS detector}

\author{The ATLAS Collaboration}
\AtlasRefCode{STDM-2016-15}
\PreprintIdNumber{CERN-EP-2017-163}

\AtlasJournalRef{Phys. Rev. D 97 (2018) 032005}
\AtlasDOI{10.1103/PhysRevD.97.032005}

\AtlasAbstract{%
Measurements of $\PZ\PZ$ production in the \llll{} channel in proton--proton collisions at $13$~\TeV{} center-of-mass energy at the Large Hadron Collider are presented. The data correspond to \intlumi{}~\ifb{} of collisions collected by the ATLAS experiment in 2015 and 2016. Here $\ell$ and $\ell'$ stand for electrons or muons. Integrated and differential $\PZ\PZ \to \llll$ cross sections with $\PZ \to \ell^+\ell^-$ candidate masses in the range of 66~\GeV{} to 116~\GeV{} are measured in a fiducial phase space corresponding to the detector acceptance and
corrected for detector effects. The differential cross sections are presented in bins of twenty observables, including several that describe the jet
activity. The integrated cross section is also extrapolated to a total phase space and to all Standard Model decays of $\PZ$ bosons with mass between
66~\GeV{} and 116~\GeV{}, resulting in a value of \extrapolatedxsectshort{}~$[$\extrapolatedxsectstatuncert{}~(stat.)\;\extrapolatedxsectsystuncert{}~(syst.)\;\extrapolatedxsectlumiuncert{}~(lumi.)$]$~pb. The measurements are found to be in good agreement with the Standard Model. A search for neutral triple gauge couplings is performed using the transverse momentum distribution of the leading \PZ{} boson candidate. No evidence for such couplings is found and exclusion limits are set on their parameters.
}

\hypersetup{pdftitle={ATLAS document},pdfauthor={The ATLAS Collaboration}}

\begin{document}
\maketitle


\section{Introduction}
\begin{figure}[h!]
	\centering
	\subfigure[]{
		\centering
		\includegraphics[width=0.27\textwidth]{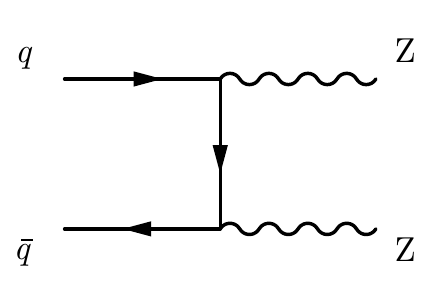}
        \label{fig:qqZZ}
	}
	\hspace{3cm}
	\subfigure[]{
		\centering
		\includegraphics[width=0.27\textwidth]{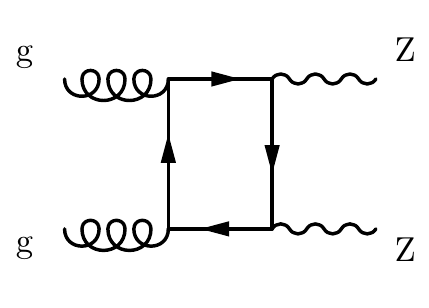}
        \label{fig:ggZZ}  
	}\\[5mm]
	\subfigure[]{
		\centering
		\includegraphics[width=0.27\textwidth]{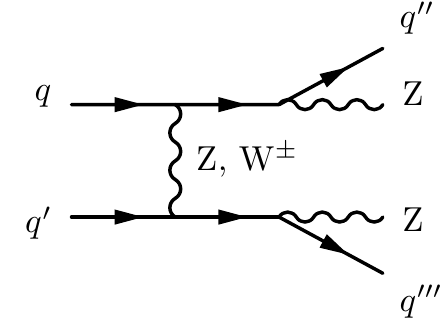}
        \label{fig:nonvbsZZ}
	}
	\hspace{3cm}
	\subfigure[]{
		\centering
		\includegraphics[width=0.27\textwidth]{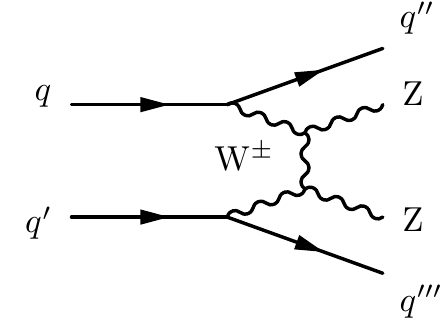}
        \label{fig:vbsZZ}  
	}
	\vspace{-0.3cm}
	\caption{Examples of leading-order SM Feynman diagrams for \ZZ{} production in proton--proton collisions: (a) $\Pquark\APquark$-initiated, (b) $\Pgluon\Pgluon$-initiated, (c) electroweak $\PZ\PZ jj$ production, (d) electroweak $\PZ\PZ jj$ production via weak-boson scattering.}
  \label{fig:ZZ}
\end{figure}

The study of the production of \PZ{} boson pairs in proton--proton ($\Pproton\Pproton$) interactions at the Large Hadron Collider (LHC) \cite{Evans:2008zzb} tests the electroweak sector of the Standard Model (SM) at the highest available energies. Example Feynman diagrams of \ZZ{} production at the LHC are shown in Figure~\ref{fig:ZZ}. In $\Pproton\Pproton$ collisions at a center-of-mass energy of $\sqs=13$~\TeV{}, \ZZ{} production is dominated by quark--antiquark ($\Pquark\APquark$) interactions, with an $\mathcal{O}(10\%)$ contribution from loop-induced gluon--gluon ($\Pgluon\Pgluon$) interactions~\cite{Grazzini:2015hta,Caola:2015psa}. The production of $\PZ\PZ$ in association with two electroweakly produced jets, denoted EW-$\PZ\PZ jj$, 
includes the rare \ZZ{} weak-boson scattering process. Study of \ZZ{} production in association with jets is an important step in searching for \ZZ{} weak-boson scattering, which has so far not been experimentally observed by itself.

\vspace{-0.2cm}

The SM \ZZ{} production can also proceed via a Higgs boson propagator, although this contribution is expected to be suppressed in the region where both \PZ{} bosons are produced nearly on-shell, as is the case in this analysis. Non-Higgs-mediated \ZZ{} production is an important background in studies of the Higgs boson properties~\cite{HIGG-2014-11,HIGG-2014-10,CMS-HIG-14-036,CMS-HIG-14-028}. It is also a major background in searches for new physics processes producing pairs of \PZ{} bosons at high invariant mass~\cite{HIGG-2013-20,CMS-HIG-13-031,EXOT-2016-01,CMS-B2G-16-004} 
and it is sensitive to anomalous triple gauge couplings (aTGCs) of neutral gauge bosons, which are not allowed in the SM~\cite{Baur:2000ae}. The SM does not have tree-level vertices coupling three neutral gauge bosons ($\PZ\PZ\PZ$, $\PZ\PZ\Pphoton$), because these would violate the underlying $\text{SU}(2)_{\text{L}} \times \text{U}(1)_{\text{Y}}$ symmetry. However, these couplings exist in some extensions of the SM, enhancing the $\PZ\PZ$ production cross section in regions where the energy scale of the interaction is high.

\begin{figure}[h!]
	\centering
	\includegraphics[width=0.27\textwidth]{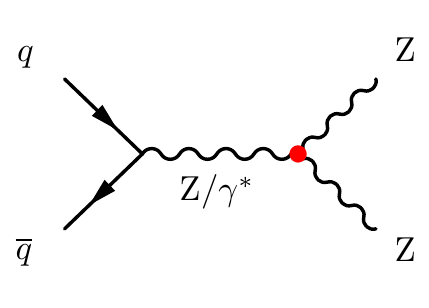}
	\label{fig:atgcZZ}  
	\vspace{-0.2cm}
	\caption{Example Feynman diagram of $\PZ\PZ$ production containing an aTGC vertex, here indicated by a red dot, which is forbidden in the SM.}
  \label{fig:atgc_feynman}
\end{figure}

An example Feynman diagram of $\PZ\PZ$ production via aTGC is shown in Figure~\ref{fig:atgc_feynman}.

\vspace{-0.2cm}

\newcommand{\coordfootnote}{ATLAS uses a right-handed coordinate system with its origin at the nominal interaction point in the center of the detector and the $z$-axis along the beam pipe. The $x$-axis points to the center of the LHC ring, and the $y$-axis points upward. Cylindrical coordinates $(r,\phi)$ are used in the transverse plane, $\phi$ being the azimuthal angle around the $z$-axis. The pseudorapidity is defined in terms of the polar angle $\theta$ as $\eta=-\ln[\tan(\theta/2)]$. Transverse momentum $\pt$ is the projection of momentum onto the transverse plane.}

Integrated and differential \ZZ{} production cross sections were previously measured  at $\sqs=7$ and 8~\TeV{} by the ATLAS and CMS collaborations~\cite{STDM-2012-02,CMS-SMP-13-005,STDM-2014-16,CMS-SMP-12-007} and found to be consistent with SM predictions. The integrated $\Pproton\Pproton \to \ZZllll$ cross section at $\sqs=13$~\TeV{} was recently measured by the ATLAS \cite{STDM-2015-13} and CMS \cite{CMS-SMP-16-001} collaborations, each analyzing data corresponding to an integrated luminosity of about 3~\ifb{}. Searches for aTGCs were previously performed at lower center-of-mass energies by ATLAS \cite{STDM-2014-16}, CMS \cite{CMS-SMP-13-005,CMS-SMP-12-016}, D0 \cite{Abazov:2007ad}, and by the LEP experiments \cite{Alcaraz:2006mx}.
This paper 
represents an extension of the ATLAS measurement, using a total of $36.1 \pm 1.1$~\ifb{} of data collected with the ATLAS detector in the years 2015 and 2016.

In this analysis, candidate events are reconstructed in the fully leptonic $\ZZllll$ decay channel where $\ell$ and $\ell'$ can be an electron or a muon. Throughout this analysis, ``\PZ{} boson'' refers to the superposition of a $\PZ$ boson and a virtual photon in the mass range from 66~\GeV{} to 116~\GeV{}, as these are not strictly distinguishable when decaying to charged leptons.
A fiducial phase space is defined, reflecting both the acceptance of the ATLAS detector~\cite{PERF-2007-01,ATLAS-TDR-19} and the selections imposed on the reconstructed leptons in this analysis. Both the integrated and differential cross sections are measured, the latter with respect to 20 different observables. Ten of these directly measure the jet activity in the events. The observed event yields are unfolded to the fiducial phase space using simulated samples to model the detector effects. The integrated cross sections are inclusive with respect to jet production. For easier comparison to other measurements, the integrated fiducial cross sections determined in different leptonic channels are combined and extrapolated to a total phase space and to all \PZ{} boson decay modes. A search for aTGCs is performed by looking for deviations of the data from the SM predictions at high values of the transverse momentum of the leading-\pt{} \PZ{} boson, which is one of the observables most sensitive to the energy scale of the interaction.\footnote{\coordfootnote}

Differential fiducial cross sections are measured with respect to the following observables:
\begin{itemize}

\item Transverse momentum of the four-lepton system, \ptfourl{};
\item Absolute rapidity of the four-lepton system, $|\yfourl{}|$;
\item Separation in azimuthal angle between the two \PZ{} boson candidates, \dphiZZ{}, defined such that it lies in the interval $[0, \pi]$;
\item Absolute difference in rapidity between the two \PZ{} boson candidates, $|\dyZZ{}|$;
\item Transverse momentum of the leading-\pt{} and the subleading-\pt{} \PZ{} boson candidates, \ptz{} and \ptzsub{};
\item Transverse momentum of each of the four leptons; 
\item Number of jets with $\pt > 30$~\GeV{} and $|\eta| < 4.5$;
\item Number of jets with $\pt > 30$~\GeV{} and $|\eta| < 2.4$;
\item Number of jets with $\pt > 60$~\GeV{} and $|\eta| < 4.5$;
\item Scalar sum of the transverse momenta of all jets in the event with $\pt > 30$~\GeV{} and $|\eta| < 4.5$;
\item Absolute pseudorapidity of the leading-\pt{} and the subleading-\pt{} jets;
\item Transverse momentum of the leading-\pt{} and the subleading-\pt{} jets;
\item Absolute difference in rapidity between the two leading-\pt{} jets, $|\dyjj{}|$; 
\item Invariant mass of the two leading-\pt{} jets, \mjj{}. 
\end{itemize}

These measurements provide a detailed description of the kinematics in $\PZ\PZ{}$ events and allow comparisons and validations of current and future predictions. Some of the differential measurements are particularly motivated: the transverse momentum of the four-lepton system directly measures the recoil against all other particles produced in the collision and therefore provides information about quantum chromodynamics (QCD) and electroweak radiation across the entire range of scales. The rapidity of the four-lepton system is sensitive to the $z$-component of the total momentum of the initial-state partons involved in the \ZZ{} production. It may therefore be sensitive to the parton distribution functions (PDFs). The azimuthal-angle separation and rapidity difference between the \PZ{} boson candidates probe their angular correlations and may help extract the contribution of double-parton-scattering \ZZ{} production. The azimuthal-angle separation is also sensitive to radiation of partons and photons produced in association with the \ZZ{} pair. The scalar sum of the transverse momenta of all jets provides a measure of the overall jet activity that is independent of their azimuthal configuration. The measurements of $|\dyjj{}|$ and \mjj{} are particularly sensitive to the  EW-$\PZ\PZ jj$ process. They both tend to have larger values in weak-boson scattering than in other \ZZ{} production channels, providing an important step towards the study of \ZZ{} production via weak-boson scattering.

\section{ATLAS detector}

The ATLAS detector \cite{PERF-2007-01,ATLAS-TDR-19} is a multipurpose particle detector with a cylindrical geometry. It consists of layers of inner tracking detectors, calorimeters, and muon chambers. The inner detector (ID) is immersed in a 2~T axial magnetic field generated by a thin superconducting solenoid and provides charged-particle tracking and momentum measurement in the pseudorapidity range $|\eta| < 2.5$. The calorimeter system covers the pseudorapidity range $|\eta| < 4.9$. Electromagnetic calorimetry is provided by 
high-granularity lead/liquid-argon calorimeters in the region $|\eta| < 3.2$ and by copper/liquid-argon calorimeters in the region 3.2 < $|\eta|$ < 4.9. Within $|\eta| < 2.47$ the finely segmented electromagnetic calorimeter, together with the ID information, allows electron identification. Hadronic calorimetry is provided by the steel/scintillator-tile calorimeter 
within $|\eta| < 1.7$ and two copper (or tungsten)/liquid-argon calorimeters within $1.7 < |\eta| < 4.9$. The muon spectrometer (MS) comprises separate trigger and high-precision tracking chambers. The precision chamber system covers the region $|\eta| < 2.7$ with three layers of monitored drift tubes, complemented by cathode strip chambers in the forward region, where the hit rate is highest. The muon trigger system covers the range $|\eta| < 2.4$ with resistive plate chambers in the central, and thin gap chambers in the forward regions. A two-level trigger system is used to select events of interest in real time \cite{Aaboud:2016leb}. The Level-1 trigger is implemented in hardware and uses a subset of detector information to reduce the event rate to a value of around 100~kHz. This is followed by a software-based high-level trigger system that reduces the event rate to about 1~kHz.

\section{Simulated samples and theoretical predictions}\label{sec:mc}
Event samples simulated with Monte Carlo (MC) event generators are used to obtain corrections for detector effects and to estimate signal and background contributions. Throughout this paper, unless stated otherwise, orders of calculations refer to perturbative expansions in the strong coupling constant \alphas{} in QCD and all calculations use the  \cten{} \cite{Lai:2010vv} PDFs with the evolution order in \alphas{} corresponding to the perturbative order in \alphas{} in the calculation. MC event generator versions are only given the first time the event generator is mentioned for each sample.

The nominal signal samples are generated with \SHERPA{}~2.2.1~\cite{Gleisberg:2008ta_atlas,Hoeche:2009rj,Gleisberg:2008fv,Schumann:2007mg,Schonherr:2008av,Cascioli:2011va,Hoeche:2012yf} using the NNPDF 3.0 NNLO PDFs~\cite{Ball:2014uwa} (with $\alphas = 0.118$ at the \PZ{} pole mass), with the $\Pquark\APquark$-initiated process simulated at next-to-leading order (NLO) for \ZZ{} plus zero or one additional jet and at leading order (LO) for two or three additional jets generated at the matrix-element level.
A \SHERPA{}~2.1.1 \ZZ{} sample is generated with the loop-induced \gluongluon-initiated process simulated at LO using NLO PDFs, including subprocesses involving a Higgs boson propagator, with zero or one additional jet. 
The \gluongluon-initiated process first enters at next-to-next-to-leading order (NNLO) and is therefore not included in the NLO sample for the $\Pquark\APquark$-initiated process. Due to different initial states, the \gluongluon-initiated process does not interfere with the $\Pquark\APquark$-initiated process at NLO. The loop-induced \gluongluon-initiated process calculated at its LO ($\alphas^2$) receives large corrections at NLO ($\alphas^3$) \cite{Caola:2015psa}. The cross section of the sample is therefore multiplied by an NLO/LO $K$-factor of $1.67 \pm 0.25$ \cite{Caola:2015psa}.
The EW-$\PZ\PZ jj$ process is simulated using \SHERPA{}~2.1.1 at its lowest contributing order in the electroweak coupling, $\alpha^6$ (including the decays of the $\PZ$ bosons). It includes the triboson subprocess $\ZZ V \to \llll jj$, where the third boson $V$ decays hadronically.
\SHERPA{} also simulates parton showering, electromagnetic radiation, the underlying event, and hadronization in the above samples, using its default set of tuned parameters (tune).
Throughout this paper, the prediction obtained by summing the above samples is referred to as the nominal \SHERPA{} setup.

An alternative prediction for the $\Pquark\APquark$-initiated process is obtained using the \POWHEG{} method and framework~\cite{Nason:2004rx,Frixione:2007vw} as implemented in \POWHEGBOX{} 2 \cite{Alioli:2010xd}, with a diboson event generator \cite{Melia:2011tj,Nason:2013ydw} used to simulate the $\PZ\PZ$ production process at NLO. The simulation of parton showering, electromagnetic radiation, the underlying event, and hadronization is performed with \PYTHIA{}~8.186 \cite{Sjostrand:2006za,Sjostrand:2007gs} using the \aznlo{} parameter tune~\cite{STDM-2012-23}. This sample is used to estimate the systematic uncertainty due to modeling differences between the event generators.

Additional samples are generated to estimate the contribution from
background events.
Triboson events are simulated at LO with \SHERPA{}~2.1.1. Samples of $\Ptop\APtop\kern0.5pt\PZ$ events are simulated at LO with \MADGRAPH~2.2.2 \cite{Alwall:2014hca} +
\py{}~8.186 using the \nnpdf{} PDFs~\cite{Ball:2012cx} and the \afourteen{} tune~\cite{ATL-PHYS-PUB-2014-021}.

In all MC samples, additional $\Pproton\Pproton$ interactions occurring in the same bunch crossing as the
process of interest or in nearby ones (pileup) are simulated with \py{} using
\mstw{} PDFs~\cite{Martin:2009iq} and the \atwo{} tune~\cite{ATL-PHYS-PUB-2012-003}. 
The samples are then passed through a simulation of the ATLAS detector~\cite{SOFT-2010-01}
based on \GEANT{} 4~\cite{Agostinelli:2002hh}. 
Weights are applied to the simulated events to correct for the
small differences from data in the reconstruction,
identification, isolation, and impact parameter efficiencies for electrons and
muons~\cite{PERF-2016-01,PERF-2015-10}. Furthermore, the lepton momentum or energy scales and resolutions are adjusted such that data and simulation match \cite{ATL-PHYS-PUB-2016-015,PERF-2015-10}.

NNLO cross sections for $\Pproton\Pproton \to \ZZllll{}$ in the fiducial and total phase space are provided by \matrixnnlo{} \cite{Grazzini:2015hta}, also in bins of the jet-inclusive unfolded distributions. They include the \gluongluon-initiated process at its lowest contributing order, which accounts for about 60\% of the cross-section increase with respect to NLO \cite{Cascioli:2014yka}. The calculation uses a dynamic QCD scale of $\mfourl / 2$ and the NNPDF 3.0 PDFs, with NNLO PDFs being used also for the \gluongluon-initiated process. It uses the $G_{\mu}$ electroweak scheme, in which the Fermi constant $G_{\mu}$ as well as the pole masses of the weak bosons are taken as independent input parameters \cite{Jegerlehner1990}.

The NNLO calculation is also used for extrapolation of the integrated cross section from the fiducial to a total phase space. The estimation of PDF uncertainties with \matrixnnlo{} is currently unfeasible, because it would require repeating the entire calculation for each PDF variation, which is too computationally expensive. Therefore, these are estimated using an NLO (LO) calculation for the $\Pquark\APquark$-initiated ($\Pgluon\Pgluon$-initiated) process from \MCFM{}~\cite{Campbell:1999ah}, taking the mass of the four-lepton system, $\mfourl$, as the dynamic QCD scale. NLO PDFs are used for the \gluongluon-initiated process and its contribution is multiplied by the NLO/LO $K$-factor of $1.67 \pm 0.25$.

Electroweak corrections at next-to-leading order (NLO EW)~\cite{Biedermann:2016lvg,Biedermann:2016yvs} are calculated in the fiducial phase space, also in bins of the jet-inclusive unfolded distributions. The $G_{\mu}$ scheme is used. The NLO/LO EW $K$-factor integrated across the entire fiducial phase space is about 0.95. The NLO EW corrections are calculated with respect to the $\Pquark\APquark$-initiated process at LO in $\alphas$, meaning that they cannot be obtained differentially in observables that are trivial at LO in $\alphas$, e.g.~the transverse momentum of the four-lepton system. Where a differential calculation is not possible, the integrated value in the fiducial phase space is used. The higher-order NNLO QCD and NLO EW corrections are applied to the predictions only where explicitly stated.

The NNLO calculations serve as the basis of a SM prediction incorporating the formally most accurate available predictions. The contribution of the \gluongluon-initiated process is multiplied by the NLO/LO $K$-factor of $1.67 \pm 0.25$. The NLO EW corrections are applied as multiplicative $K$-factors, differentially in the observable of interest if available, otherwise integrated over the fiducial phase space. They are never applied to the $\Pgluon\Pgluon$-initiated loop-induced process, as its topology is considered too different from the LO QCD predictions of the $\Pquark\APquark$-initiated process for which the NLO EW corrections are calculated. The cross section of the EW-$\ZZ jj$ process calculated with \SHERPA{} is added to the signal prediction.

\section{Fiducial definition}
\subsection{Fiducial phase space}
\label{sec:fiducial}
The fiducial phase space is defined using final-state particles, meaning particles whose average lifetime $\tau_0$ 
satisfies $c\tau_0 > 10~\text{mm}$ \cite{ATL-PHYS-PUB-2015-013}. A prompt lepton, photon, or neutrino refers to a final-state particle that does not originate from the decay of a hadron or $\tau$ lepton, or any material interaction (such as Bremsstrahlung or pair production) \cite{ATL-PHYS-PUB-2015-013}. Particles other than leptons, photons, and neutrinos are never considered prompt in this analysis.

The requirements used to define the fiducial phase space mirror the selections applied to the reconstructed leptons. This is done to ensure that the extrapolation from the observed data to the fiducial phase space is as model-independent as possible, ideally depending only on detector effects.

Events in the fiducial phase space contain at least four prompt electrons and/or prompt muons. The four-momenta of all prompt photons within $\Delta R = \sqrt{(\Delta\eta)^2 + (\Delta\phi)^2} = 0.1$ of a lepton are added to the four-momentum of the closest lepton. This dressing is done to emulate the effects of quasi-collinear electromagnetic radiation from the charged leptons on their experimental reconstruction in the detector \cite{ATL-PHYS-PUB-2015-013}. Each dressed lepton is required to have transverse momentum $\pt > 5$~\GeV{} and absolute pseudorapidity $|\eta| < 2.7$.

All possible pairs of same-flavor opposite-charge dileptons are formed, referred to as quadruplets. In each quadruplet, the three highest-$\pt$ leptons must satisfy $\pt > 20$~\GeV{}, 15~\GeV{}, and 10~\GeV{}, respectively. If multiple selected quadruplets are present, the quadruplet minimizing $|m_{\ell\ell}^a - m_{\PZ}| + |m_{\ell\ell}^{b} - m_{\PZ}|$ is selected, where $m_{\ell\ell}^{a,\,b}$ is the mass of a given same-flavor opposite-charge dilepton and $m_{\PZ} = 91.1876$~\GeV{} is the \PZ{} boson pole mass \cite{Olive:2016xmw}. All remaining requirements are applied to the leptons in the final selected quadruplet. Any two same-flavor (different-flavor) leptons $\ell_i$, $\ell^{(\prime)}_j$ must be separated by $\Delta R (\ell_i,\ell^{(\prime)}_j) > 0.1$~(0.2). All possible same-flavor opposite-charge dileptons must have an invariant mass greater than $5$~\GeV{}, to match the same requirement in the selection of reconstructed events, which is introduced to reduce the background from leptonically decaying hadrons. If all leptons are of the same flavor, the dilepton pairing that minimizes $|m_{\ell\ell}^a - m_{\PZ}| + |m_{\ell\ell}^{b} - m_{\PZ}|$ is chosen. 
The selected dileptons are defined as the \PZ{} boson candidates. Each is required to 
have an invariant mass between 66~\GeV{} and 116~\GeV{}. Based on the leptons in the chosen quadruplet, events are classified into three signal channels: \eeee{}, \mmmm{}, and \eemm{}.

Jets are used for several differential cross sections. They
are clustered from all final-state particles except prompt leptons, prompt neutrinos, and prompt photons using the anti-$k_t$ algorithm \cite{cacciari08} with radius parameter 0.4, implemented in \textsc{FastJet} \cite{Cacciari:2011ma}. Jets are required to have $\pt > 30$~\GeV{} and $|\eta| < 4.5$. Jets within $\Delta R = 0.4$ of any selected fiducial lepton (as defined above) are rejected. 

The fiducial selection is summarized in Table~\ref{tab:fiducial_selection}.

\begin{table}[!htbp]
\centering
\begin{tabular}{ll}
\toprule
\textbf{Type} & \textbf{Input or requirement}\\
\midrule
Leptons ($e$, \Pmu{}) & Prompt\\
	& Dressed with prompt photons within $\Delta R = 0.1$ (added to closest prompt lepton)\\
	& $\pt > 5~\GeV$\\
	& $|\eta|<2.7$\\
\midrule
Quadruplets & Two same-flavor opposite-charge lepton pairs\\
	& Three leading-\pt{} leptons satisfy $\pt > 20$~\GeV{}, 15~\GeV{}, 10~\GeV{}\\
\midrule
Events & Only quadruplet minimizing $|m_{\ell\ell}^a - m_{\PZ}| + |m_{\ell\ell}^{b} - m_{\PZ}|$ is considered\\
	& Any same-flavor opposite-charge dilepton has mass $m_{\ell\ell} > 5$~\GeV{}\\
	& $\Delta R > 0.1$ (0.2) between all same-flavor (different-flavor) leptons\\
	& Dileptons minimizing $|m_{\ell\ell}^a - m_{\PZ}| + |m_{\ell\ell}^{b} - m_{\PZ}|$ are taken as \PZ{} boson candidates\\
	& \PZ{} boson candidates have mass $66~\GeV{} < m_{\ell\ell} < 116$~\GeV{}\\ 
\midrule
Jets & Clustered from all non-prompt particles\\
	& Anti-$k_t$ algorithm with $R = 0.4$\\
	& $\pt > 30~\GeV$\\
	& $|\eta| < 4.5$\\
	& Rejected if within $\Delta R = 0.4$ of a fiducial lepton\\
\bottomrule
\end{tabular}
\caption{Summary of the selection criteria defining the fiducial phase space.}
\label{tab:fiducial_selection}
\end{table}

\subsection{Signal-process definition}
\label{sec:signal_process_definition}
Some SM processes can pass the fiducial selection but are still excluded from the signal. They are considered irreducible backgrounds and are subtracted from the sample of selected candidate events. Any events containing four prompt leptons plus any additional leptons, neutrinos, or photons are considered irreducible backgrounds. An example is the triboson process $\PZ\PZ\PWplus \to \ell^+\ell^-\ell^{\prime +}\ell^{\prime -}\ell^+\nu_{\ell}$. In practice, predictions only exist for a subset of such processes. The irreducible backgrounds that are subtracted are discussed in Section~\ref{sec:bkg}. They are very small, approximately~1\% of the predicted signal. 

The fiducial phase space is inclusive with respect to jets, independently of their origin. Triboson (and higher boson-multiplicity) processes producing a $\PZ\PZ$ pair decaying leptonically with any additional electroweak bosons decaying hadronically are included in the signal, as are any other SM processes of the pattern $(\PZ\PZ \to \ell^+\ell^-\ell^{\prime +}\ell^{\prime -}) + (X \to \text{jets})$. In practice, only the process $\PZ\PZ V \to \llll jj$ is included in the theoretical predictions, in the EW-$\ZZ jj$ sample generated with \SHERPA{}.

Production via double parton scattering in the same $\Pproton\Pproton$ collision is included in the signal. Its contribution is not included in the theoretical predictions, but is expected to be smaller than 1\% of the total signal yield. 
This estimate assumes incoherent double parton scattering and is based on a measurement of the effective area parameter at $\sqs=7$~\TeV{} \cite{STDM-2011-29}.
Various other measurements of the effective area parameter were made~\cite{ALITTI1991145, Abe:1993rv, Abe:1997xk, Abazov:2009gc, STDM-2015-08, Aaij:2012dz,Chatrchyan:2013xxa,Abazov:2014fha} and suggest no significant dependence on the center-of-mass energy nor the final state from which the area parameter was measured.

\py{} is used to calculate the fraction of produced events that fall in the fiducial phase space.

\section{Event selection}
\label{sec:selection}
The event selection begins with trigger and data-quality requirements.
Candidate events are preselected by single-, di-, or trilepton triggers~\cite{Aaboud:2016leb}, with a combined efficiency very close to 100\%. They must have at least one primary vertex~\cite{ATL-PHYS-PUB-2015-026} with two or more associated tracks with $\pt > 400$~\MeV{}. Events must satisfy cleaning criteria \cite{ATLAS-CONF-2015-029} designed to reject events with excessive noise in the calorimeters. The data are subjected to quality requirements to reject events in which detector components were not operating correctly.

Following this preselection, muons, electrons, and jets are selected in each event as described below. Based on these, the best lepton quadruplet is selected and required to satisfy further selection criteria.

\subsection{Selection of muons, electrons, and jets}
\label{sec:object_selection}

A muon is reconstructed by matching a track (or track segment) reconstructed in the MS to a track reconstructed in the ID \cite{PERF-2015-10}. Its four-momentum is calculated from the curvature of the track fitted to the combined detector hits in the two systems, correcting for energy deposited in the calorimeters. In regions with limited coverage from the MS ($|\eta| < 0.1$) or outside the ID acceptance ($2.5 < |\eta| < 2.7$), muons can also be reconstructed by matching calorimeter signals consistent with muons to ID tracks (calorimeter-tagged muons) or standalone in the MS, respectively. 
Quality requirements and the loose identification criteria are applied as described in Ref.~\cite{PERF-2015-10}.
Muons are required to have $|\eta| < 2.7$ and $\pt > 5$~\GeV{}. Calorimeter-tagged muons must have $\pt > 15$~\GeV{}.

An electron is reconstructed from an energy deposit (cluster) in the electromagnetic calorimeter matched to a high-quality track in the ID. Its momentum is computed from the cluster energy and the direction of the track and calibrated \cite{ATL-PHYS-PUB-2016-015}. Electrons are required to have $|\eta| < 2.47$ and $\pt > 7$~\GeV{}.

Electrons can be distinguished from other particles using several identification criteria that rely on the shapes of electromagnetic showers as well as tracking and track-to-cluster matching quantities. Following the description in Ref.~\cite{PERF-2016-01}, the output of a likelihood function taking these quantities as input is used to identify electrons, choosing the loose working point.

Leptons are required to originate from the hard-scattering vertex, defined as the reconstructed vertex~\cite{ATL-PHYS-PUB-2015-026} with the largest sum of the $\pta^{2}$ of the associated tracks. The longitudinal impact parameter of each lepton track, calculated with respect to the hard-scattering vertex and multiplied by $\sin\theta$ of the track, is required to be less than 0.5~mm. Furthermore, muons must have a transverse impact parameter calculated with respect to the beam line less than 1~mm in order to reject muons originating from cosmic rays. 
The significance of the transverse impact parameter\footnote{Defined as the absolute measured transverse impact parameter divided by its uncertainty.} calculated with respect to the beam line is required to be less than three (five) for muons (electrons).
Standalone muons are exempt from all three requirements, as they do not have an ID track.

Leptons are required to be isolated from other particles using both ID-track and calorimeter-cluster information.
Muons (electrons) with transverse momentum $\pt$ are removed if the summed transverse momentum of other ID tracks within $\Delta R = \min[0.3, 10~\GeV / \pt]$ ($\min[0.2, 10~\GeV / \pt]$) of the lepton exceeds $0.15\, \pt$, or if the summed transverse energy of other topological clusters~\cite{Aad:2016upy} within $\Delta R = 0.2$ of the lepton exceeds $0.3\, \pt$ ($0.2\, \pt$).

Jets \cite{ATL-PHYS-PUB-2015-036} are clustered from topological clusters in the calorimeters using the anti-$k_t$ algorithm \cite{cacciari08} with radius parameter 0.4. Their energy is calibrated as described in Ref.~\cite{Aaboud:2017jcu}. 
They are required to have $|\eta| < 4.5$ and $\pt > 30$~\GeV{}, as in the fiducial definition. In order to reject jets originating from pileup interactions, they must either pass a jet vertex tagging selection \cite{ATLAS-CONF-2014-018,ATL-PHYS-PUB-2015-034} or have $\pt > 60$~\GeV{}.

In order to avoid the reconstruction of multiple electrons, muons, and/or jets from the same detector signature, all but one such overlapping objects are removed.
Electron candidates sharing an ID track with a selected muon are rejected, except if the muon is only calorimeter-tagged, in which case the muon is rejected instead. Electron candidates sharing their track or calorimeter cluster with a selected higher-\pt{} electron are rejected. Jets within $\Delta R = 0.4$ of a selected lepton are rejected.

\subsection{Quadruplet selection}
As in the fiducial definition (Section~\ref{sec:fiducial}), events must contain at least one quadruplet. All possible quadruplets in a given event are considered for further selection.
At most one muon in each quadruplet may be a calorimeter-tagged or standalone muon. The three highest-$\pt$ leptons in each quadruplet must satisfy $\pt > 20$~\GeV{}, 15~\GeV{}, 10~\GeV{}, respectively.
If multiple selected quadruplets are present, the best quadruplet is chosen as in the fiducial phase-space selection. 
Only the best quadruplet is considered further and the following requirements are imposed on the leptons in that quadruplet.
Any two same-flavor (different-flavor) leptons $\ell_i$, $\ell^{(\prime)}_j$ must be separated by $\Delta R (\ell_i,\ell^{(\prime)}_j) > 0.1$~(0.2). All possible same-flavor opposite-charge dileptons must have an invariant mass greater than $5$~\GeV{}, to reduce background from leptonic hadron decays. 
The two \PZ{} boson candidates, formed as in the fiducial definition, are required to have an invariant mass between 66~\GeV{} and 116~\GeV{}. Figure~\ref{fig:2d_mass_plots} shows the distribution of invariant masses of the $\PZ$ boson candidates in the selected data events. Based on the leptons in the chosen quadruplet, events are classified into the \eeee{}, \mmmm{}, and \eemm{} signal channels.

\begin{figure}[h!]
\centering
\subfigure[]{\includegraphics[width=0.48\textwidth]{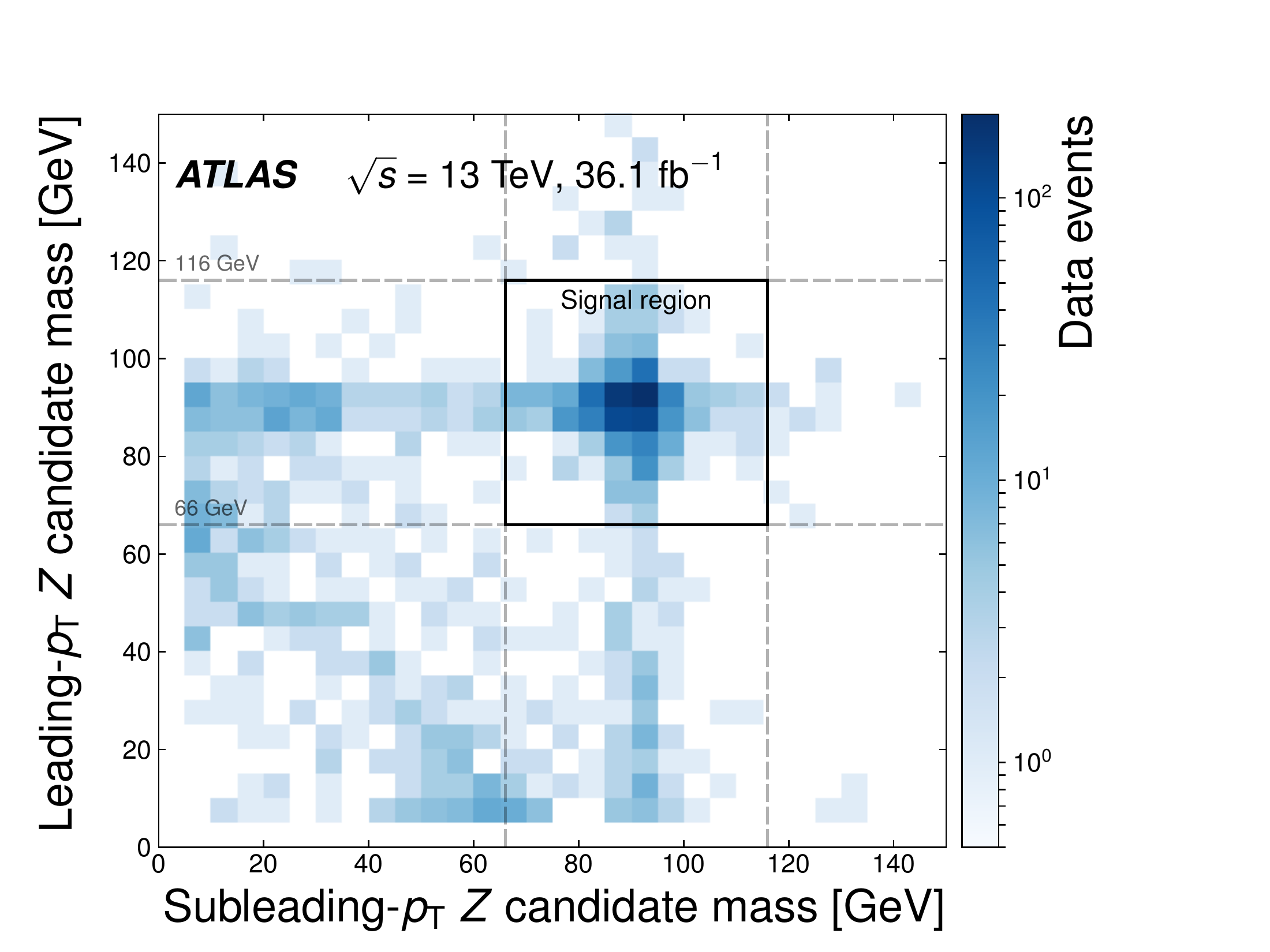}}
\hspace{1mm}
\subfigure[]{\includegraphics[width=0.48\textwidth]{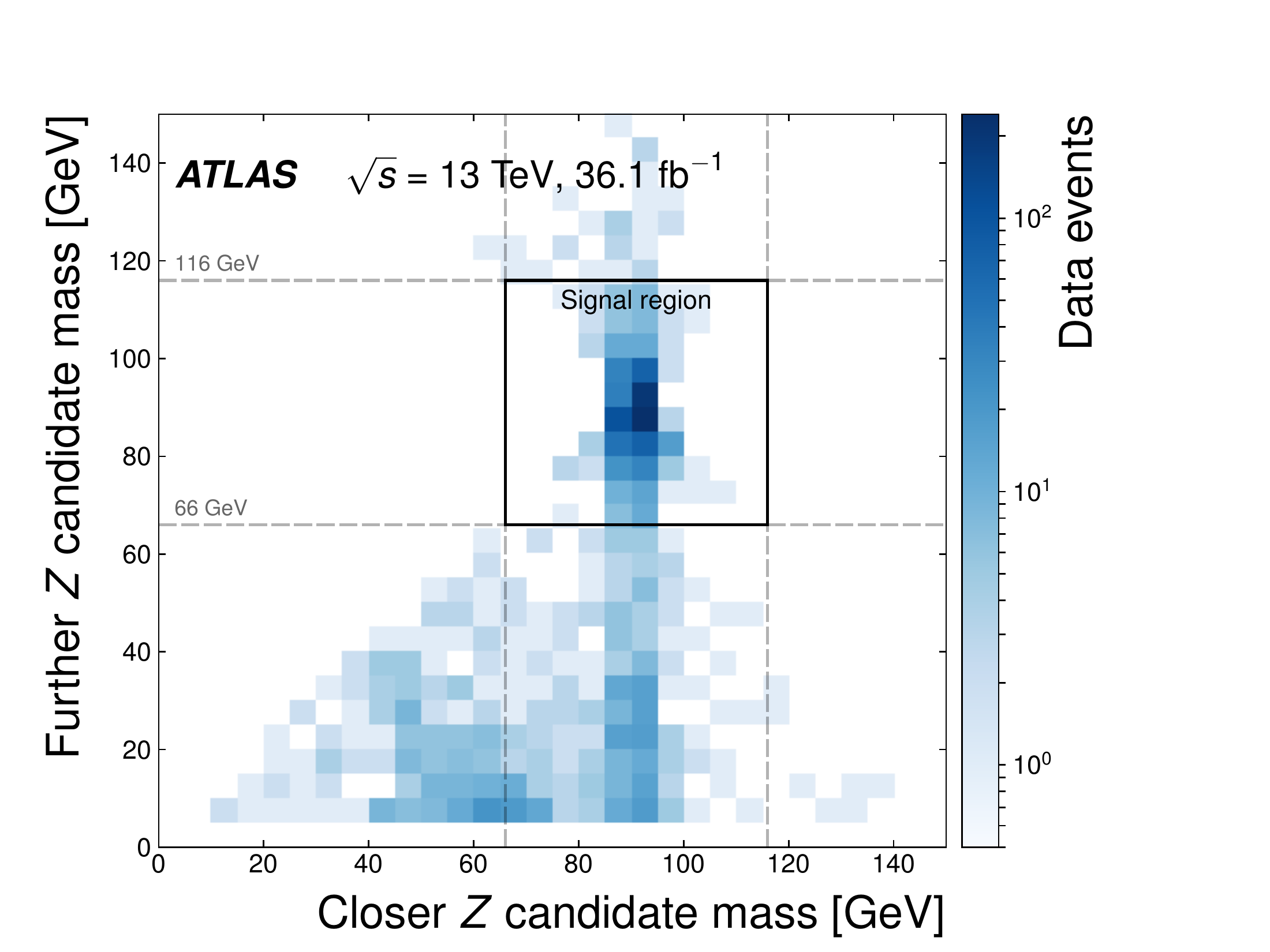}}

\caption{Invariant mass of one selected \PZ{} boson candidate dilepton vs.~the other, in the selected data events before the \PZ{} boson candidate mass requirement. All other selections have been applied. (a) shows the \PZ{} boson candidates arranged by transverse momentum. 
(b) shows the \PZ{} boson candidates arranged by proximity of their mass to the \PZ{} boson pole mass. The solid rectangle shows the signal region. Dashed gray lines mark the \PZ{} boson candidate mass requirements for each pair, 66~\GeV{} to 116~\GeV{}. Only data are shown.}
\label{fig:2d_mass_plots}
\end{figure}

\section{Background estimation}
\label{sec:bkg}
The expected total background is very small, approximately 2\% of the total predicted yield in each decay channel. 

Irreducible backgrounds from processes with at least four prompt leptons in the final state are estimated with the simulated samples described in Section~\ref{sec:mc}, including uncertainties from the cross-section predictions, luminosity measurement, and experimental effects, described in Section~\ref{sec:uncertainties}. Non-hadronic triboson processes (15\% of the total background estimate) and $\Ptop\APtop\kern1pt\PZ$ processes with leptonic $\PW/\PZ$ boson decays (19\%) are considered. Simulated samples are also used to estimate the background from \ZZ{} processes where at least one \Z{} boson decays to $\tau$ leptons (8\%), which is not an irreducible background as defined in Section~\ref{sec:signal_process_definition}.

Events from processes with two or three prompt leptons, e.g. \PZ{}, \WW{}, \WZ{}, $\Ptop\APtop$, and \ZZ{} events where one \PZ{} boson decays hadronically, can pass the event selection if associated jets, non-prompt leptons, or photons are misidentified as prompt leptons.
This background is estimated using a data-driven technique as follows. A lepton selection that is orthogonal to the nominal selection in Section~\ref{sec:object_selection} is defined by reversing some of its requirements. Muons must fail the transverse impact parameter requirement or the isolation requirement, or both. Electrons must fail either the isolation requirement or the likelihood-based identification, but not both.
A high-purity data sample of events containing a $\PZ$ boson candidate decaying to a pair of electrons or muons is selected. The leptons forming the \PZ{} candidate must pass tight selection criteria, different from those used anywhere else in this analysis. Any additional reconstructed leptons in this sample are assumed to be misidentified, after the approximately~4\% contamination from genuine third leptons from $\PW\PZ$ and $\PZ\PZ$ production has been subtracted using MC simulation. Using the observed rates of third leptons passing the nominal or the reversed selection, $n_l$ and $n_r$, transfer factors $f$ are defined as

\vspace{-1\baselineskip}
\begin{equation*}\label{eq:transfer_factor}
	f = \frac{n_l}{n_r}
\end{equation*}
and measured in bins of $\pt$ and $\eta$ of the third leptons.
A background control sample of data events is then selected, satisfying all the $\PZ\PZ$ selection criteria described in Section~\ref{sec:selection}, except that one or two leptons in the final selected quadruplet are required to only satisfy the reversed criteria and not the nominal criteria. The number of observed events with one lepton (two leptons) satisfying only the reversed criteria is denoted $N_{lllr}$ ($N_{llrr}$). The events originate predominantly from processes with two or three prompt leptons. Using MC simulation, the contamination from genuine $\PZ\PZ$ events is estimated to be approximately~36\% of $N_{lllr}$ and approximately~1\% of $N_{llrr}$. The number of background events with one or two misidentified leptons can be calculated as

\vspace{-1\baselineskip}
\begin{equation}\label{eq:ddbkg}
	N_{\text{misid.}} = \sum_{i}^{N_{lllr}} f_i - \sum_{i}^{N_{lllr}^{\PZ\PZ}} w_i f_i - \sum_{i}^{N_{llrr}} f_i f^{\prime}_i + \sum_{i}^{N_{llrr}^{\PZ\PZ}} w_i f_i f^{\prime}_i,
\end{equation}
where the superscript $\PZ\PZ$ indicates the MC-simulated contributing events from $\PZ\PZ$ production, $w_i$ indicates the simulated weight of the $i$th event,\footnote{The simulated weights are products of cross-section weights of the generated events and factors correcting for differences in selection efficiencies between simulation and data.} and $f_i$ and $f^{\prime}_i$ are the transfer factors depending on $\pt$ and $\eta$ of the leptons passing the reversed selection. 
In differential distributions, the yields in Eq.~\eqref{eq:ddbkg} are considered separately in each bin.
Systematic uncertainties are applied to account for statistical fluctuations of the measured transfer factors, and for the simplification that the origins, fractions and selection efficiencies of misidentified leptons are assumed equal in the sample where the transfer factors are determined and in the background control sample. The latter uncertainties are estimated using transfer factors obtained from simulation of the different background processes and taking the difference between the result and the nominal method as the uncertainty. An additional uncertainty due to the modeling of the $\PZ\PZ$ contamination in the background control sample is estimated by varying $N_{lllr}^{\PZ\PZ}$ and $N_{llrr}^{\PZ\PZ}$ up and down by 50\%. The final total uncertainty is $100\%$ ($71\%$, $95\%$) in the \eeee{} (\eemm{}, \mmmm{}) channel. The misidentified-lepton background amounts to 58\% of the total background estimate. As a cross-check, the background is also estimated using an independent method in which $\PZ\PZ$ events with one same-flavor same-charge lepton pair as one of the \PZ{} boson candidates are selected. The results are found to agree well with the nominal method, differing by less than one standard deviation in all channels.

Background from two single \PZ{} bosons produced in different $\Pproton\Pproton$ collisions in the same bunch crossing is estimated by considering the \PZ{} boson production cross sections and the probability of the two primary vertices lying so close to each other that the detector cannot resolve them as separate vertices. It is found to be negligible ($< 0.1\%$ of the total signal prediction).

The observed and predicted event yields for signal and background are shown in Table~\ref{tab:yields}. The prediction uncertainties are discussed in Section~\ref{sec:uncertainties}. Figure~\ref{fig:data_mc_plots} shows the distributions of data and predictions for the mass and transverse momentum of the four-lepton system, the transverse momentum of the leading \PZ{} boson candidate, and the jet multiplicity. The data and the nominal \SHERPA{} prediction agree well. The prediction using \POWHEGpy{} to simulate the $\Pquark\APquark$-initiated process tends to underestimate the normalization slightly, which can be understood from its lack of higher-order real-emission corrections that \SHERPA{} implements. \POWHEGpy{} also provides a worse description of high jet multiplicities, as it only describes one parton emission at matrix-element level.

\begin{table}[h!]
\centering
{\small
\begin{tabular}{lllll}
\toprule
\textbf{Contribution} & \textbf{\eeee{}} & \textbf{\eemm{}} & \textbf{\mmmm{}} & \textbf{Combined}\\
\midrule
Data & 249 & 465 & 303 & 1017\\
\midrule
Total prediction (\SHERPA{}) & $198\phantom{.0\pm}^{+16}_{-14}$ & $469\phantom{.0\pm}^{+35}_{-31}$ & $290\phantom{.0\pm}^{+22}_{-21}$ & \phantom{0}$958\phantom{.0\pm}^{+70}_{-63}$\\
\midrule

Signal ($\Pquark\APquark$-initiated) & $168\phantom{.0\pm}^{+14}_{-13}$ & $400\phantom{.0\pm}^{+31}_{-28}$ & $246\phantom{.0\pm}^{+19}_{-18}$ & \phantom{0}$814\phantom{.0\pm}^{+63}_{-57}$\\
Signal ($\Pgluon\Pgluon$-initiated) & $\phantom{0}21.3 \phantom{0}\pm 3.5$ & $\phantom{0}50.2\phantom{0}\pm 8.2$ & $\phantom{0}29.7\phantom{0}\pm 4.9$ & \phantom{0}$101\phantom{.00}\pm 17$\\
Signal (EW-$\PZ\PZ jj$) & $\phantom{00}4.36\pm 0.42$ & $\phantom{0}10.23\pm 0.72$ & $\phantom{00}6.43\pm 0.55$ & \phantom{0}$\phantom{0}21.0\phantom{0}\pm1.2$\\
$\PZ\PZ\to \tau^{+}\tau^{-}[\ell^{+}\ell^{-}, \tau^{+}\tau^{-}]$ & $\phantom{00}0.59 \pm 0.09$ & $\phantom{00}0.55 \pm 0.08$ & $\phantom{00}0.55 \pm 0.09$ & \phantom{0}$\phantom{00}1.69 \pm 0.16$\\
Triboson & $\phantom{00}0.68 \pm 0.21$ & $\phantom{00}1.50 \pm 0.46$ & $\phantom{00}0.96 \pm 0.30$ & \phantom{0}$\phantom{00}3.14 \pm 0.30$\\
$\Ptop\APtop\kern1pt\PZ$ & $\phantom{00}0.81 \pm 0.25$ & $\phantom{00}1.86 \pm 0.56$ & $\phantom{00}1.42 \pm 0.43$ & \phantom{0}$\phantom{00}4.1\phantom{0} \pm 1.2$\\
Misid.~lepton background & $\phantom{00}2.1\phantom{0}\pm 2.1$ & $\phantom{00}4.9\phantom{0}\pm 3.9$ & $\phantom{00}5.3\phantom{0}\pm 5.2$ & \phantom{0}$\phantom{0}12.3\phantom{0}\pm 8.3$\\
\midrule
Total prediction (\matrixnnlo{} + corrections) & $197\phantom{.0\pm}^{+15}_{-14}$ & $470\phantom{.0\pm}^{+34}_{-31}$ & $286\phantom{.0\pm}^{+22}_{-21}$ & $\phantom{0}953\phantom{.0\pm}^{+69}_{-64}$\\ 
Total prediction (\POWHEGpy{} & $193 \phantom{.00}\pm 11$ & $456 \phantom{.00}\pm 24$ & $286 \phantom{.00}\pm 17$ & \phantom{0}$934 \phantom{.00}\pm 50$\\
with higher-order corrections, \SHERPA{}) & \\
\bottomrule
\end{tabular}
}
\caption{Observed and predicted yields, using the nominal \SHERPA{} setup for the signal predictions. All statistical and systematic uncertainties are included in the prediction uncertainties. An alternative total prediction is given, using \SHERPA{} reweighted to the total NNLO prediction from \matrixnnlo{} with NLO EW corrections, adding the contribution of the EW-$\ZZ jj$ process generated with \SHERPA{}, to predict the signal yield. A second alternative total prediction, identical to the nominal \SHERPA{} setup, except using \POWHEGpy{} with NNLO QCD and NLO EW corrections applied event by event to simulate the $\Pquark\APquark$-initiated process, is shown at the bottom.}
\label{tab:yields}
\end{table}

\begin{figure}[p]
\centering
\subfigure[]{\includegraphics[width=0.48\textwidth]{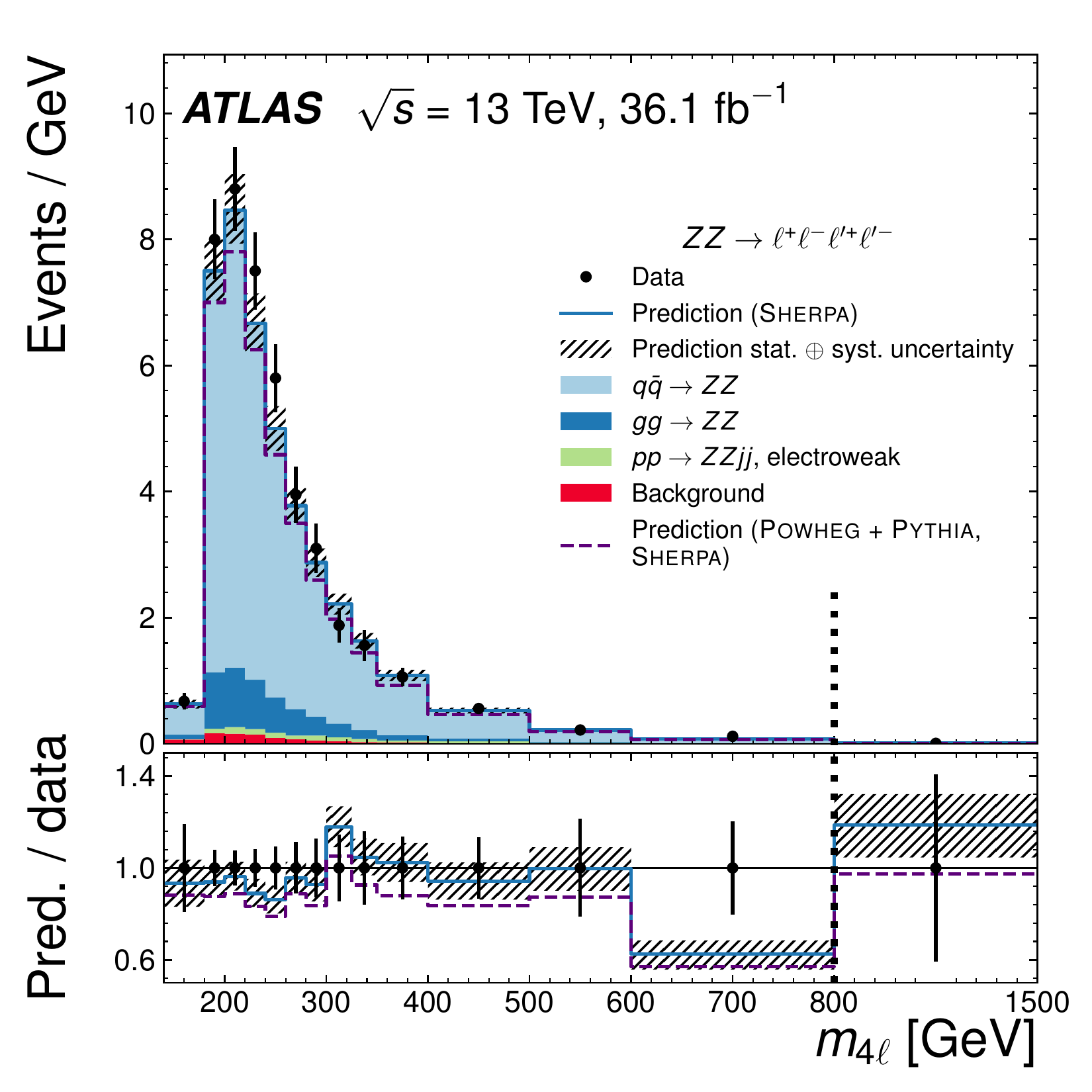}}
\hspace{1mm}
\subfigure[]{\includegraphics[width=0.48\textwidth]{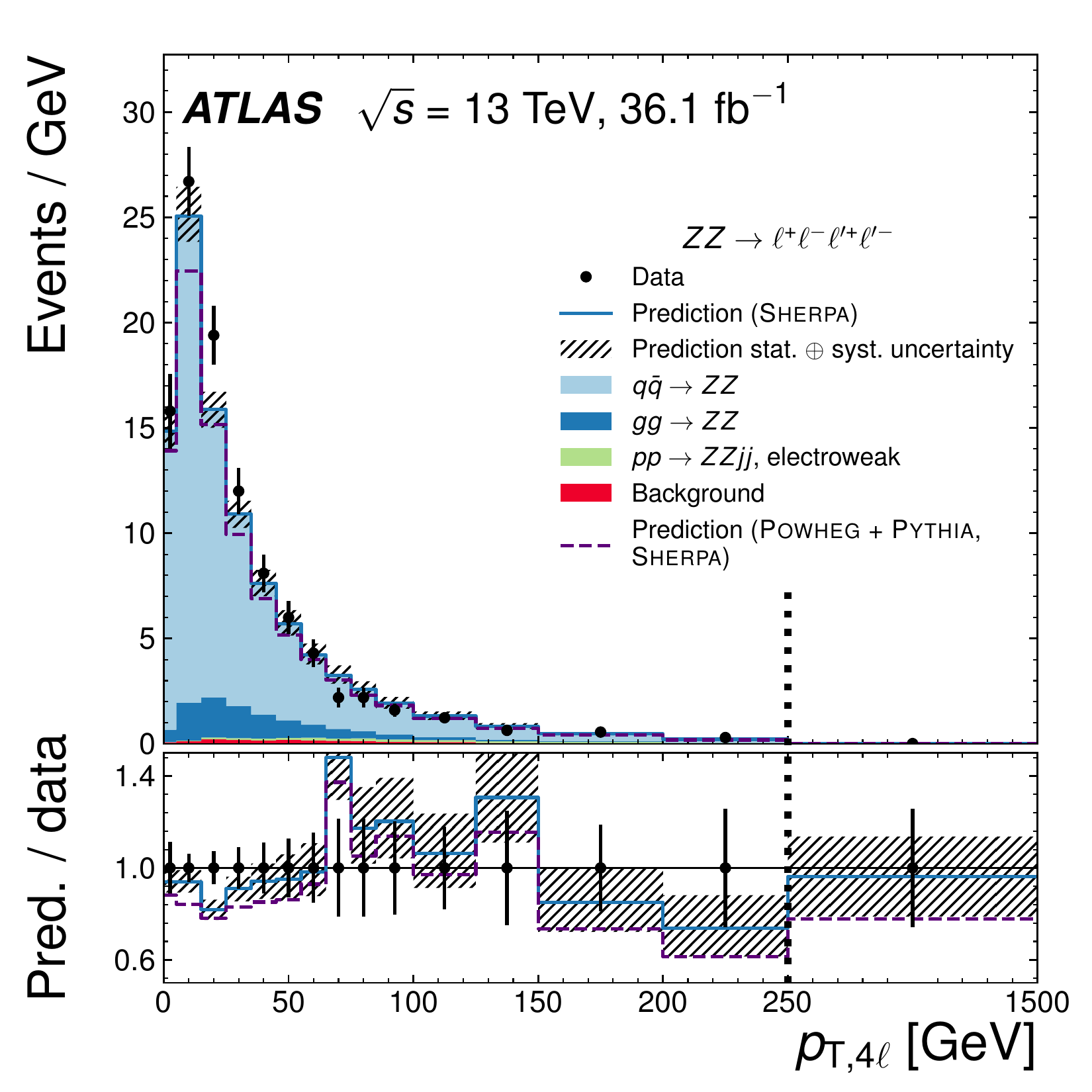}}
\subfigure[]{\includegraphics[width=0.48\textwidth]{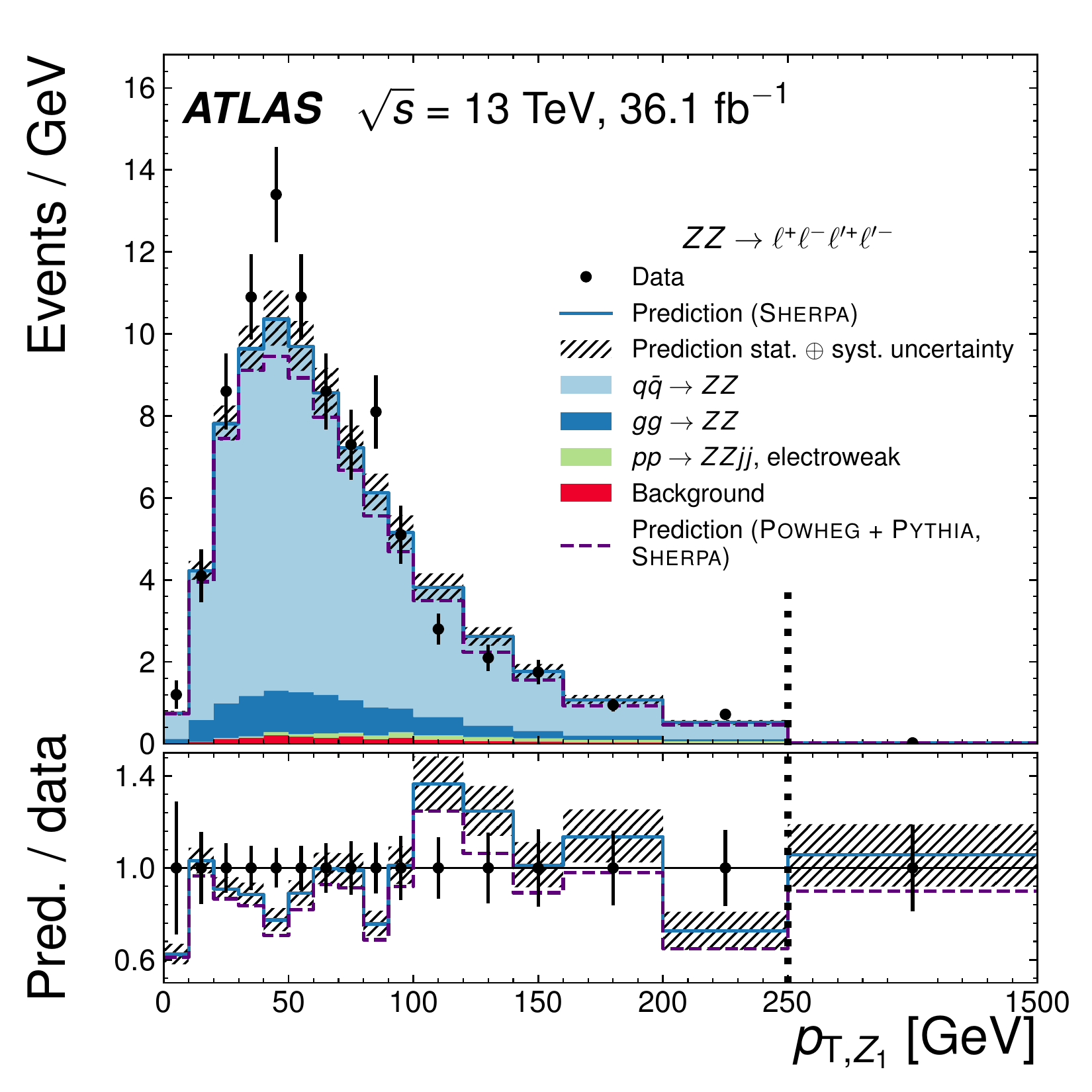}}
\hspace{1mm}
\subfigure[]{\includegraphics[width=0.48\textwidth]{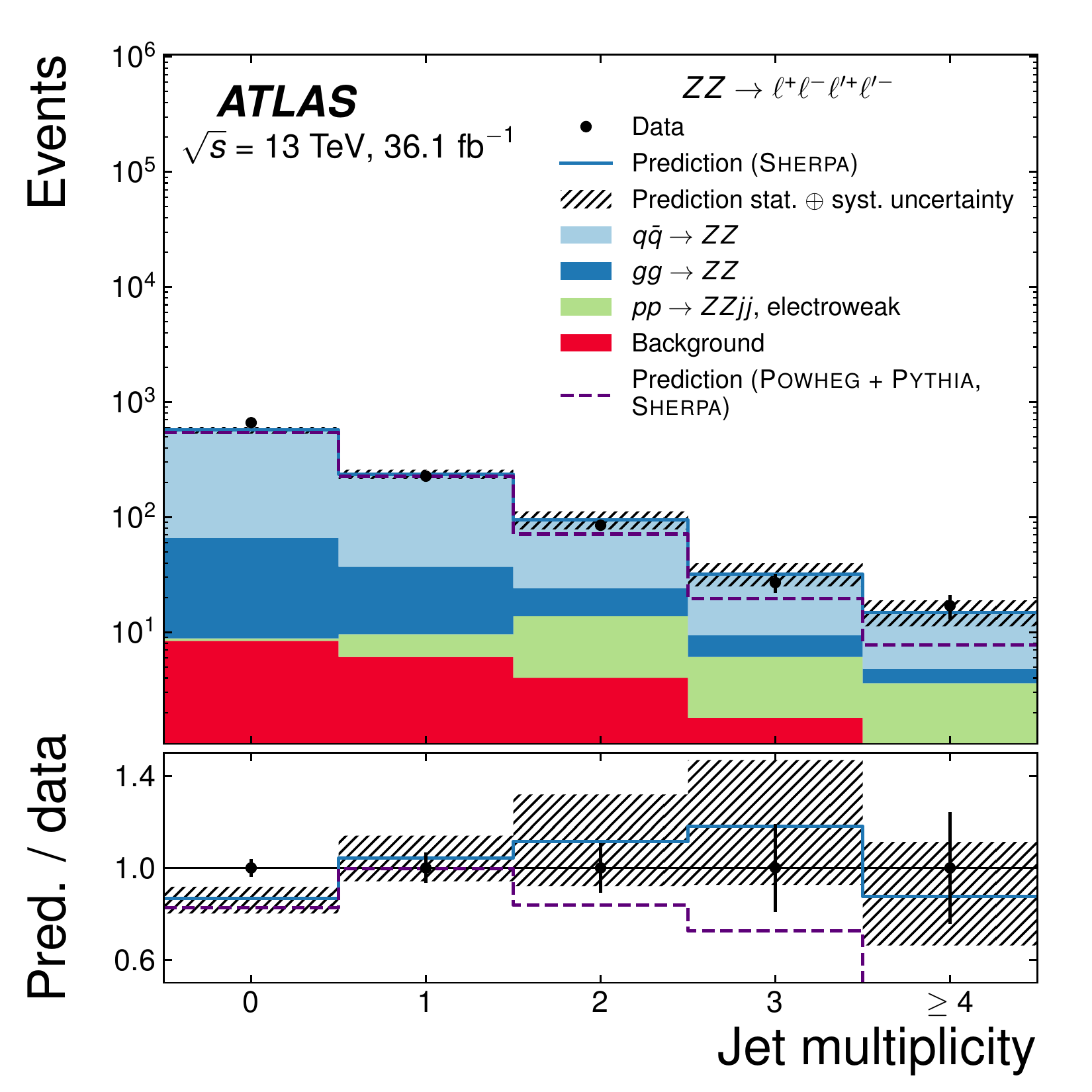}}
\caption{Measured distributions of the selected data events along with predictions in bins of (a) the four-lepton mass, (b) the four-lepton transverse momentum, (c) the transverse momentum of the leading \PZ{} boson candidate, and (d) the multiplicity of jets selected according to the least restrictive criteria used in this analysis ($|\eta| < 4.5$ and $\pt > 30$~\GeV{}). The main prediction uses the nominal \SHERPA{} setup. The prediction uncertainty includes the statistical and systematic components, all summed in quadrature. Different signal contributions and the background are shown, as is an alternative prediction that uses \POWHEGpy{} to generate the $\Pquark\APquark$-initiated subprocess. In (a), (b), and (c), the last bin is shown using a different $x$-axis scale for better visualization. The scale change is indicated by the dashed vertical line.}
\label{fig:data_mc_plots}
\end{figure}

\clearpage
\section{Systematic uncertainties}
\label{sec:uncertainties}
The sources of systematic uncertainty are introduced below. Their effects on the predicted integrated signal yields after event selection are shown in Table~\ref{tab:yield_uncertainties}.

For leptons and jets, uncertainties of the momentum or energy scale and resolution are considered \cite{ATL-PHYS-PUB-2016-015,PERF-2015-10,Aaboud:2017jcu}. Uncertainties of the lepton reconstruction and identification efficiencies \cite{PERF-2016-01,PERF-2015-10} as well as the efficiency of the jet vertex tagging requirements \cite{ATLAS-CONF-2014-018,ATL-PHYS-PUB-2015-034} in the simulation are taken into account. All of the above depend on the \pt{} and $\eta$ of the lepton or jet. The electron efficiency uncertainties contain contributions associated with the basic reconstruction, the identification, and the isolation. In addition to correlated components, each is split into $\mathcal{O}(10)$ uncorrelated components to take into account the partial decorrelation between individual electrons in different $\eta$--$\pt$ regions.
For muons, the efficiency uncertainties associated with individual muons are treated as fully correlated. This leads to a larger uncertainty for muons than for electrons.
As the selection is fully jet-inclusive, jet uncertainties do not affect the integrated yields and are therefore not shown in Table~\ref{tab:yield_uncertainties}.

The pileup modeling uncertainty is assessed by varying the number of simulated pileup interactions. The variations are designed to cover the uncertainty of the ratio of the predicted and measured cross section of non-diffractive inelastic events producing a hadronic system of mass $m_X > 13$~\GeV{} \cite{STDM-2015-05}. 
The uncertainty of the integrated luminosity is $3.2\%$. It is derived from a preliminary calibration of the luminosity scale using a pair of $x$--$y$ beam-separation scans performed in August 2015 and May 2016, following a methodology similar to that detailed in Ref.~\cite{DAPR-2011-01}.
QCD scale uncertainties of predicted cross sections are evaluated by varying the factorization scale $\mu_{\text{f}}$ and renormalization scale $\mu_{\text{r}}$ up and down independently by a factor of two, but ignoring the extreme variations ($2 \mu_{\text{f}}$, $0.5 \mu_{\text{r}}$) and ($0.5 \mu_{\text{f}}$, $2 \mu_{\text{r}}$), and taking the largest deviations from the nominal value as the systematic uncertainties.
PDF uncertainties of predicted cross sections are evaluated considering the uncertainty of the used set, as well as by comparing to two other reference sets \cite{Butterworth:2015oua}. The reference sets are MMHT 2014 \cite{Harland-Lang:2014zoa} and NNPDF 3.0 (CT10), if CT10 (NNPDF 3.0) is the nominal set. The envelope of the nominal set's uncertainty band and the deviation of the reference sets from the nominal set is used as the uncertainty estimate.
The theoretical uncertainties due to PDFs and QCD scales along with the luminosity uncertainty dominate the total uncertainty of the integrated yields, as shown in Table~\ref{tab:yield_uncertainties}.
A predicted theoretical modeling uncertainty is taken into account in the unfolding of differential cross sections. It is estimated by using \POWHEGpy{} instead of \SHERPA{} to generate the $\Pquark\APquark$-initiated subprocess, and taking the absolute deviation of the result obtained with this setup from the one obtained with the nominal \SHERPA{} setup as an uncertainty, symmetrizing it with respect to the nominal value. This contribution is not shown in Table~\ref{tab:yield_uncertainties}, because it is never applied to yields, where it would be dominated by cross-section normalization differences rather than differences in the reconstruction efficiencies. A further source of uncertainty are statistical fluctuations in the MC samples. The associated uncertainty of the measured differential cross sections is $< 1\%$ in most bins, reaching $3\%$ in rare cases.

In the search for aTGCs, an additional uncertainty due to the factorization approximation of NNLO QCD and NLO EW corrections is applied as follows. Following a criterion motivated in Ref.~\cite{Gieseke:2014gka}, events are classified as having high QCD activity if $\left|\sum_{i} \vec{p}_{\mathrm{T},\,i}\right| > 0.3 \sum_{i} |\vec{p}_{\mathrm{T},\,i}|$, where the sums are over fiducial leptons. In events with high QCD activity, the NLO EW $K$-factors are in turn not applied and applied with doubled deviation from unity, as $1 + 2 (\text{$K$-factor} - 1)$. The deviations from the nominal result are taken as uncertainties.

The uncertainty of the misidentified-lepton background is described in Section~\ref{sec:bkg}. A 30\% normalization uncertainty is applied for triboson and $\Ptop\APtop\kern1pt\PZ$ backgrounds with four genuine leptons to account for the cross-section uncertainty. The background uncertainties are considered uncorrelated with other sources.

The propagation of uncertainties in the unfolding as well as the estimation of unfolding-specific uncertainties is described in Section~\ref{sec:unfolding}.

\begin{table}[h!]
\centering
\begin{tabular}{lr}
\toprule
\textbf{Source} &   \textbf{Effect on total predicted yield [\%]}\\
\midrule
MC statistical uncertainty    & $0.4$\\ 
Electron efficiency            & $0.9$\\
Electron energy scale \& resolution   & $< 0.1$\\
Muon efficiency                & $1.7$\\
Muon momentum scale \& resolution       & $< 0.1$\\
Pileup modeling           & $1.2$\\
Luminosity				& $3.2$\\
QCD scales           & $^{+5.2}_{-4.7}$\\
PDFs                      & $^{+2.7}_{-1.7}$\\
Background prediction & $0.9$\\
\midrule
Total                          & $^{+7.4}_{-6.6}$\\
\bottomrule
\end{tabular}
\caption{Relative uncertainties in percent of the predicted integrated signal yields after event selection, derived using the nominal \SHERPA{} setup. All uncertainties are rounded to one decimal place.}
\label{tab:yield_uncertainties}
\end{table}

\section{Integrated cross section}

The integrated fiducial cross section \sigmafid{} is determined by a maximum-likelihood fit in each channel separately as well as for all channels combined. The expected yield in each channel $i$ is given by

\vspace{-1\baselineskip}
\begin{equation*}
	N_{\text{exp}}^i = \intL \czz^i \sigmafid^i + N_{\text{bkg}}^i
\end{equation*}
where $\intL$ is the integrated luminosity, and $N_{\text{bkg}}$ is the expected background yield. The factor \czz{} is applied to correct for detector inefficiencies and resolution effects. It relates the background-subtracted number of selected events to the number in the fiducial phase space. \czz{} is defined as the ratio of generated signal events satisfying the selection criteria using reconstructed objects to the
number satisfying the fiducial criteria using the particle-level objects defined in Section~\ref{sec:fiducial}. It is determined with the nominal \sherpa{} setup. 
The \czz{} value and its total uncertainty is determined to be $0.494 \pm 0.015$ ($0.604 \pm 0.017$, $0.710 \pm 0.027$) in the \eeee{} (\eemm{}, \mmmm{}) channel. The dominant systematic uncertainties come from the uncertainties of the lepton reconstruction and identification efficiencies in the simulation, the choice of MC event generator, QCD scales and PDFs, and the modeling of pileup effects. Other smaller uncertainties come from the scale and resolution of the lepton momenta as well as statistical fluctuations in the MC sample. Table~\ref{tab:czz_uncert_overview} gives a breakdown of the systematic uncertainties.

\begin{table}[h!]
\centering
\begin{tabular}{lrrr}
\toprule
\textbf{Source} &   \textbf{\eeee{}} & \textbf{\eemm{}} & \textbf{\mmmm{}} \\
\midrule
MC statistical uncertainty    & $0.4$ & 0.2 & 0.1 \\
Electron efficiency            & $2.0$ & 1.0 & -- \\
Electron energy scale \& resolution   & $0.1$ & $< 0.1$ & -- \\
Muon efficiency                & -- & 1.6 & 3.2 \\
Muon momentum scale \& resolution       & -- & $< 0.1$ & $0.1$ \\
Pileup modeling           & $1.3$ & 0.8 & 2.0 \\
QCD scales \& PDFs           & $^{+0.4}_{-0.8}$ & $^{+0.3}_{-0.4}$ & $^{+0.3}_{-0.6}$ \\
Event generator                      & $1.8$ & 1.8 & $0.2$ \\
\midrule
Total                          & $3.1$ & $2.8$ & $3.8$ \\
\bottomrule
\end{tabular}
\caption{Relative uncertainties of the correction factor \czz{} by channel, given in percent. All uncertainties are rounded to one decimal place. Uncertainties that do not apply in a given channel are marked with a dash (--). They are either exactly zero or very close to zero.}
\label{tab:czz_uncert_overview}
\end{table}

The likelihood function to be minimized in the cross-section fit is defined as

\vspace{-1\baselineskip}
\begin{equation}\label{eq:likelihood}
	\mathcal{L} = \mathcal{L}_{\text{stat}}\; \mathcal{L}_{\text{corr}}\; \mathcal{L}_{\text{uncorr}},
\end{equation}
where

\vspace{-1\baselineskip}
\begin{equation*}
	\mathcal{L}_{\text{stat}} = \text{Poisson}(N_{\text{obs}} | N_{\text{exp}})
\end{equation*}
is the probability of observing $N_{\text{obs}}$ events given that the yield follows a Poisson distribution with mean $N_{\text{exp}}$, and $\mathcal{L}_{\text{corr}}$ and $\mathcal{L}_{\text{uncorr}}$ are products of Gaussian nuisance parameters corresponding to the uncertainties of $\intL$, \czz{}, and $N_{\text{bkg}}$. The term $\mathcal{L}_{\text{corr}}$ contains the nuisance parameters that are fully correlated between channels, i.e.~all except the statistical uncertainties, while $\mathcal{L}_{\text{uncorr}}$ contains those that are uncorrelated, i.e.~the statistical uncertainties of \czz{} and $N_{\text{bkg}}$ in each channel. Nuisance parameters corresponding to different sources of systematic uncertainty are considered uncorrelated.
In the combined cross-section fit, the product over channels is taken in the likelihood function shown in Eq.~\eqref{eq:likelihood}, fixing the relative contributions of the signal channels to their theoretically predicted values.

Table~\ref{tab:integrated_cross_sections} shows the integrated fiducial cross sections for each channel as well as all channels combined, along with a theoretical prediction. Measurements and predictions agree within approximately one standard deviation, except for the \eeee{} channel, where they agree within approximately 2.5 standard deviations. The sum of the \eeee{} and \mmmm{} cross sections is not equal to the \eemm{} cross section. This is because of interference in the \eeee{} and \mmmm{} channels, the bias caused by the pairing prescription in the fiducial definition, as well as other small differences in the fiducial selection (different $\Delta R(\ell_i\ell^{(\prime)}_j)$ requirement, $m_{\ell\ell}$ > 5~\GeV{} for any same-flavor opposite-charge pair). Figure~\ref{fig:fiducial_cross_sections} shows the ratio of measured over predicted cross sections. The goodness of the combined cross-section fit is assessed, taking as hypothesis that the relative contributions of the channels are as predicted. This assumes lepton universality in $\PZ \to \ell^+ \ell^-$, which is experimentally confirmed to high precision \cite{ALEPH:2005ab,Aaboud:2016btc}. Using the maximum likelihood for the observed yields, $\mathcal{L}_{\text{obs}}$, and for the expected yields, $\mathcal{L}_{\text{exp}}$, the ratio $-2 \ln(\mathcal{L}_{\text{obs}} / \mathcal{L}_{\text{exp}})$ is found to be 8.7. The $p$-value is calculated as the fraction of $10^5$ MC pseudoexperiments giving a larger ratio than the fit to data, and is found to be $2.3\%$. This relatively low $p$-value is driven by the compatibility of the $\eeee$ channel with the other two channels.

\begin{table}[h!]
\centering
\begin{tabular}{lll}
\toprule
\textbf{Channel}			& \textbf{Measurement [fb]} & \textbf{Prediction [fb]}\\
\midrule
$\eeee$	& $13.7^{+1.1}_{-1.0}$ $\big[\pm 0.9$ (stat.) $\pm 0.4$        (syst.) $^{+0.5}_{-0.4}$ (lumi.)$\big]$			& $10.9^{+0.5}_{-0.4}$\\
$\eemm$	& $20.9^{+1.4}_{-1.3}$ $\big[\pm 1.0$ (stat.) $\pm 0.6$        (syst.) $^{+0.7}_{-0.6}$ (lumi.)$\big]$	        & $21.2^{+0.9}_{-0.8}$\\
$\mmmm$	& $11.5^{+0.9}_{-0.9}$ $\big[\pm 0.7$ (stat.) $\pm 0.4$        (syst.) $\pm 0.4$        (lumi.)$\big]$			& $10.9^{+0.5}_{-0.4}$\\
\midrule
Combined & $46.2^{+2.5}_{-2.3}$ $\big[\pm 1.5$ (stat.) $^{+1.2}_{-1.1}$(syst.) $^{+1.6}_{-1.4}$ (lumi.)$\big]$		    & $42.9^{+1.9}_{-1.5}$\\
\bottomrule
\end{tabular} 
\caption{Measured and predicted integrated fiducial cross sections. The prediction is based on an NNLO calculation from \matrixnnlo{} \cite{Grazzini:2015hta} with the $\Pgluon\Pgluon$-initiated contribution multiplied by a global NLO correction factor of $1.67$ \cite{Caola:2015psa}. A global NLO EW correction factor of $0.95$ \cite{Biedermann:2016lvg,Biedermann:2016yvs} is applied, except to the $\Pgluon\Pgluon$-initiated loop-induced contribution, and the contribution of around 2.5\% from EW-$\ZZ jj$ generated with \sherpa{} is added. For the prediction, the QCD scale uncertainty is shown.
}
\label{tab:integrated_cross_sections}
\end{table}

\begin{figure}[h!]
\centering
\includegraphics[width=0.68\textwidth]{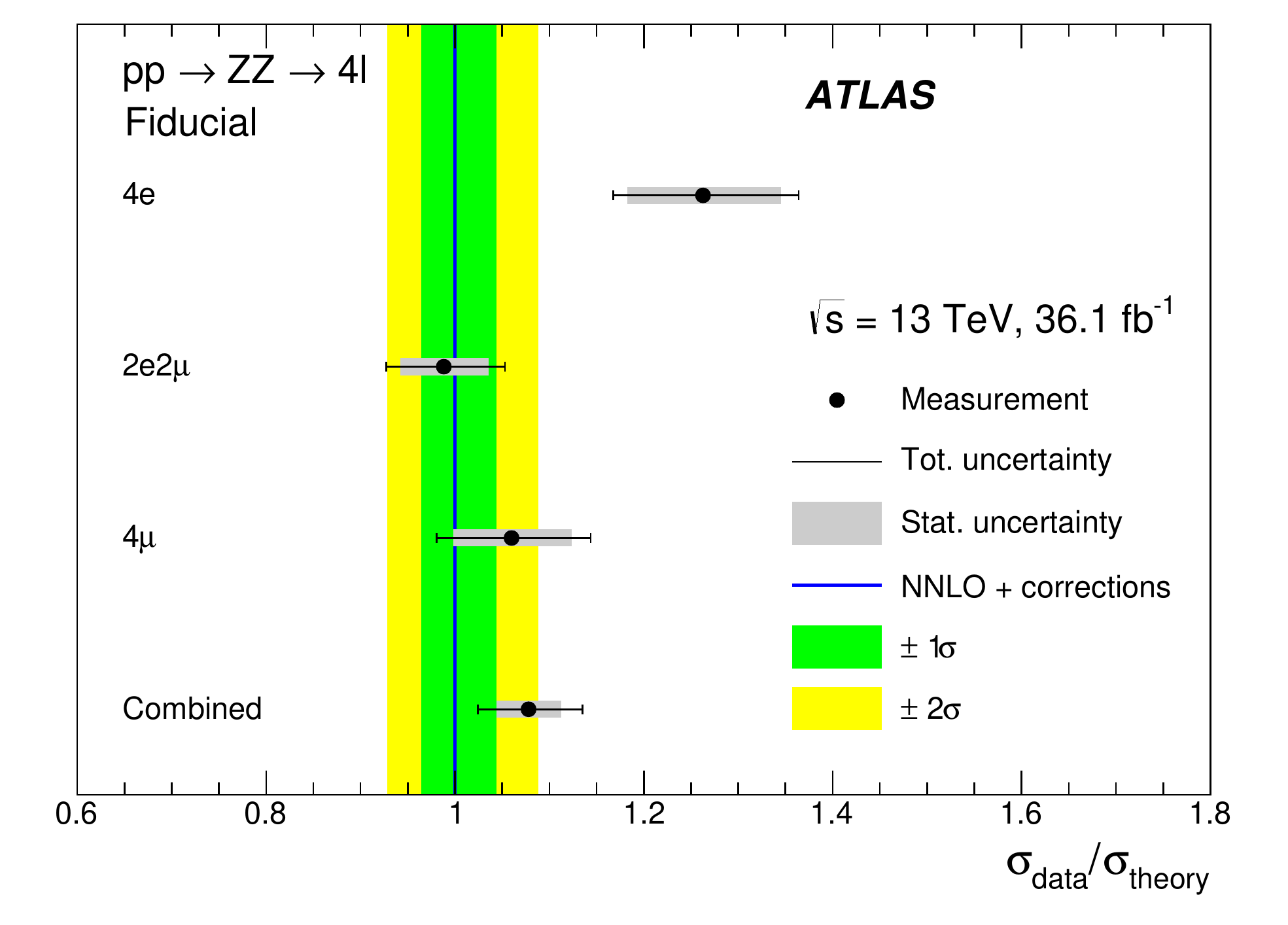}
\caption{Comparison of measured integrated fiducial cross sections to a SM prediction based on an NNLO calculation from \matrixnnlo{} with the $\Pgluon\Pgluon$-initiated contribution multiplied by a global NLO correction factor of $1.67$. A global NLO EW correction factor of $0.95$ is applied, except to the $\Pgluon\Pgluon$-initiated loop-induced contribution, and the contribution of around 2.5\% from EW-$\ZZ jj$ generated with \sherpa{} is added. For the prediction, the QCD scale uncertainty is shown as one- and two-standard-deviation bands.}
\label{fig:fiducial_cross_sections}
\end{figure}

\subsection{Extrapolation to total phase space and all $\PZ$ boson decay modes}
Extrapolation of the cross section is performed to a total phase space for \PZ{} bosons with masses in the range from 66~\GeV{} to 116~\GeV{} and any SM decay. The total phase space is the same as the fiducial phase space (Section~\ref{sec:fiducial}), except that no $\pt$, $\eta$, and $\Delta R$ requirements are applied to the leptons. The ratio of the fiducial to total phase-space cross section is determined using the \matrixnnlo{} setup described in Section~\ref{sec:mc} and found to be $\azz = 0.58\pm 0.01$, 
where the uncertainty includes the following contributions. A similar value is found when the calculation is repeated with the nominal \sherpa{} setup, and the difference between these (1.0\% of the nominal value) is included in the uncertainty of \azz{}. Other included uncertainties are derived from PDF variations (0.4\%, calculated with \MCFM{}) and QCD scale variations (0.8\%).

To calculate the extrapolated cross section, the combined fiducial cross section is divided by \azz{} and by the leptonic branching fraction $4 \times (3.3658\%)^{2}$ \cite{Olive:2016xmw}, where the factor of four accounts for the different flavor combinations of the decays. The result is obtained using the same maximum-likelihood method as for the combined fiducial cross section, but now including the uncertainties of \azz{} as additional nuisance parameters. The used leptonic branching fraction value excludes virtual-photon contributions. Based on a calculation with \PYTHIA{}, including these would increase the branching fraction $\ZZllll$ by about 1--2\%. 

The extrapolated cross section is found to be \extrapolatedxsectshort{}~$[$\extrapolatedxsectstatuncert{}~(stat.)\;\extrapolatedxsectsystuncert{}~(syst.)\;\extrapolatedxsectlumiuncert{}~(lumi.)$]$~pb. The NNLO prediction from \matrixnnlo{}, with the $\Pgluon\Pgluon$-initiated process multiplied by a global NLO correction factor of $1.67$ \cite{Caola:2015psa} is $16.9^{+0.6}_{-0.5}$~pb, where the uncertainty is estimated by performing QCD scale variations. A comparison of the extrapolated cross section to the NNLO prediction as well as to previous measurements is shown in Figure~\ref{fig:extrapolated_cross_sections}.

\begin{figure}[h!]
\centering
\includegraphics[width=0.68\textwidth]{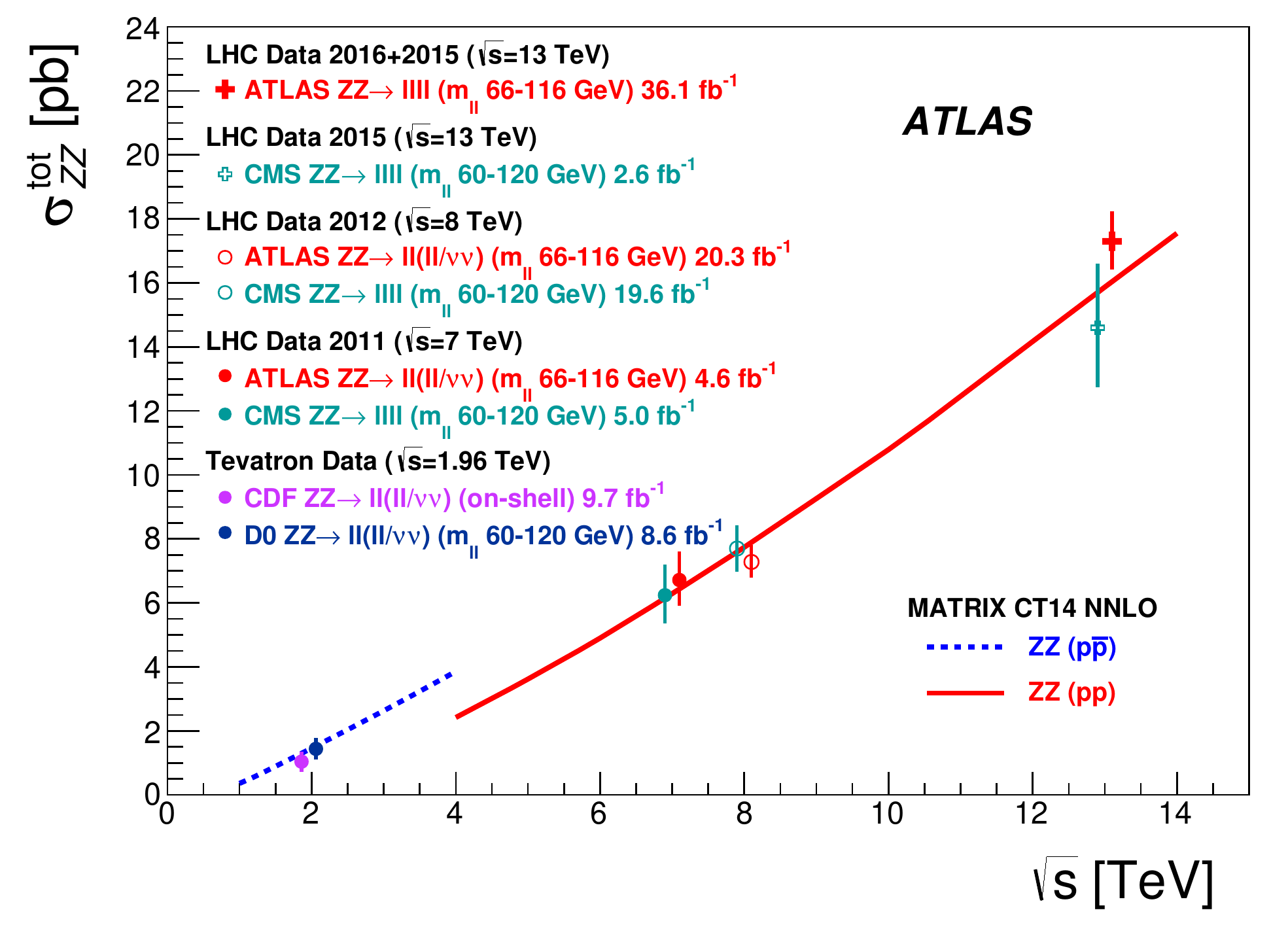}
\caption{Extrapolated cross section compared to other measurements at various center-of-mass energies by ATLAS, CMS, CDF, and D0~\cite{STDM-2012-02,CMS-SMP-13-005,CMS-SMP-12-007,Aaltonen:2014yfa,Abazov:2012cj,Khachatryan:2016txa}, and to pure NNLO predictions from \matrixnnlo{} (with no additional higher-order corrections applied). The total uncertainties of the measurements are shown as bars. Some data points are shifted horizontally to improve readability.}
\label{fig:extrapolated_cross_sections}
\end{figure}

\section{Differential cross sections}
\label{sec:unfolding}

Differential cross sections are obtained by counting candidate events in each bin of the studied observable, subtracting the expected background, and unfolding to correct for detector effects. The unfolding takes into account events that pass the selection but are not in the fiducial phase space (which may occur due to detector resolution or misidentification), bin migrations due to limited detector resolution, as well as  detector inefficiencies. To minimize the model dependence of the measurement, the unfolding corrects and extrapolates the measured distributions to the fiducial phase space, rather than extrapolating to non-fiducial regions. For each given observable distribution, all of the above detector effects are described by a response matrix $R$ whose elements $R_{ij}$ are defined as the probability of an event in true bin $j$ being observed with the detector in bin $i$. The response matrix therefore relates the true distribution $t$ and the background-subtracted measured distribution $m$,

\vspace{-1\baselineskip}
\begin{equation*}\label{eq:response_matrix_definition}
	m_i = R_{ij} t_j.
\end{equation*}
Two examples of response matrices are shown in Figure~\ref{fig:response_matrices}. The purity, defined as the fraction of events that are reconstructed in their true bin, is greater than 70\% for jet-inclusive observables, except in very few bins. In jet-exclusive observables, the purity is greater than 60\% in most bins, but drops to as low as 35\% in some bins. This is due to contribution from jets originating from or contaminated by pileup interactions, as well as worse jet energy resolution and poorer knowledge of the jet energy scale than is the case for leptons.

The unfolding is performed by computing the inverse of the response matrix, using regularization to numerically stabilize the solutions, decreasing their statistical uncertainty at the cost of a small regularization bias. An iterative unfolding method based on Bayes' theorem \cite{DAgostini:1994fjx} is used, which combines the measured distribution with the response matrix to form a likelihood and takes the predicted true distribution as prior. It applies Bayes' theorem iteratively, using the posterior distribution as prior for the next iteration, each iteration decreasing the dependence on the initial prior. Depending on the observable, either two or three unfolding iterations are performed. In each case, the number is optimized to minimize the overall uncertainty. More iterations lead to higher statistical uncertainty and fewer iterations to higher unfolding method uncertainty due to stronger dependence on the theoretical prediction of the underlying distribution.

The nominal response matrices, corrections, and priors are obtained using the nominal \sherpa{} setup. Reconstructed objects in the MC simulation are not required to have a corresponding generated object. 

\begin{figure}[h!]
\centering
\subfigure[Transverse momentum of the leading-\pt{} \PZ{} boson candidate.]{\includegraphics[width=0.48\textwidth]{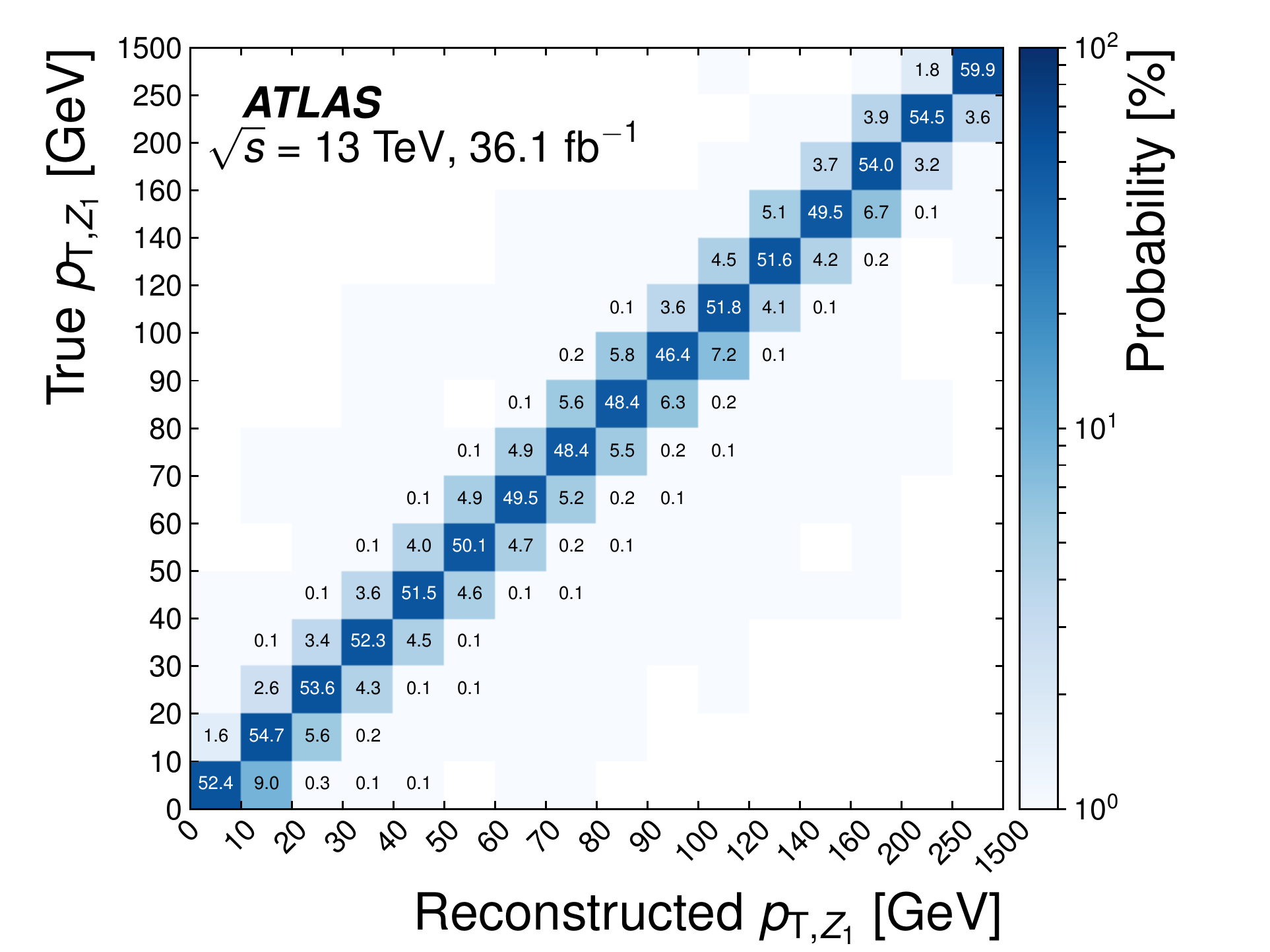}}
\subfigure[Jet multiplicity, considering all selected jets.]{\includegraphics[width=0.48\textwidth]{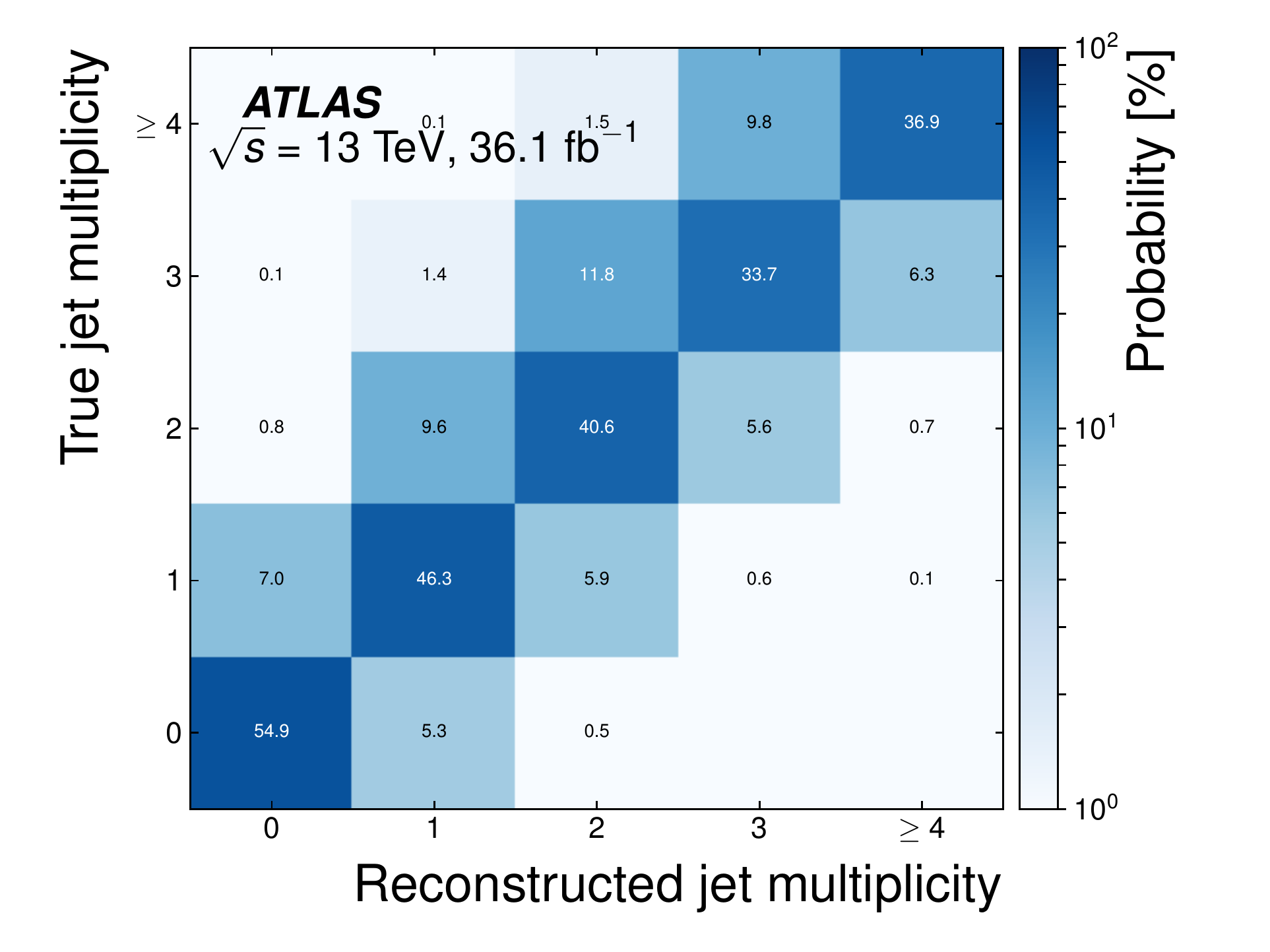}}
\caption{Example response matrices used in the unfolding for two different observables, obtained with the nominal \SHERPA{} setup.}
\label{fig:response_matrices}
\end{figure}

The statistical uncertainty due to fluctuations in the data is estimated by generating 2000 sets of random pseudodata following a Poisson distribution in each bin whose expectation value is the number of observed data events in that bin. The unfolding is repeated with the sets of pseudodata, taking the root mean square of the deviation of the resulting unfolded spectrum from the actual unfolded data as the statistical uncertainty in each bin.  Another uncertainty due to statistical fluctuations in the MC simulations used to obtain the response matrix is obtained the same way, repeating the unfolding using randomly fluctuated copies of the response matrix.

Experimental and theoretical-modeling uncertainties are estimated by repeating the unfolding with the varied response matrix and taking the deviation from the nominal of the resulting unfolded distribution as the uncertainty. Background uncertainties are estimated by subtracting the varied background predictions from the data before unfolding.

The uncertainty due to imperfect modeling of the observable by MC simulation as well as the inherent bias of the unfolding stemming from regularization is estimated using a data-driven method \cite{Malaescu:2009dm}. The initial priors are reweighted by a smooth polynomial function such that there is very good agreement between the prior folded with the response matrix and the observed data. The folded reweighted prior is unfolded using the nominal response matrix. The deviations of the obtained unfolded distribution from the reweighted prior are used as the unfolding bias uncertainty in each bin. This uncertainty is smaller than 1\% of the cross section in almost all bins, but reaches 22\% in individual bins (such as the first bin of the mass of the two leading-\pt{} jets, where the modelling of the data is poor).

The unfolding is repeated using \POWHEGpy{} instead of \sherpa{} to model the $\Pquark\APquark$-initiated process and the difference between the unfolded distributions obtained in this way is assigned as an additional systematic event generator uncertainty.

The statistical uncertainty of the data is typically in the range of 5--40\% of the cross section. 
It dominates the total uncertainty in most bins. In jet-inclusive observables, the largest systematic uncertainty comes from the modeling of the response matrix (up to approximately~25\%). In jet-exclusive observables, the jet energy scale uncertainty is an additional large contribution (3--23\%). Figure~\ref{fig:unfolding_uncertainties} shows detailed bin-by-bin uncertainties for selected observables.

\begin{figure}[h!]
\centering
\subfigure[Transverse momentum of the four-lepton system.]{\includegraphics[width=0.48\textwidth]{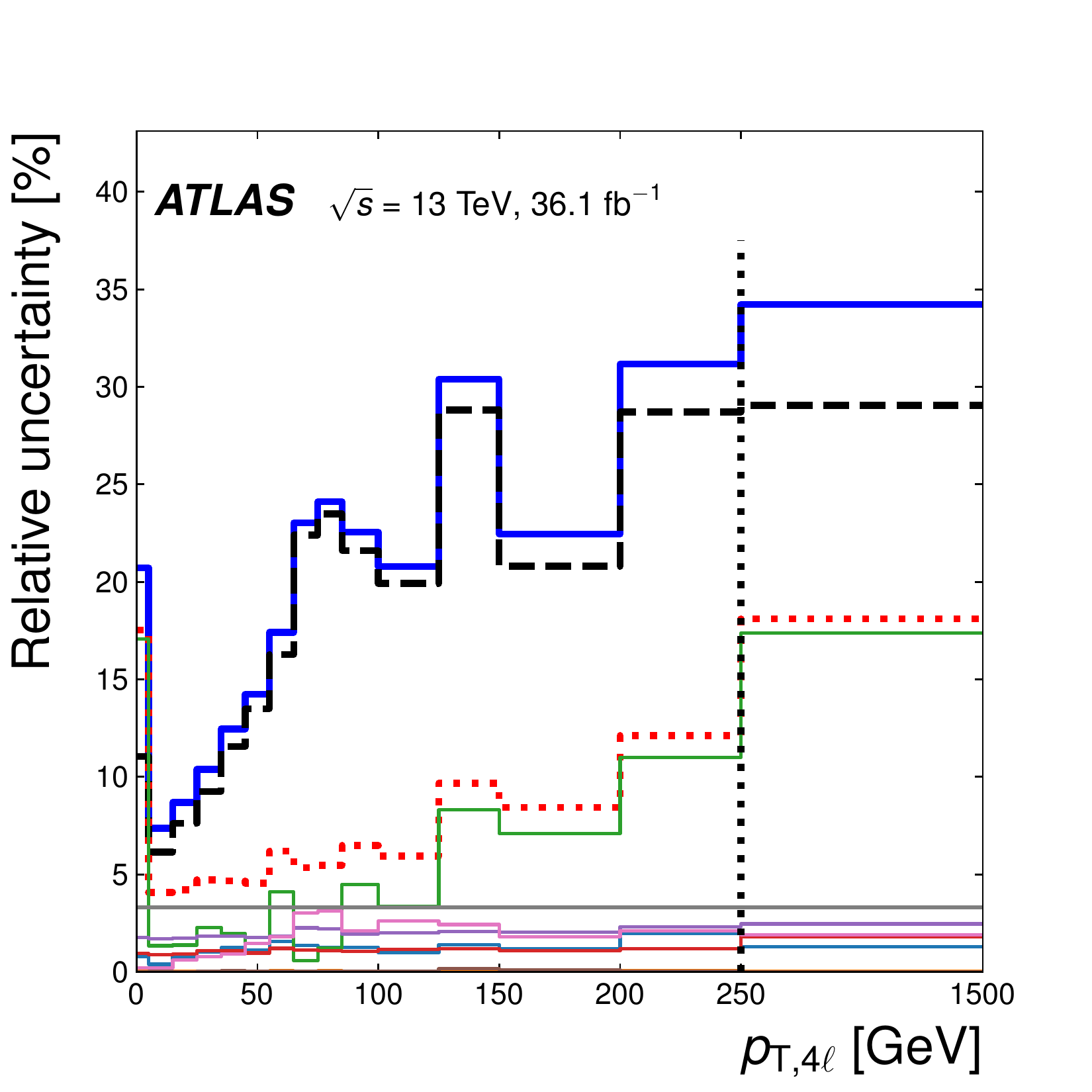}}
\subfigure[Jet multiplicity, considering all selected jets.]{\includegraphics[width=0.48\textwidth]{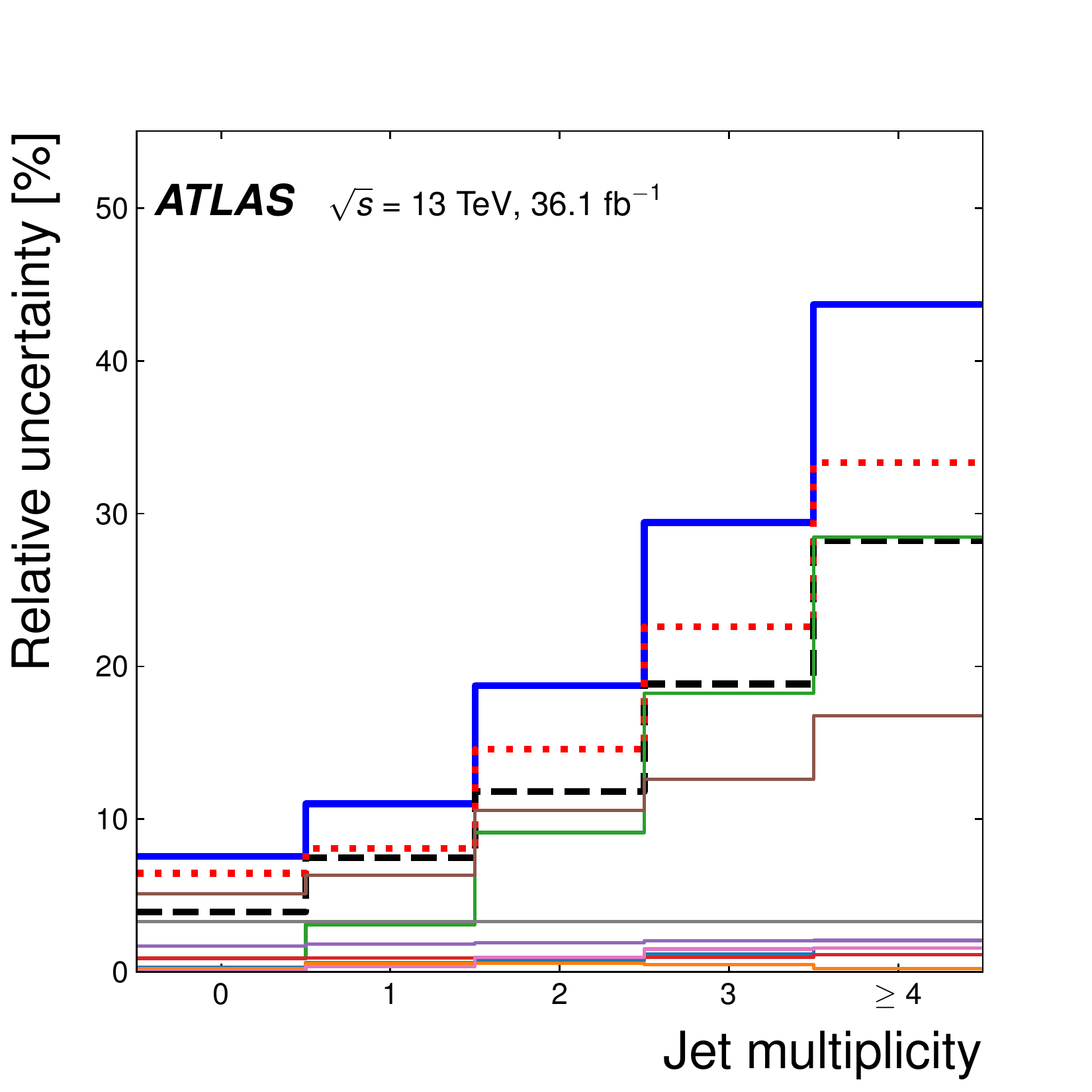}}
\subfigure[Invariant mass of the two leading-\pt{} jets.]{\includegraphics[width=0.48\textwidth]{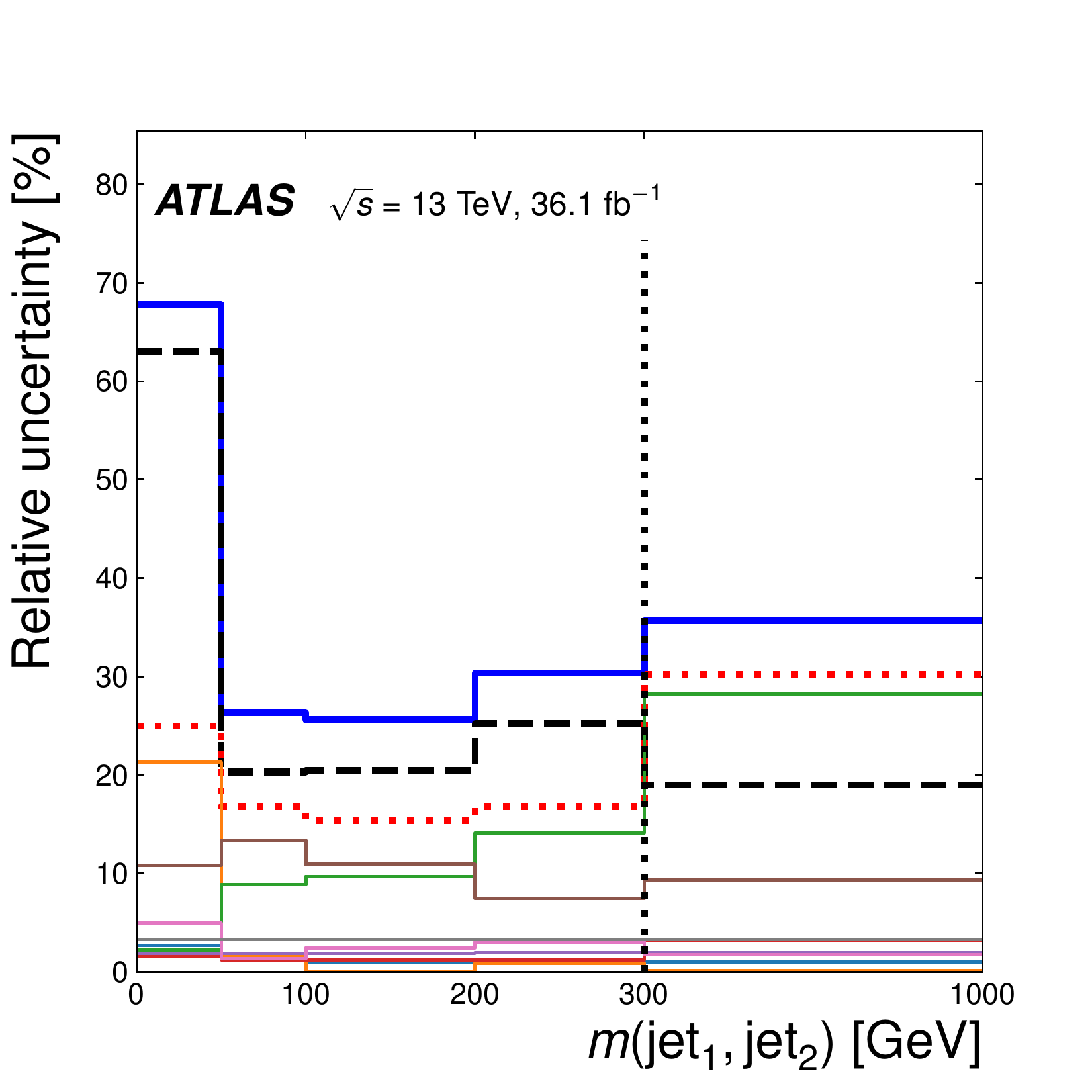}}
\subfigure{\includegraphics[width=0.48\textwidth]{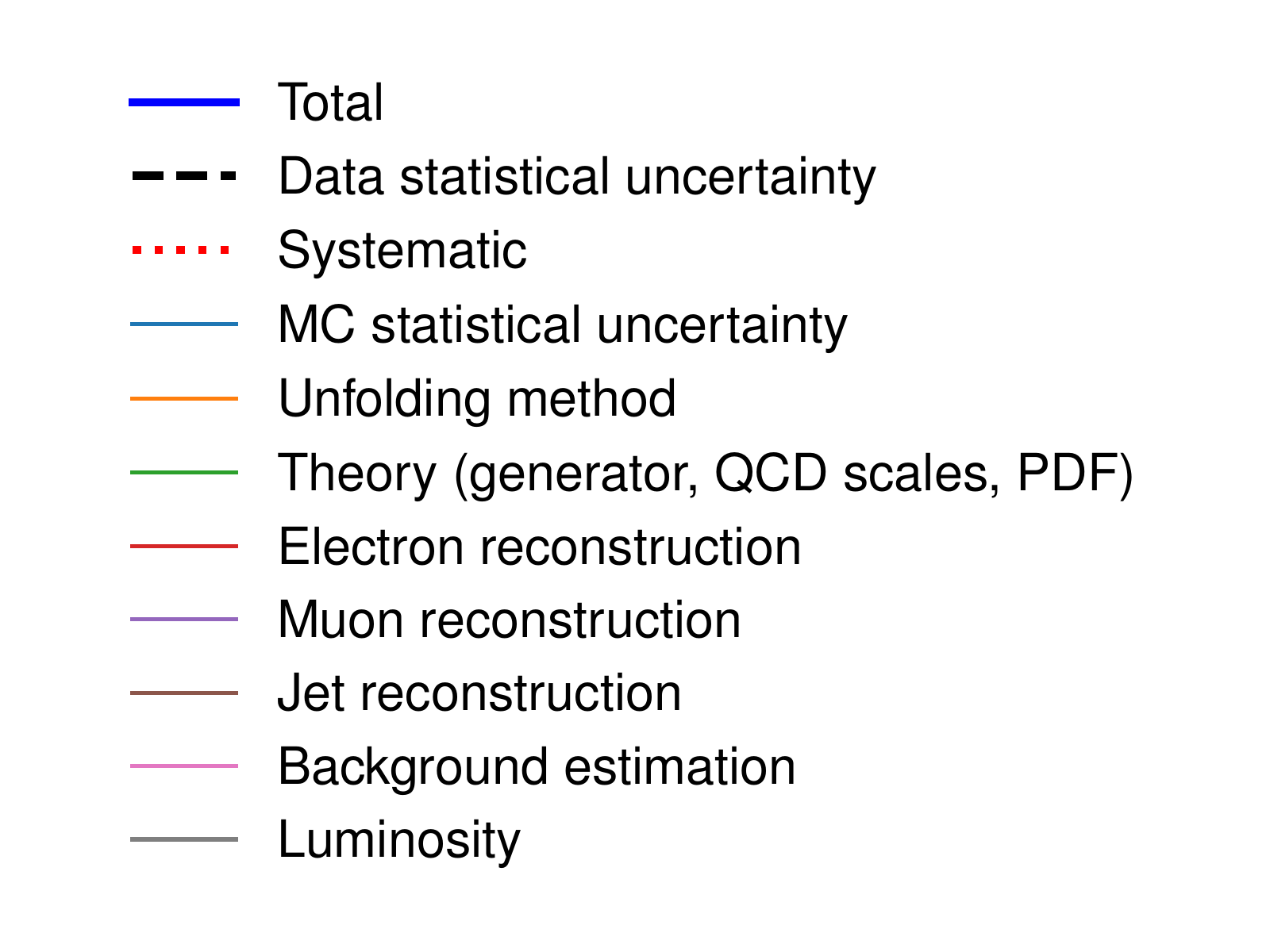}}
\caption{Uncertainty contributions after unfolding in each bin of three representative observables. The total systematic uncertainty contains all uncertainties except the statistical uncertainty of the data, summed in quadrature. The theory uncertainty enters the cross-section measurements via the modeling of the detector response. It is evaluated by considering different event generators, QCD scales, and PDFs. For better visualization, the last bin is shown using a different $x$-axis scale where indicated by the dashed vertical line.} 
\label{fig:unfolding_uncertainties}
\end{figure}

Figures~\ref{fig:zz_kinematics}--\ref{fig:dijets} present the unfolded cross sections, along with comparisons to various fixed-order and parton-showered theoretical predictions. Reasonable agreement of the various predictions with the data is observed, within the statistical and systematic precision of the measurements. 

\newcommand{\brokenaxiscomment}{For better visualization, the last bin is shown using a different $x$-axis scale. The scale change is indicated by the dashed vertical line.}
\newcommand{\datauncertcomment}{The statistical uncertainty of the measurement is shown as error bars, and shaded bands indicate the systematic uncertainty and the total uncertainty obtained by summing the statistical and systematic components in quadrature. The ratio plots only show the total uncertainty.}

\clearpage
Figure~\ref{fig:zz_kinematics}(a) shows the transverse momentum of the four-lepton system, $\ptfourl$. The cross section has a peak around 10~\GeV{} and drops rapidly towards both lower and higher values. At low \ptfourl{}, the resummation of low-\pt{} parton emissions is important and fixed-order descriptions are inadequate. For this reason, the fixed-order predictions are not shown in the first two bins, 0--5~\GeV{} and 5--15~\GeV{}. The region below $\ptfourl = 60~\GeV{}$ is modeled slightly better by predictions that include a parton shower, again suggesting the importance of resummation. Above 60~\GeV{}, the fixed-order NNLO predictions describe the data slightly better. Figure~\ref{fig:zz_kinematics}(b) shows the absolute rapidity of the four-lepton system, which drops gradually towards high values. This distribution is potentially sensitive to a different choice of PDF, describing the momentum distribution of the incoming partons. Fixed-order calculations and predictions including a parton shower model this observable reasonably well, within the statistical and systematic uncertainties. The predictions tend to slightly underestimate the cross sections for small values of $|\yfourl{}|$.

\begin{figure}[h!]
\centering
\subfigure[]{\includegraphics[width=0.48\textwidth]{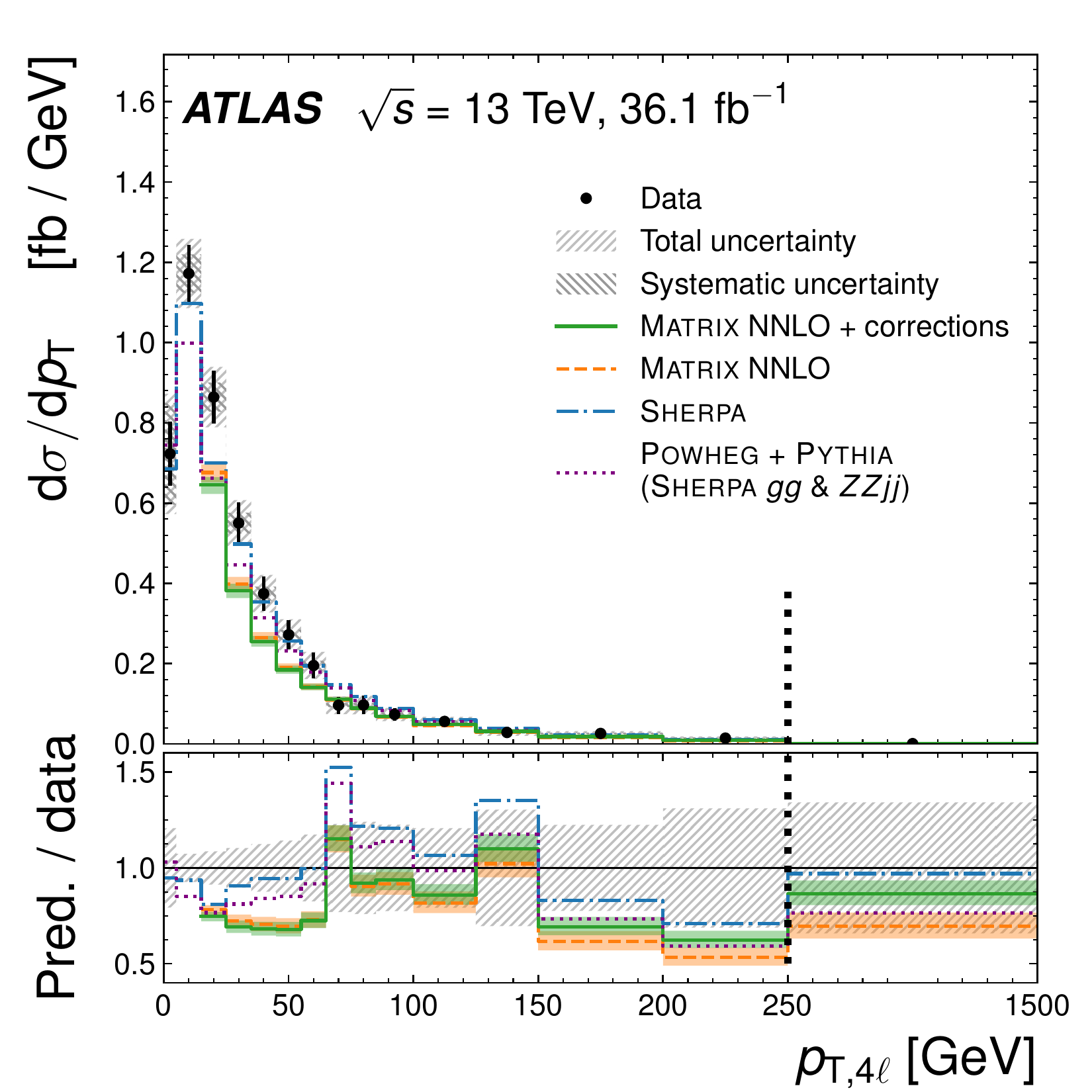}} 
\subfigure[]{\includegraphics[width=0.48\textwidth]{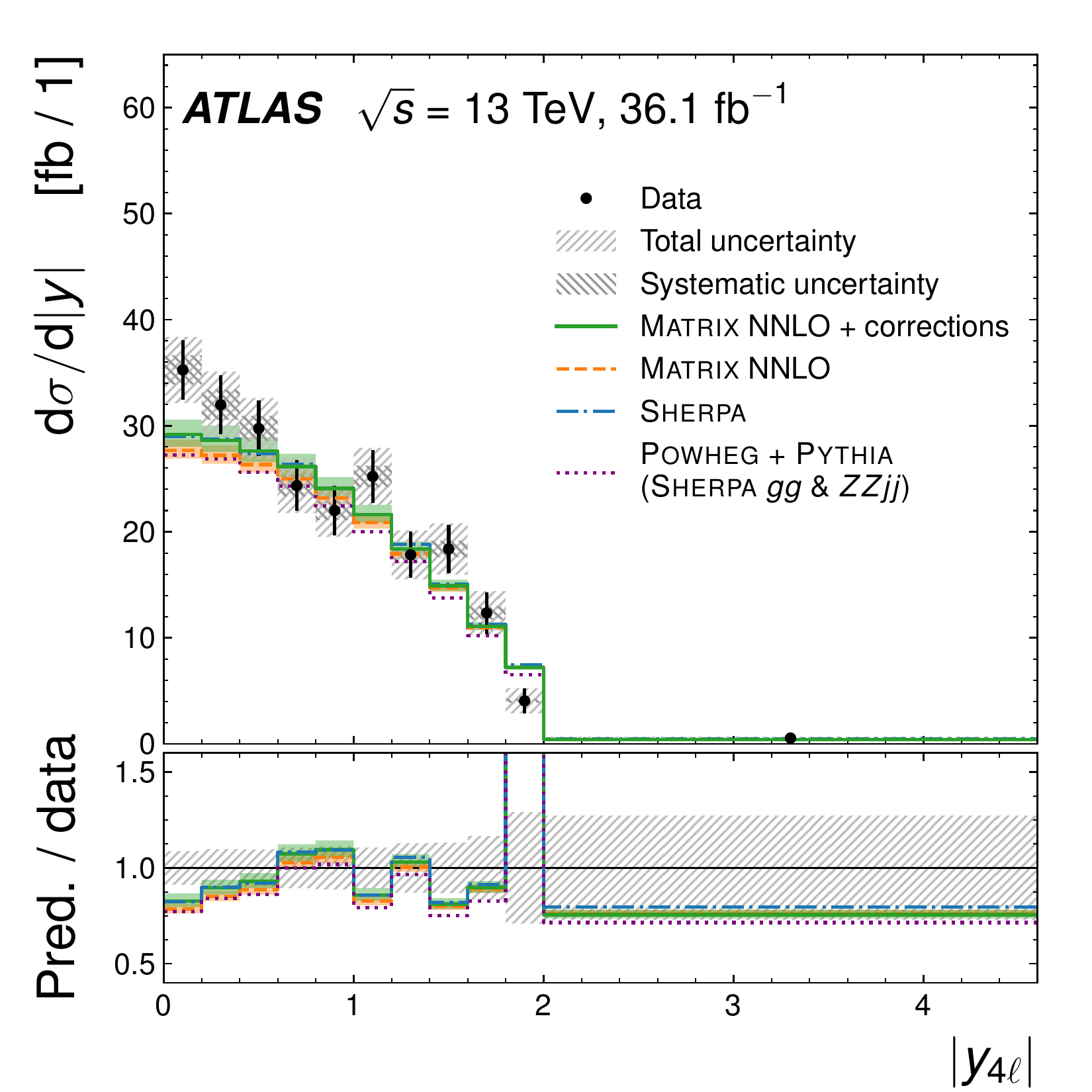}} 
\caption{Measured and predicted differential cross sections for (a) the transverse momentum and (b) the absolute rapidity of the four-lepton system. \datauncertcomment{} A pure NNLO calculation from \matrixnnlo{} is shown with no additional corrections applied. The best SM prediction is based on this NNLO calculation, with the $\Pgluon\Pgluon$-initiated contribution multiplied by a global NLO correction factor of $1.67$. For the $\ptfourl$ distribution in (a), the NLO EW correction is applied as a global factor of $0.95$ as a differential calculation is not available. For the $|\yfourl|$ distribution in (b), an NLO EW correction factor is applied in each bin. The contribution from EW-$\ZZ jj$ generated with \sherpa{} is added. For the fixed-order predictions, the QCD scale uncertainty is shown as a shaded band. Parton-showered \SHERPA{} and \POWHEGpy{} predictions are also shown. For better visualization, the last bin is shown using a different $x$-axis scale where indicated by the dashed vertical line.}
\label{fig:zz_kinematics}
\end{figure}

\clearpage
Figure~\ref{fig:zz_angles}(a) presents the azimuthal angle separation between the two \PZ{} boson candidates. The fixed-order predictions only describe the shape of the $\gluongluon$-initiated process at LO and therefore predict a distribution that is more peaked at $\pi$ than those from \SHERPA{} and \POWHEGpy{}, where the parton shower shifts some events towards lower values. Figure~\ref{fig:zz_angles}(b) shows the distribution of the absolute rapidity difference of the two \PZ{} boson candidates, which drops towards high values and is modeled by all calculations to within the uncertainties.

\begin{figure}[h!]
\centering
\subfigure[]{\includegraphics[width=0.48\textwidth]{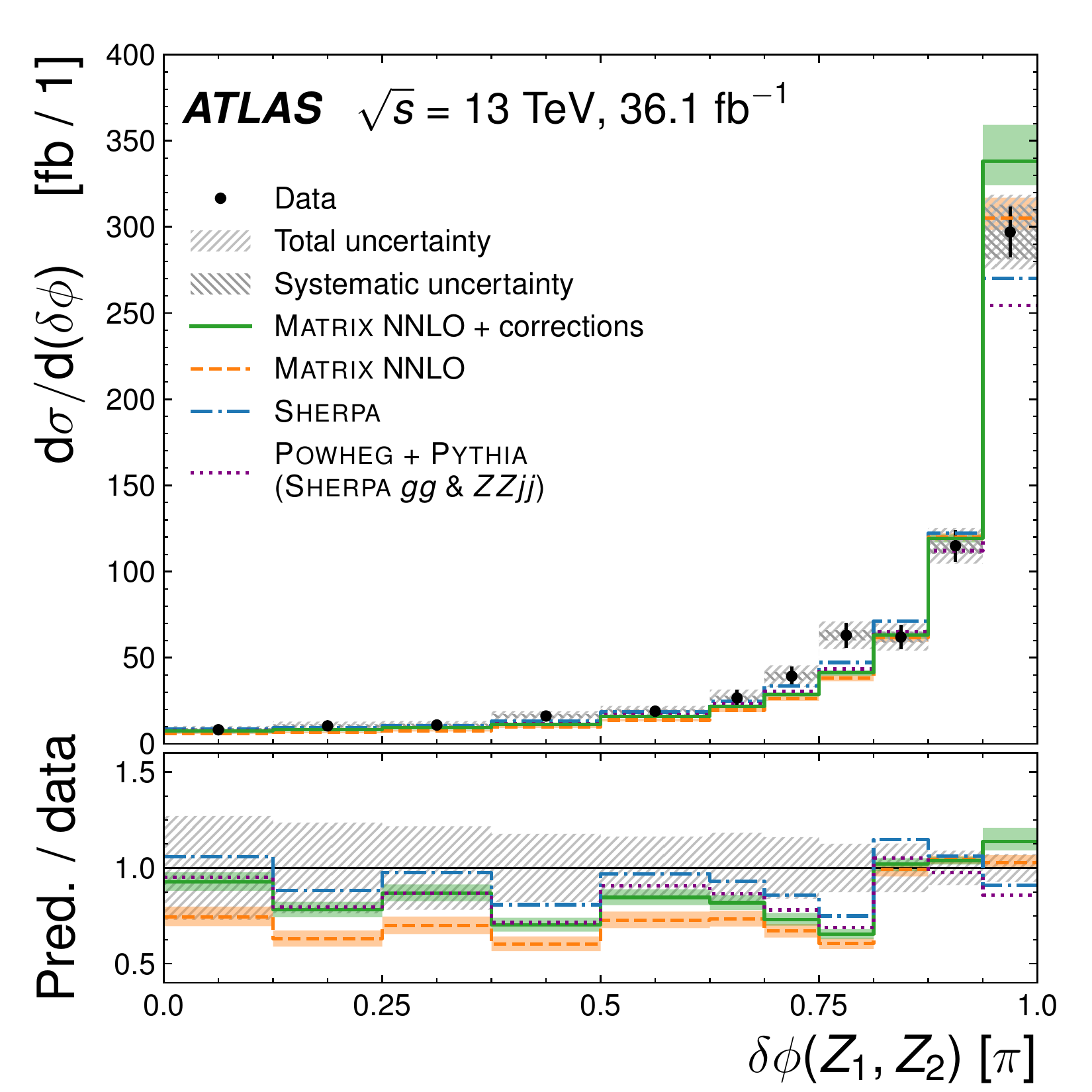}} 
\subfigure[]{\includegraphics[width=0.48\textwidth]{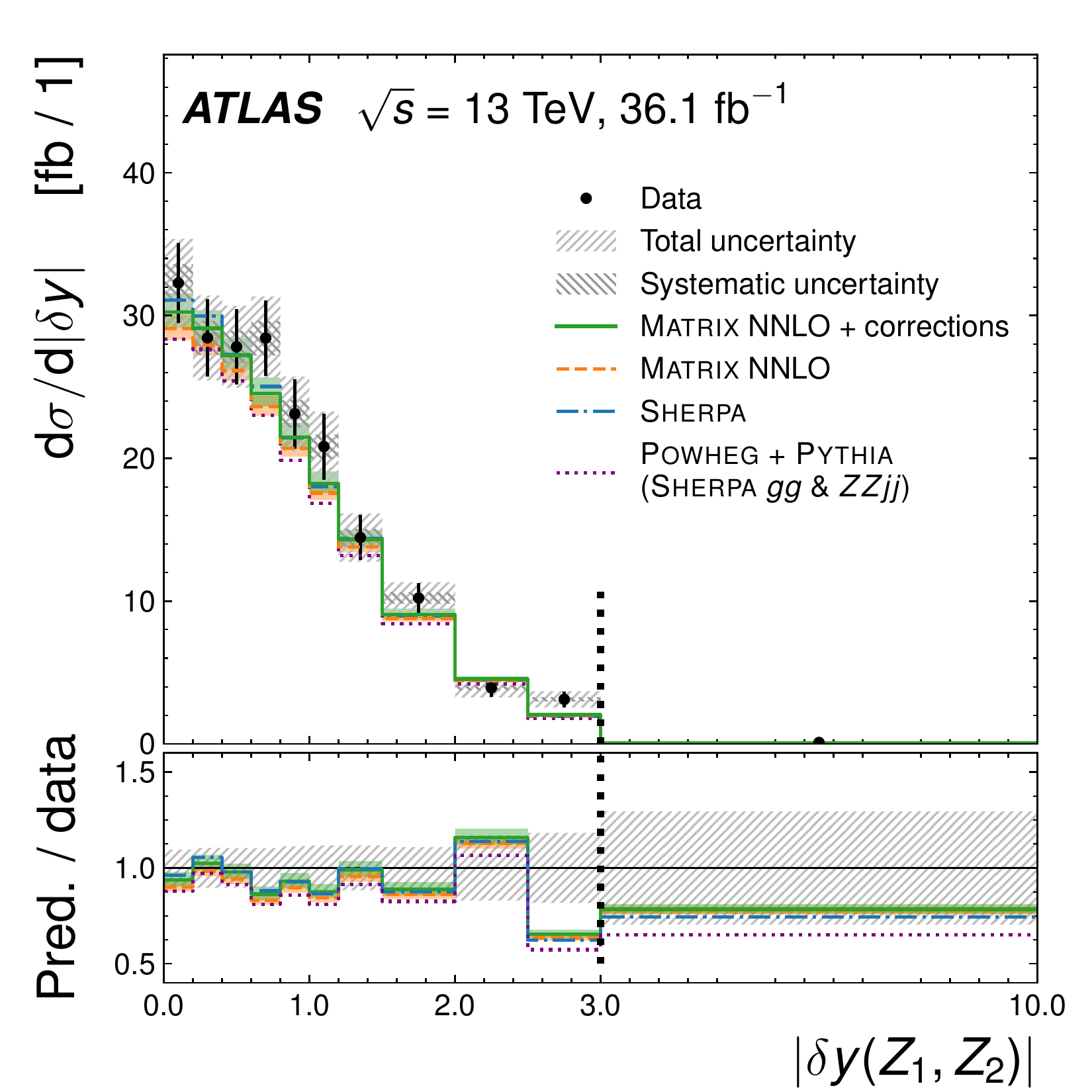}} 
\caption{Measured and predicted differential cross sections for (a) the azimuthal angle separation and (b) the absolute rapidity difference between the two $\PZ$ boson candidates. \datauncertcomment{} A pure NNLO calculation from \matrixnnlo{} is shown with no additional corrections applied. The best SM prediction is based on this NNLO calculation, with the $\Pgluon\Pgluon$-initiated contribution multiplied by a global NLO correction factor of $1.67$. For the $\dphiZZ$ distribution in (a), the NLO EW correction is applied as a global factor of $0.95$ as a differential calculation is not available. For the $|\dyZZ|$ distribution in (b), an NLO EW correction factor is applied in each bin. The contribution from EW-$\ZZ jj$ generated with \sherpa{} is added. For the fixed-order predictions, the QCD scale uncertainty is shown as a shaded band. Parton-showered \SHERPA{} and \POWHEGpy{} predictions are also shown. For better visualization, the last bin is shown using a different $x$-axis scale where indicated by the dashed vertical line.}
\label{fig:zz_angles}
\end{figure}

\clearpage
Figure~\ref{fig:z_pts} shows the transverse momentum of the leading-\pt{} and subleading-\pt{} \PZ{} boson candidates, exhibiting a wide peak around 50~\GeV{} and 30~\GeV{}, respectively. Anomalous triple gauge couplings (as discussed in Section~\ref{sec:anomalous}) would manifest as an excess in the cross section at large values of the transverse momentum of the \PZ{} bosons, which is not observed in these differential cross-section distributions (the last bin in each distribution is consistent with the SM predictions). The discrepancies at \pt{} of about 50~\GeV{}, 90~\GeV{} in the leading \PZ{} boson candidate are related to the excesses seen in Figure~\ref{fig:data_mc_plots}(c). The local significance of these excesses with respect to the \SHERPA{} prediction is estimated to be 2.3 and 2.0 standard deviations respectively. This estimate is based on the corresponding bins in the measured distribution before unfolding, as the statistical treatment is simpler due to the statistical uncertainties being Poissonian. In the estimation, both the predicted and observed yields are taken to be Poissonian and the systematic uncertainties of the prediction are taken into account.

\begin{figure}[h!]
\centering
\subfigure[]{\includegraphics[width=0.48\textwidth]{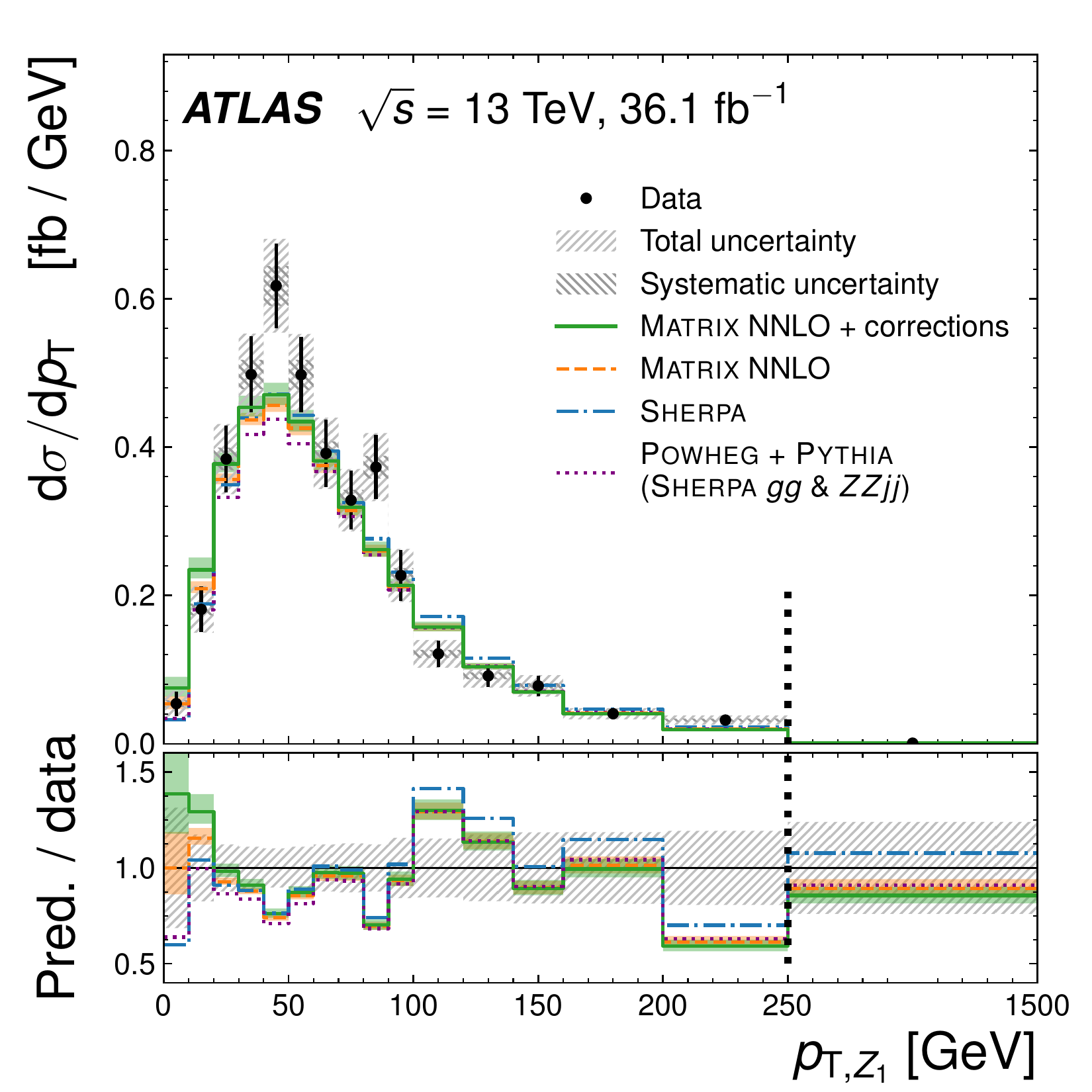}}
\subfigure[]{\includegraphics[width=0.48\textwidth]{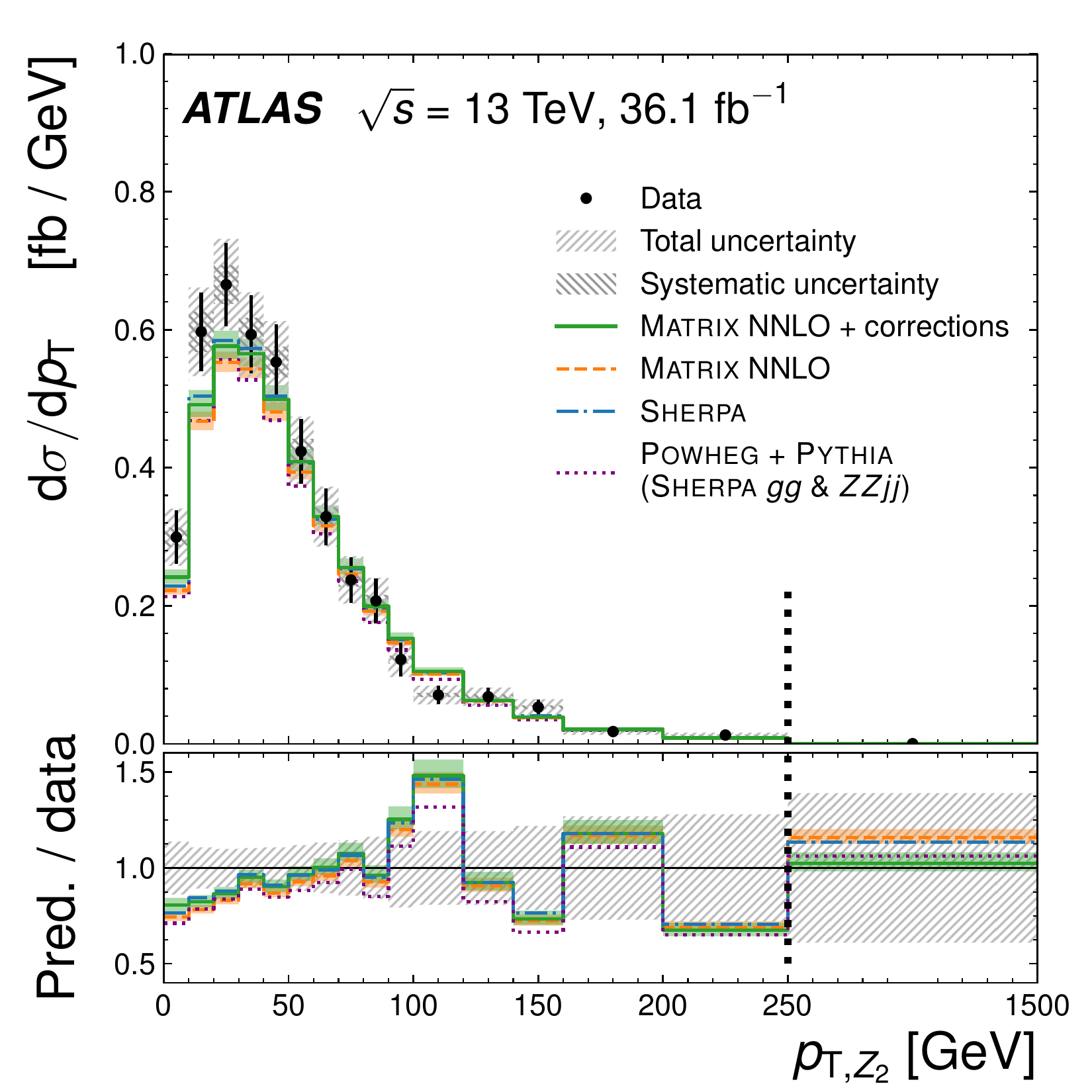}} 
\caption{Measured and predicted differential cross sections for the transverse momentum of (a) the leading-\pt{} and (b) the subleading-\pt{} \PZ{} boson candidates. \datauncertcomment{} A pure NNLO calculation from \matrixnnlo{} is shown with no additional corrections applied. The best SM prediction is based on this NNLO calculation, with the $\Pgluon\Pgluon$-initiated contribution multiplied by a global NLO correction factor of $1.67$. An NLO EW correction factor is applied in each bin. The contribution from EW-$\ZZ jj$ generated with \sherpa{} is added. For the fixed-order predictions, the QCD scale uncertainty is shown as a shaded band. Parton-showered \SHERPA{} and \POWHEGpy{} predictions are also shown. \brokenaxiscomment{}}
\label{fig:z_pts}
\end{figure}

\clearpage
Figure~\ref{fig:lepton_pts} presents the transverse momenta of the leptons in the final selected quadruplet. From the highest-\pt{} to the lowest-\pt{} lepton, the distribution becomes less peaked and more symmetric about the peak, while the position of the peak shifts from $\sim$60~\GeV{} to $\sim$50~\GeV{}, then $\sim$35~\GeV{}, and finally $\sim$25~\GeV{}. All lepton \pt{} distributions agree well with the predictions. 
\vspace{-5mm}

\begin{figure}[h!]
\centering
\subfigure[]{\includegraphics[width=0.48\textwidth]{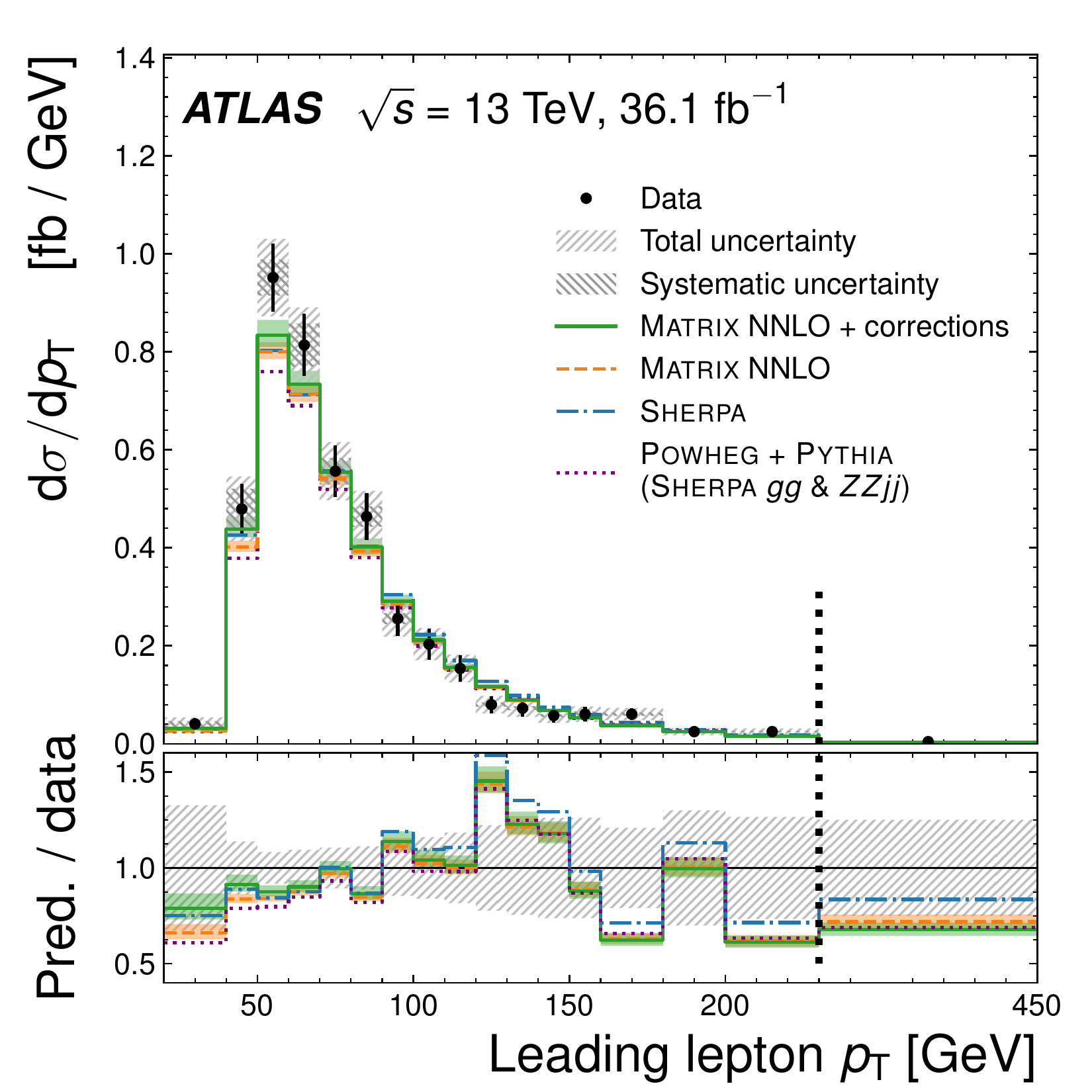}}
\subfigure[]{\includegraphics[width=0.48\textwidth]{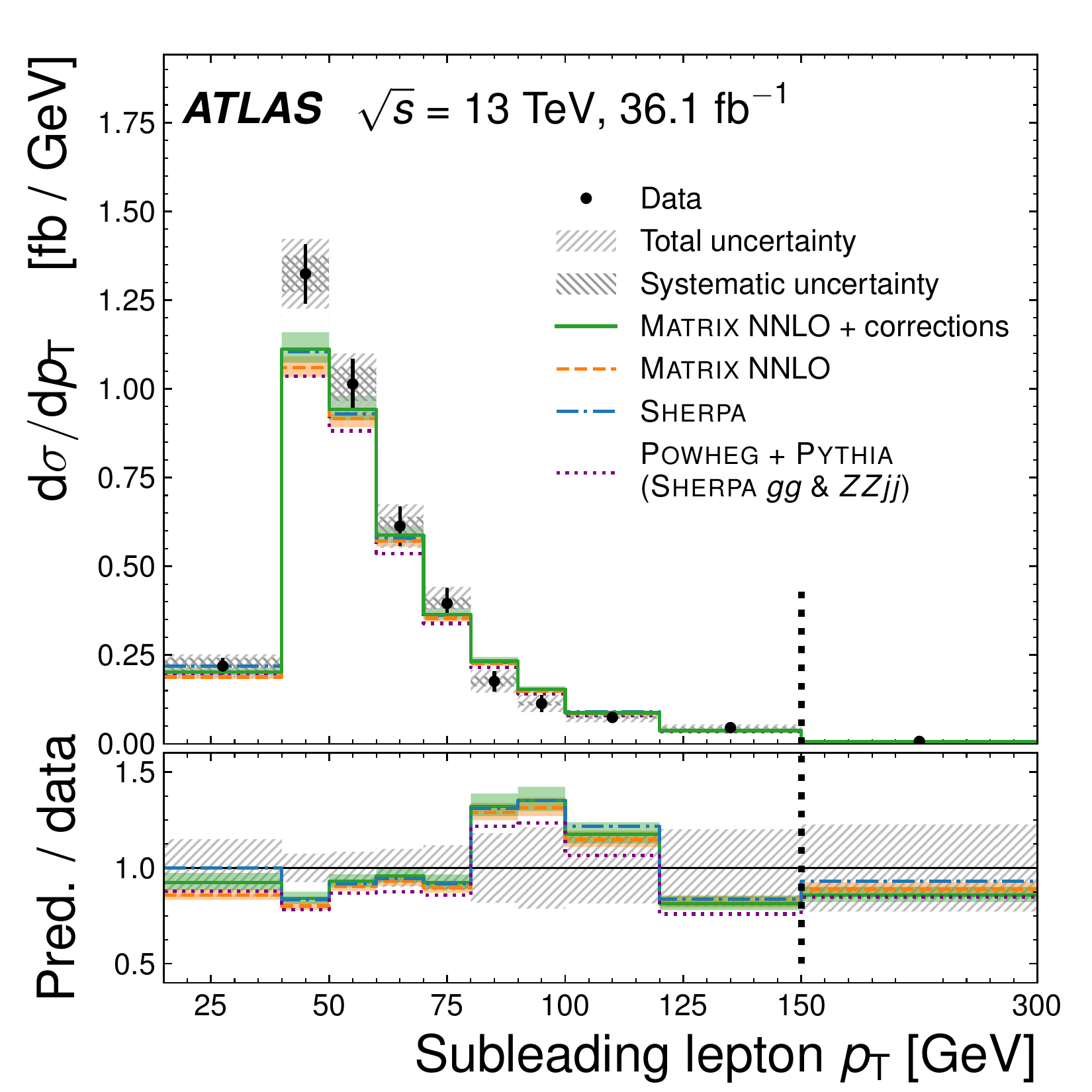}}
\vspace{-5mm}
\subfigure[]{\includegraphics[width=0.48\textwidth]{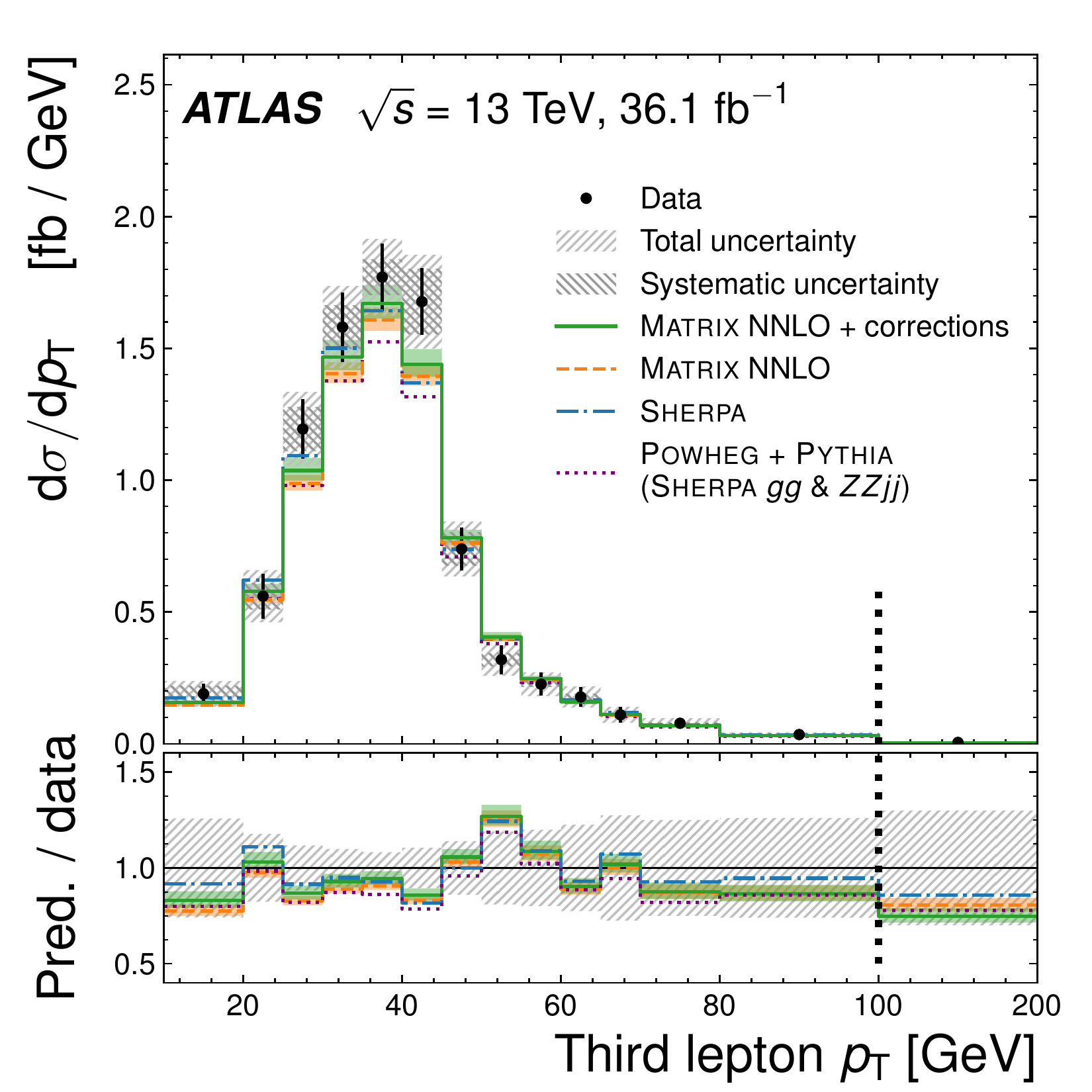}}
\subfigure[]{\includegraphics[width=0.48\textwidth]{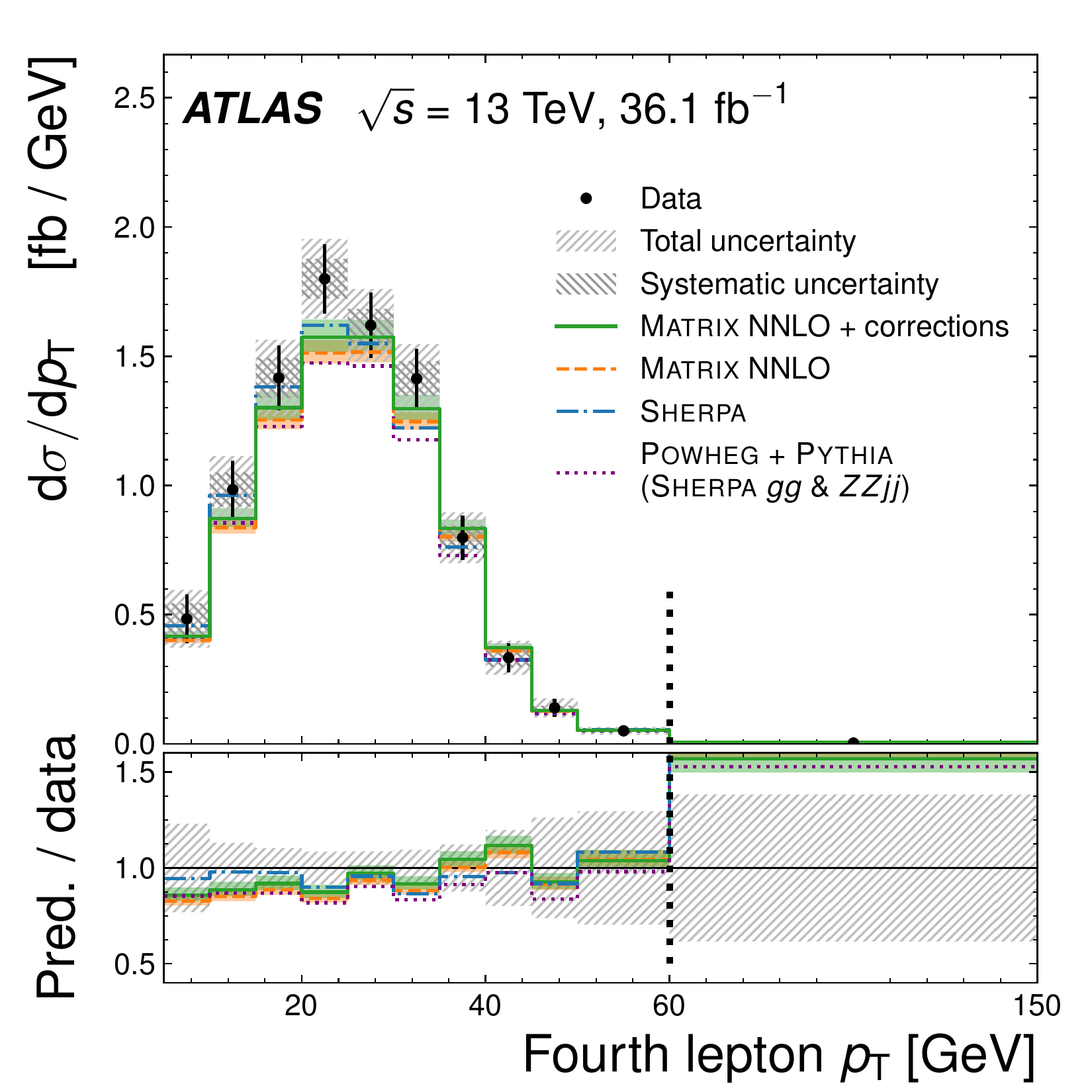}}
\caption{Measured and predicted differential cross sections with respect to the transverse momenta of the leptons in the final selected quadruplet, in descending order of transverse momentum. A pure NNLO calculation from \matrixnnlo{} is shown with no additional corrections applied. The best SM prediction is based on this NNLO calculation, with the $\Pgluon\Pgluon$-initiated contribution multiplied by a global NLO correction factor of $1.67$. An NLO EW correction factor is applied in each bin. The contribution from EW-$\ZZ jj$ generated with \sherpa{} is added. For the fixed-order predictions, the QCD scale uncertainty is shown as a shaded band. Parton-showered \SHERPA{} and \POWHEGpy{} predictions are also shown. \brokenaxiscomment{}}
\label{fig:lepton_pts}
\end{figure}
\clearpage

Figure~\ref{fig:multijets} shows the jet multiplicity distributions as well as the scalar sum of the transverse momenta of all selected jets. \POWHEGpy{} shows a clear trend towards underestimating the cross section at jet multiplicities greater than one and large jet scalar~\pt{} sum, which is expected, because in \POWHEGpy{} only the hardest parton emission is included at the matrix-element level. \SHERPA{}, however, includes up to three parton emissions at the matrix-element level, and exhibits good agreement with the measurements for these higher jet multiplicities. The central-jet multiplicity in Figure~\ref{fig:multijets}(b) is an exception, as \POWHEGpy{} describes it slightly better than \SHERPA{}.
The most significant observed disagreement is the deficit in the bin $60~\GeV{} < \sum\pt < 90~\GeV{}$ of the jet scalar~\pt{} sum. It has a local significance of 2.3 standard deviations with respect to the \SHERPA{} prediction, estimated from the corresponding bins in the measured distribution before unfolding.

\begin{figure}[htbp!]
\centering
\subfigure[]{\includegraphics[width=0.45\textwidth]{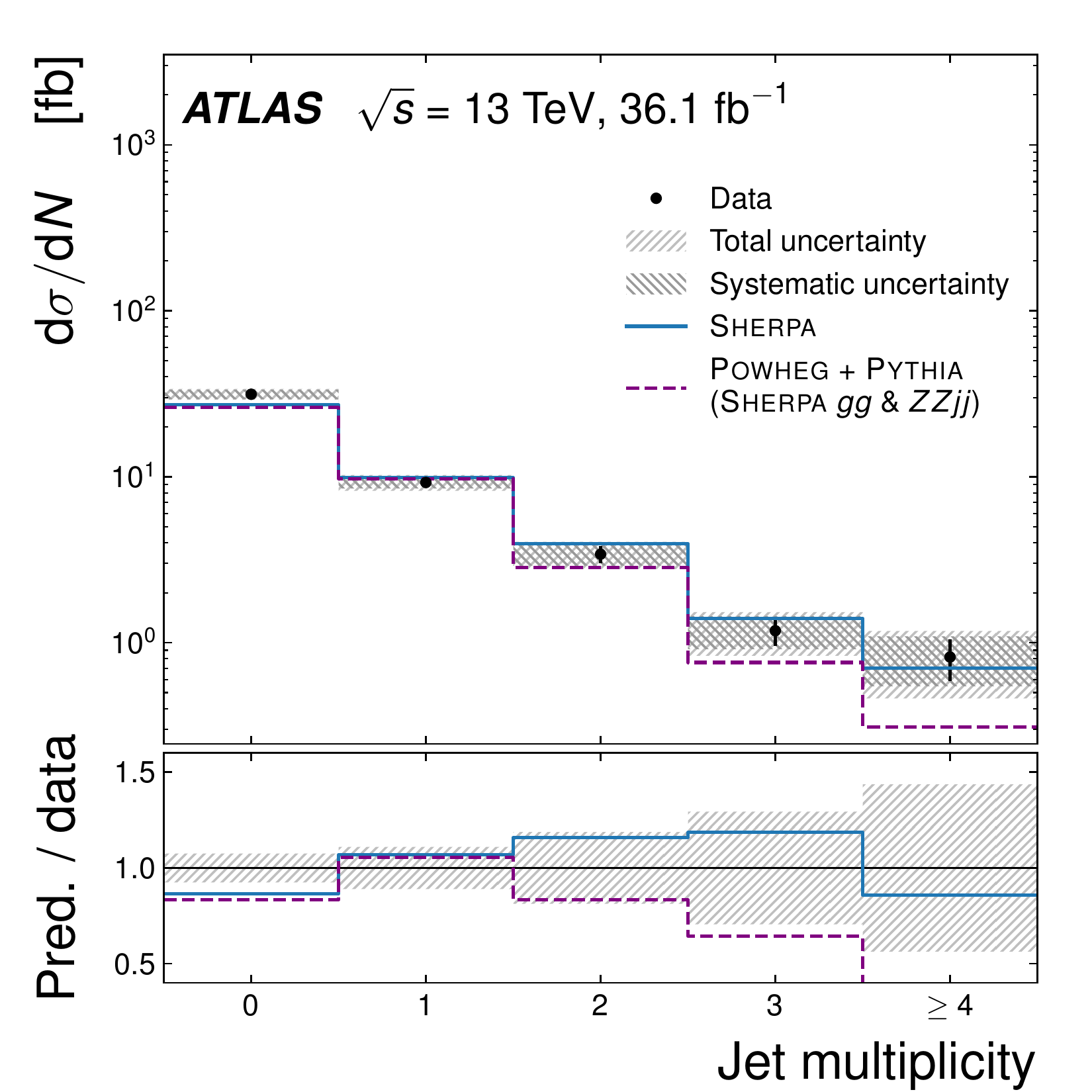}}
\subfigure[]{\includegraphics[width=0.45\textwidth]{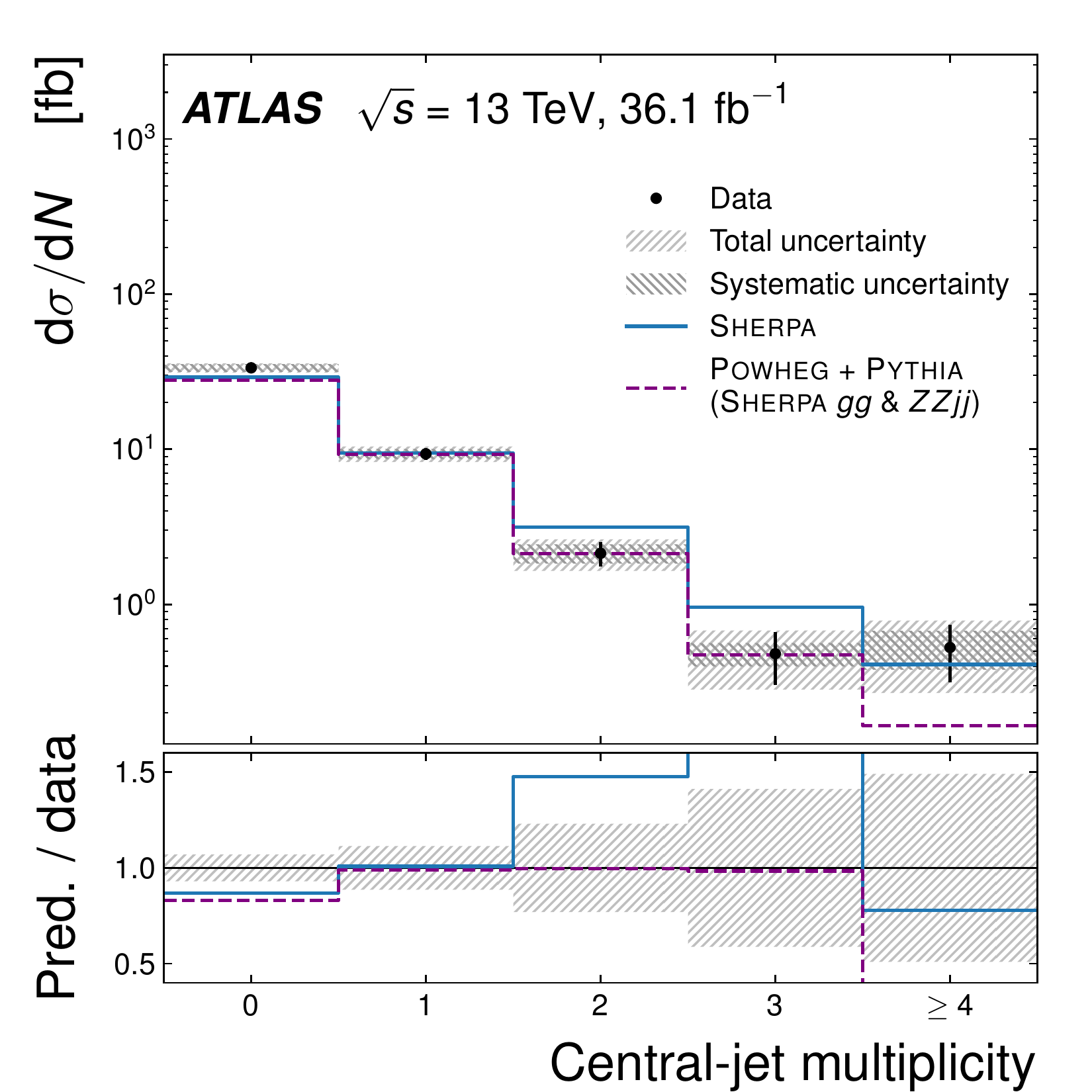}}

\subfigure[]{\includegraphics[width=0.45\textwidth]{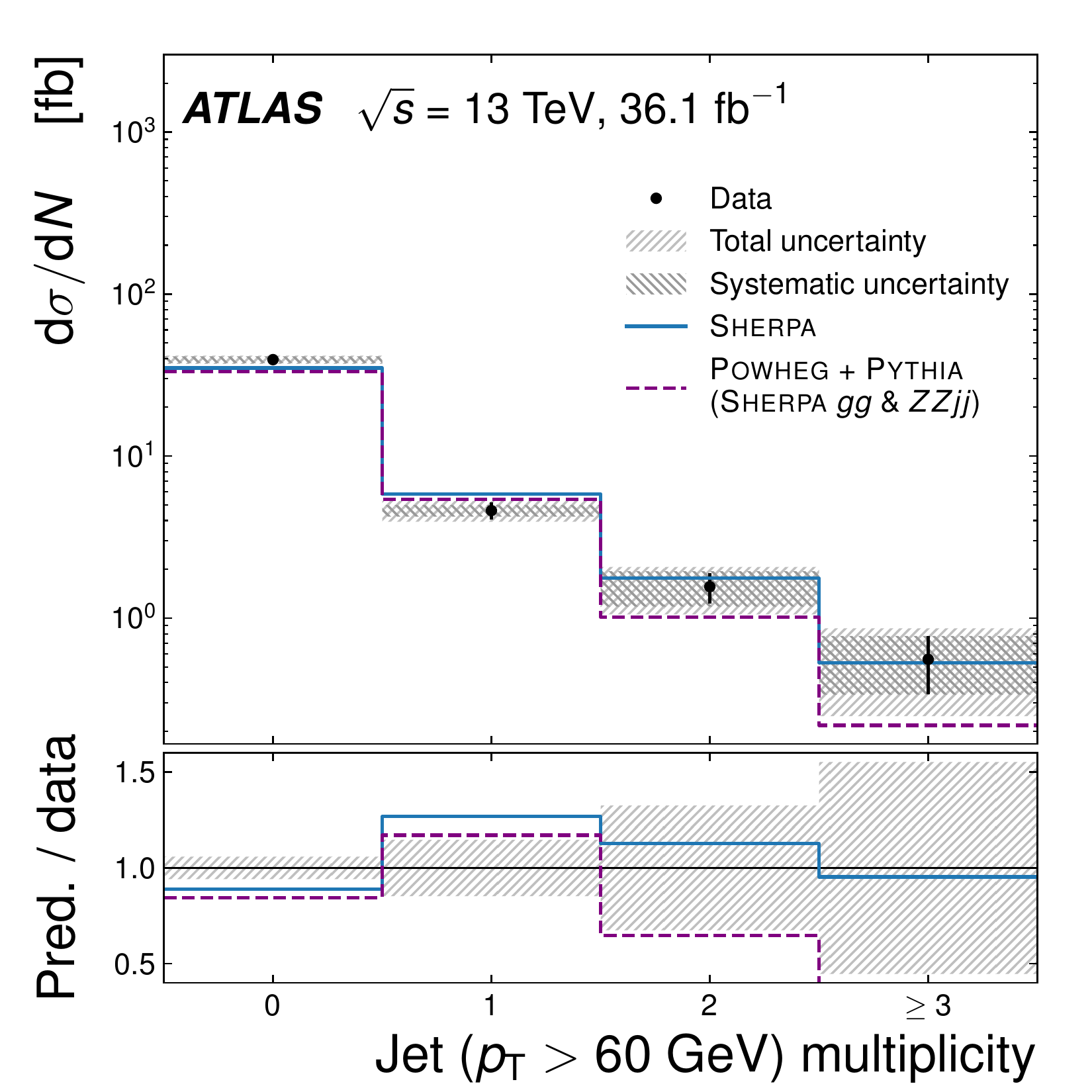}}
\subfigure[]{\includegraphics[width=0.45\textwidth]{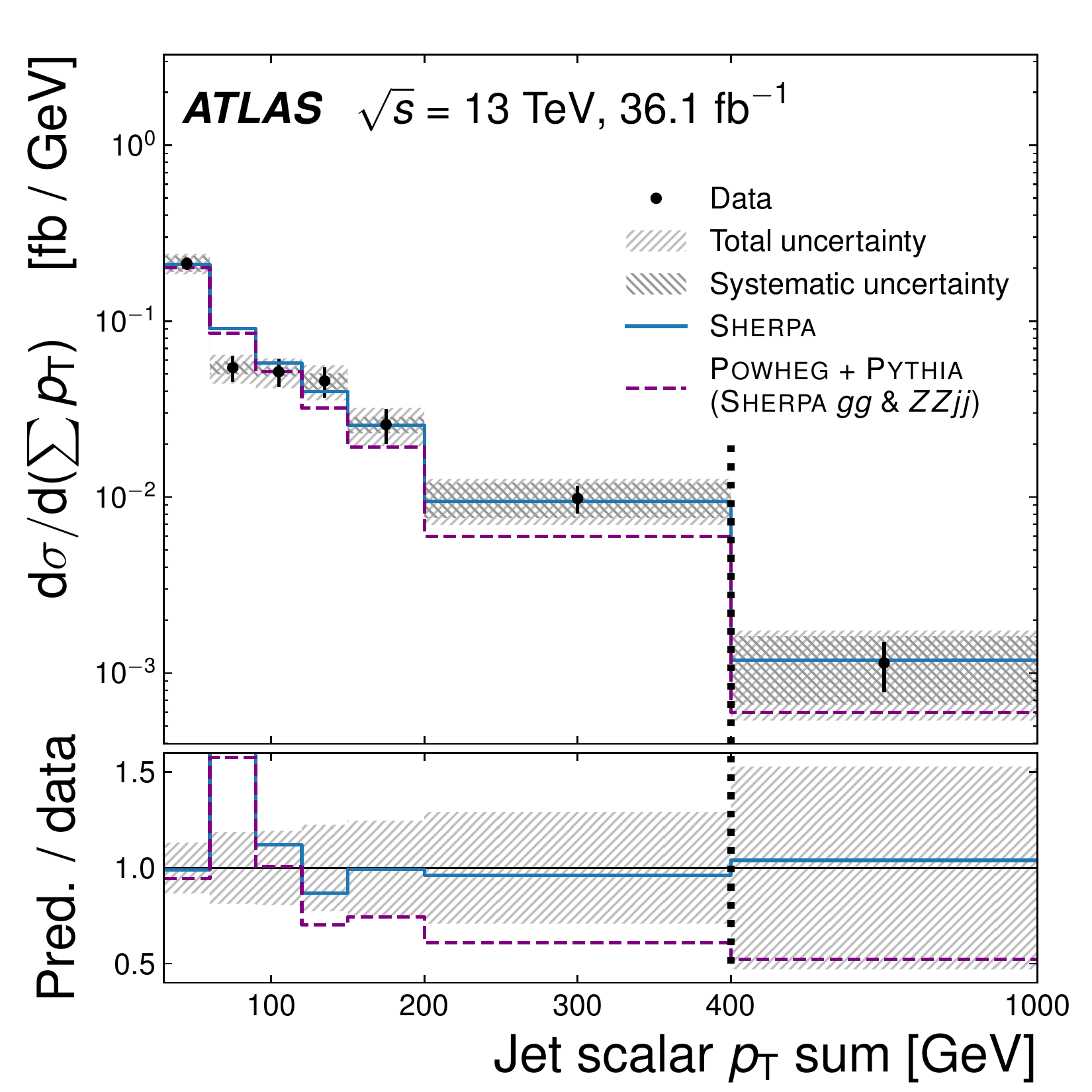}}
\caption{Measured and predicted differential cross sections for (a) the jet multiplicity (all jets), (b) the central-jet multiplicity  (jets with $|\eta| < 2.4$), (c) the jet multiplicity considering jets with $\pt > 60$~\GeV{}, and (d) the scalar sum of the transverse momenta of all selected jets. \datauncertcomment{} Parton-showered \SHERPA{} and \POWHEGpy{} predictions are shown. For better visualization, the last bin is shown using a different $x$-axis scale where indicated by the dashed vertical line.}
\label{fig:multijets}
\end{figure}

\clearpage
Figure~\ref{fig:jet_kinematics} shows the transverse momentum and absolute pseudorapidity of the leading-\pt{} and subleading-\pt{} jets. Within the relatively large uncertainties, \SHERPA{} provides a good description of the kinematics. \POWHEGpy{} also describes the shapes of the $|\eta|$ distributions well, while its normalization is too low for the subleading-\pt{} jet. \POWHEGpy{} does not describe the \pt{} distribution of either the leading-\pt{} or subleading-\pt{} very well. A deficit of events is observed in the bin $2.0 < |\eta| < 2.5$ of the subleading-\pt{} jet. Its local significance with respect to the \SHERPA{} prediction is estimated to be 3.2 standard deviations, based on the corresponding bins in the measured distribution before unfolding.
\vspace{-6mm}

\begin{figure}[h!]
\centering
\subfigure[]{\includegraphics[width=0.48\textwidth]{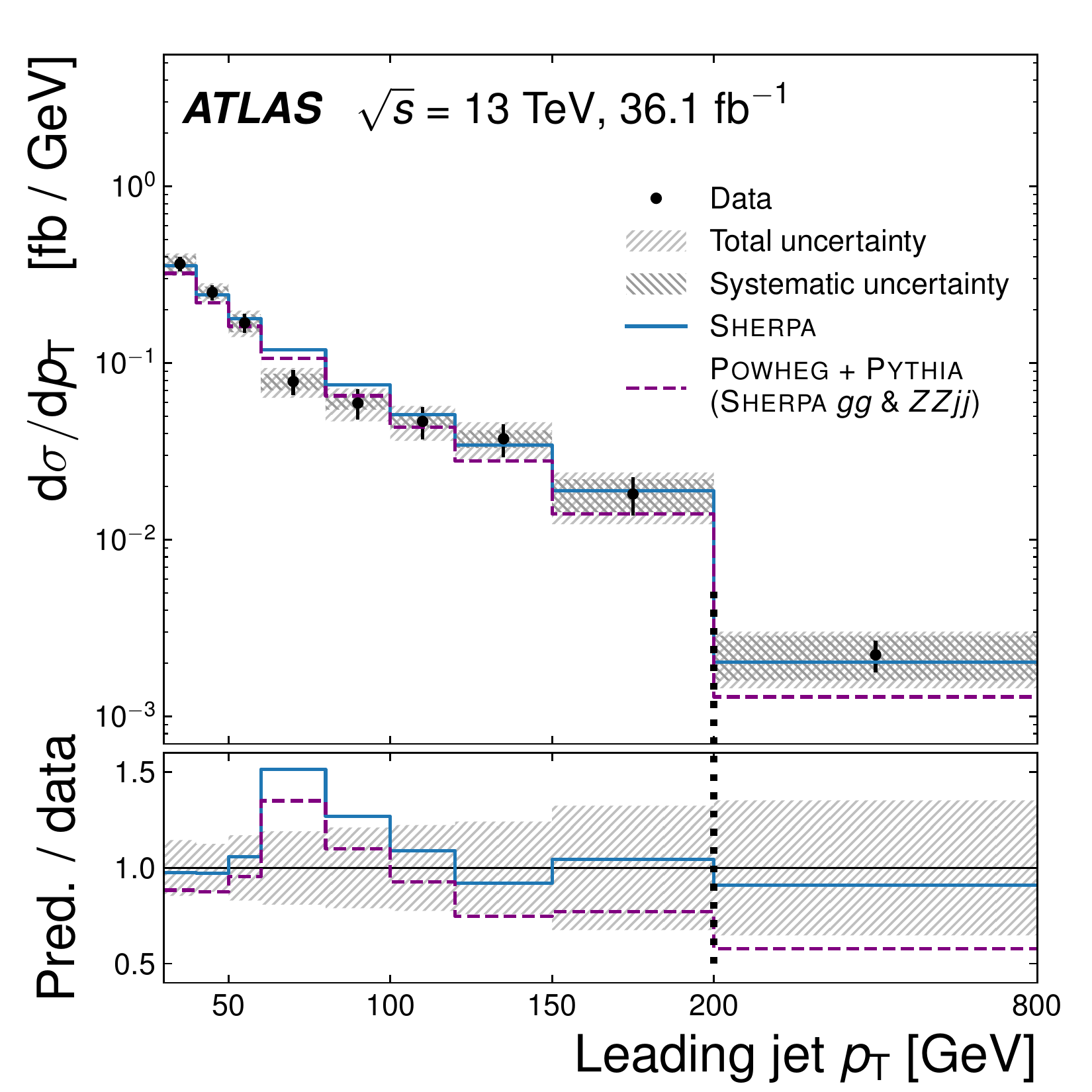}}
\subfigure[]{\includegraphics[width=0.48\textwidth]{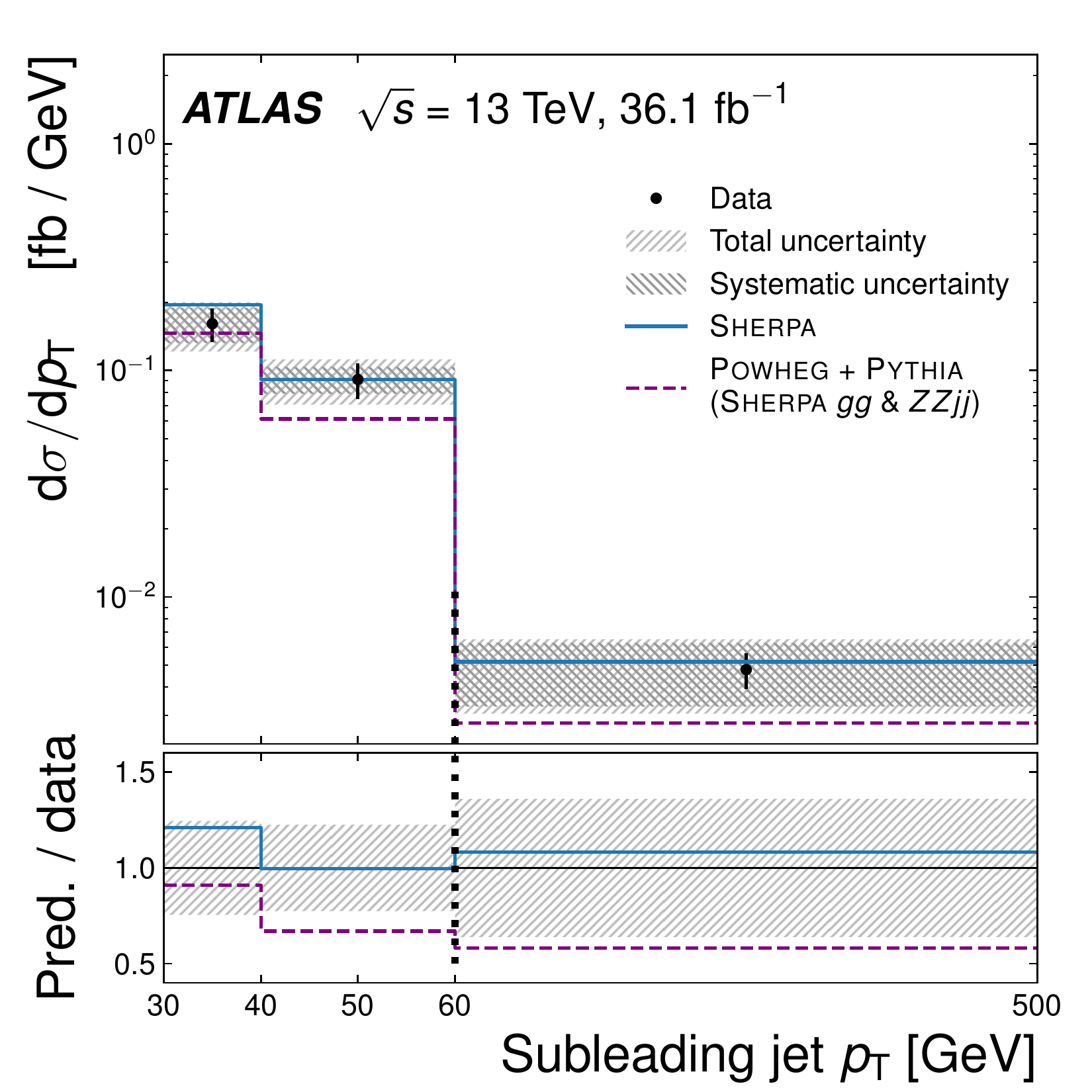}}
\vspace{-5mm}
\subfigure[]{\includegraphics[width=0.48\textwidth]{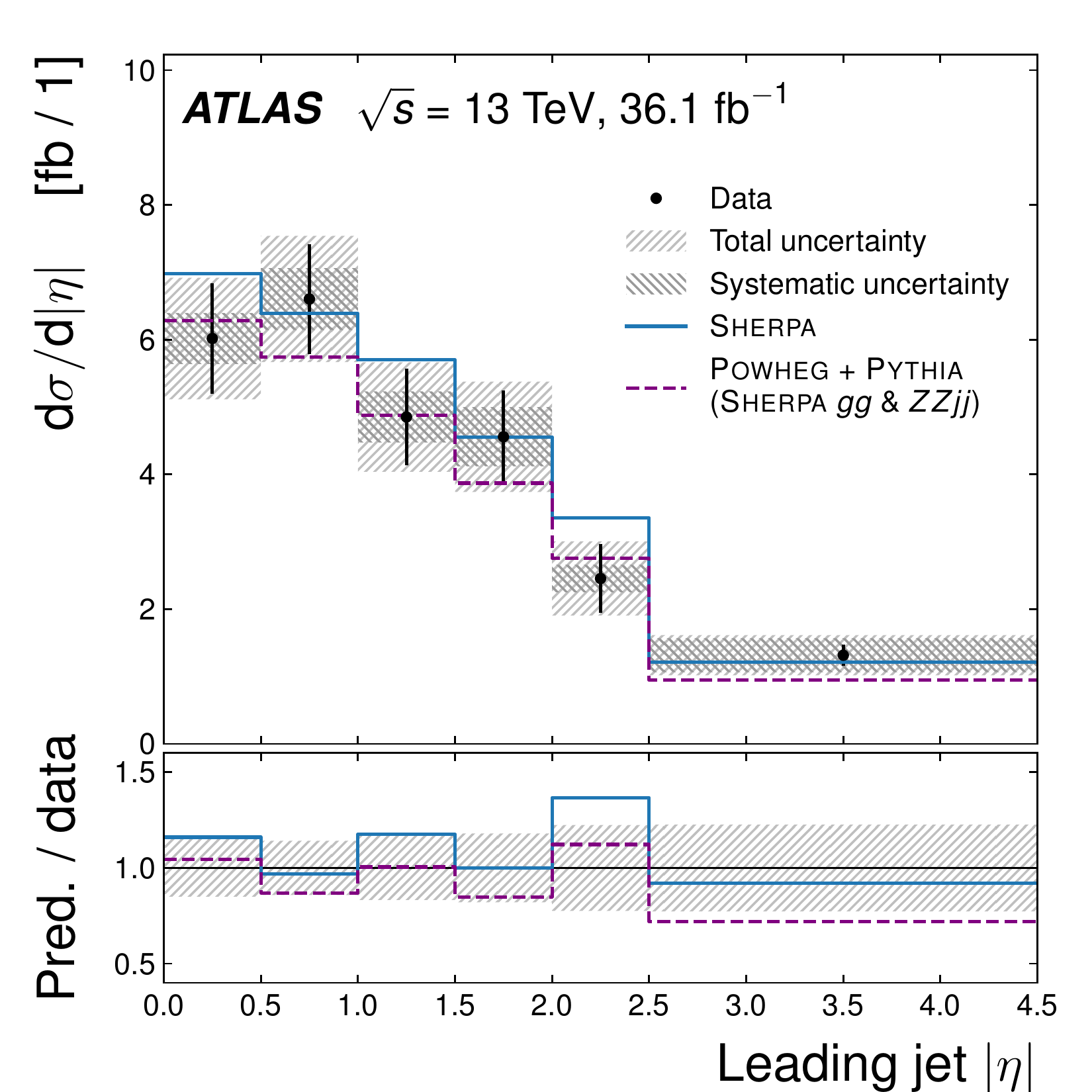}}
\subfigure[]{\includegraphics[width=0.48\textwidth]{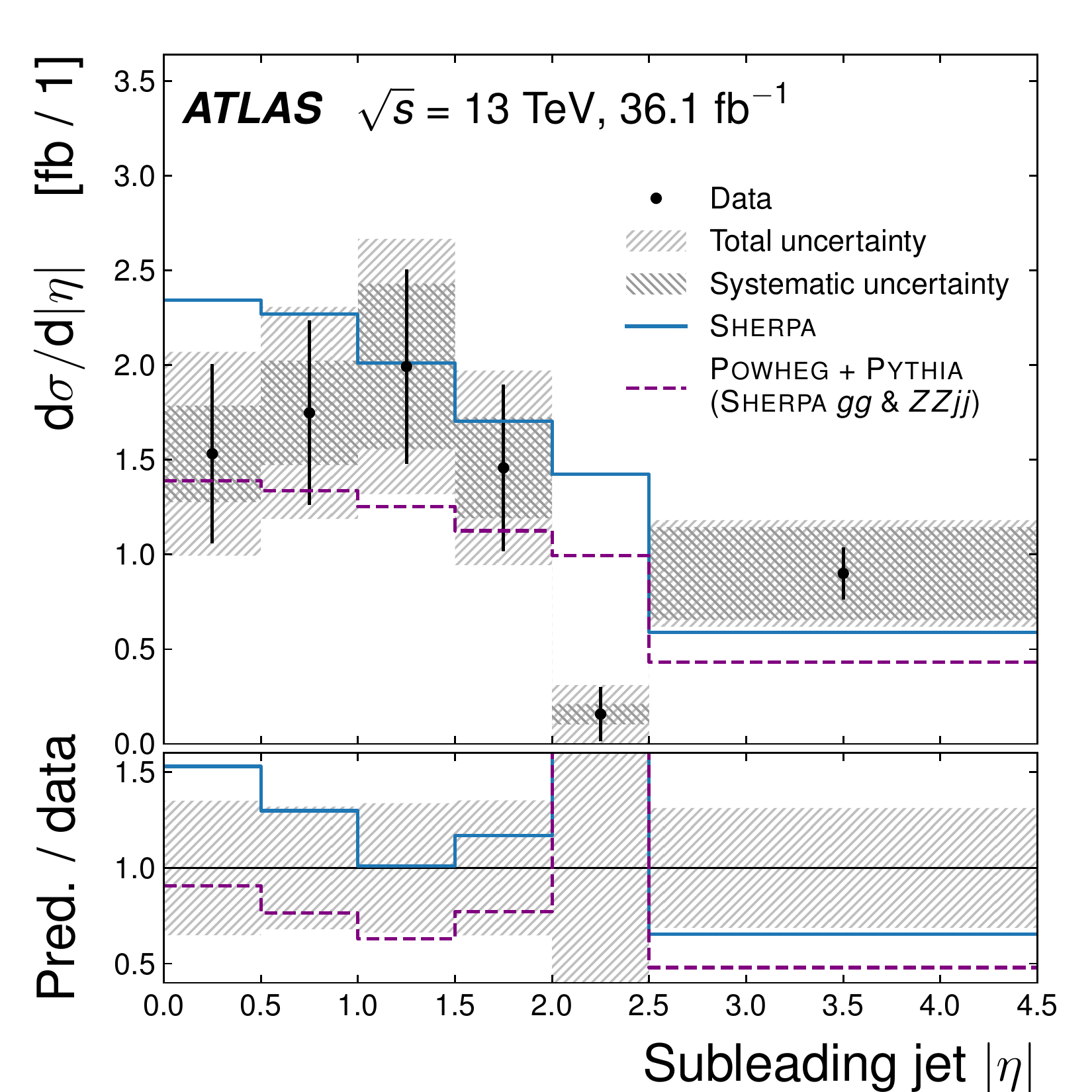}}
\caption{Measured and predicted differential cross sections for the transverse momentum of the (a) leading-\pt{} and (b) subleading-\pt{} jets, as well as the absolute pseudorapidity of the (c) leading-\pt{} and (d) subleading-\pt{} jets. \datauncertcomment{} Parton-showered \SHERPA{} and \POWHEGpy{} predictions are shown. \brokenaxiscomment{}}
\label{fig:jet_kinematics}
\end{figure}

\clearpage
Figure~\ref{fig:dijets} shows the rapidity difference and invariant mass of the two leading-\pt{} jets. The EW-$\ZZ jj$ production process predicted by \SHERPA{} is shown separately, in addition to the process-inclusive predictions from \SHERPA{} and \POWHEGpy{}. This contribution falls much less steeply towards higher values of the presented observables. The contribution from this process in the last bins in each distribution improves the agreement between prediction and measurement, demonstrating the importance of this process at these ends of the kinematic phase space.

\begin{figure}[h!]
\centering
\subfigure[]{\includegraphics[width=0.48\textwidth]{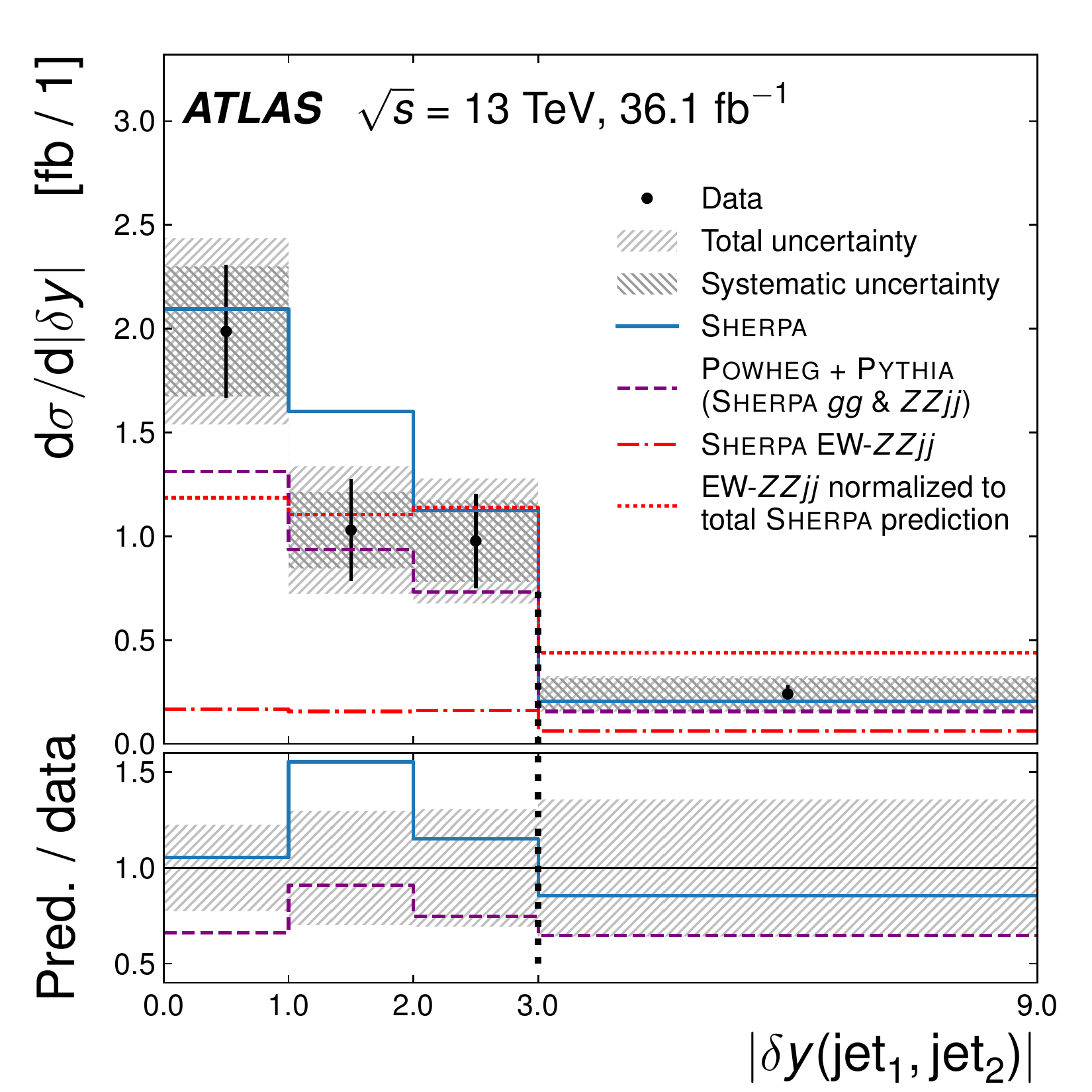}}
\subfigure[]{\includegraphics[width=0.48\textwidth]{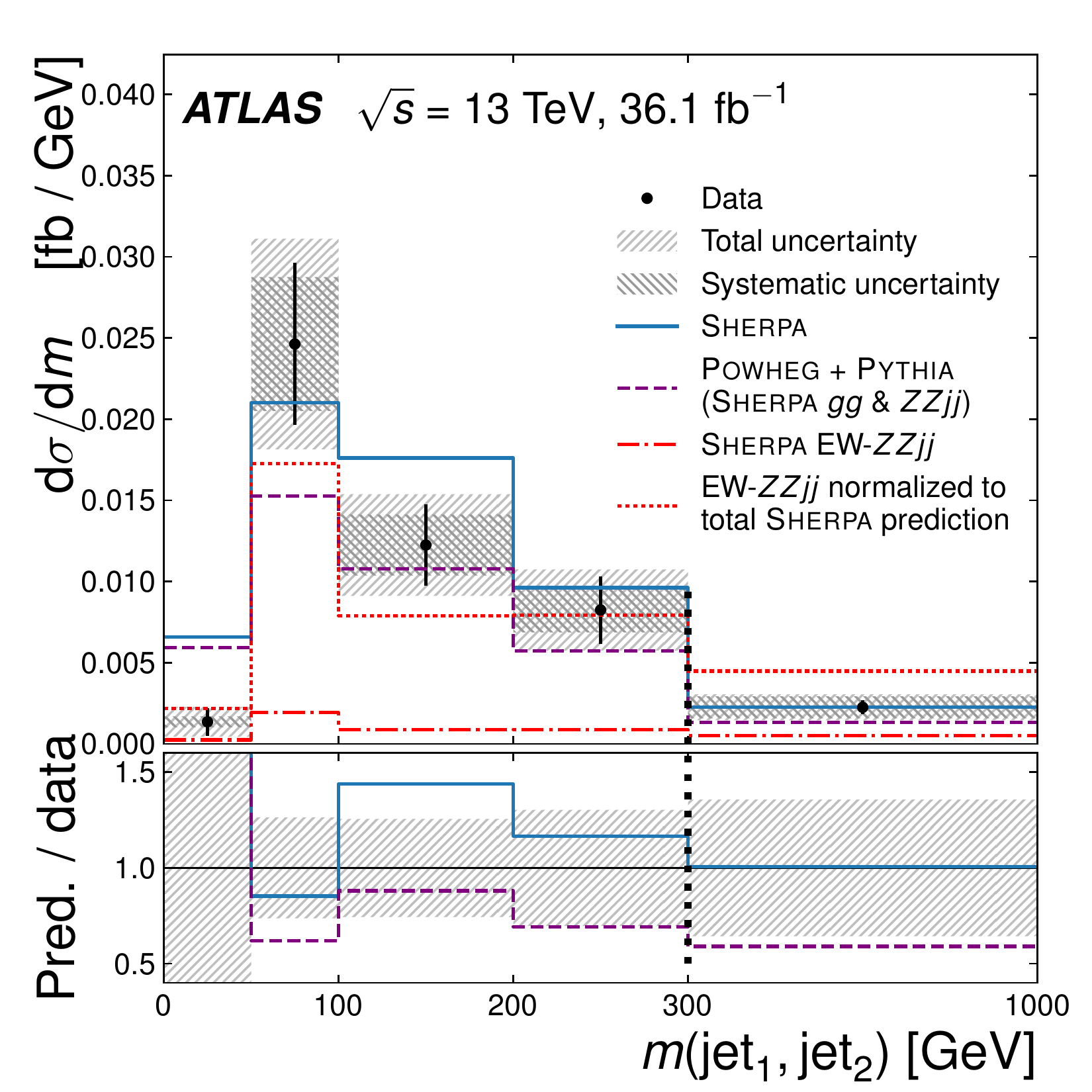}}
\caption{Measured and predicted differential cross sections for (a) the absolute difference in rapidity between the two leading-\pt{} jets and (b) the invariant mass of the two leading-\pt{} jets. \datauncertcomment{} Parton-showered \SHERPA{} and \POWHEGpy{} predictions are shown. In addition to the process-inclusive predictions, the EW-$\ZZ jj$ production process predicted by \SHERPA{} is shown separately. It is also shown normalized to the process-inclusive \SHERPA{} prediction to facilitate shape comparisons between the EW-$\ZZ jj$ subprocess and the process-inclusive $\ZZ$ production. \brokenaxiscomment{}}
\label{fig:dijets}
\end{figure}

\clearpage

\section{Search for anomalous triple gauge couplings}
\label{sec:anomalous}
The search for aTGCs uses the reconstructed transverse momentum of the leading-$\pt$ \PZ{} boson candidate (\ptz{}) to look for deviations of the data from the SM, as this variable is found to provide the highest sensitivity to their predicted effects.
The four-lepton mass provides similar sensitivity, but is not used, because no dedicated calculation of NLO EW corrections for $\Pproton\Pproton \to \ZZllll$ production binned in the four-lepton mass is available for the analysis.

The considered aTGC signal model uses an effective vertex function approach~\cite{Baur:2000cx}. It includes two coupling strengths that violate charge--parity (CP) symmetry, \fyfour{} and \fzfour{}, as well as two CP-conserving ones, \fyfive{} and \fzfive{}. No unitarizing form factor is used, as the sensitivity of the measurement is well within the unitarity bounds.

The expected number of events $N$ in the aTGC search is parameterized in terms of the coupling strengths, on which it depends both linearly and quadratically,

\vspace{-1\baselineskip}
\begin{equation}\label{eq:atgc_yield_parametrization}
\begin{split}
N\left(\fyfour, \fzfour, \fyfive, \fzfive\right) &= N_{\text{SM}} + \fyfour N_{01} + \fzfour N_{02} + \fyfive N_{03} + \fzfive N_{04}\\
&+ \left(\fyfour\right)^2 N_{11} + \fyfour \fzfour N_{12} + \fyfour \fyfive N_{13} + \fyfour \fzfive N_{14}\\
&+ \left(\fzfour\right)^2 N_{22} + \fzfour\fyfive N_{23} + \fzfour\fzfive N_{24}\\
&+ \left(\fyfive\right)^2 N_{33} + \fyfive \fzfive N_{34}\\
&+ \left(\fzfive\right)^2 N_{44},
\end{split}
\end{equation}
where $N_{\text{SM}}$ is the SM expectation and the $N_{ij}$ are yield coefficients that depend on the final-state particle momenta.
To determine the coefficients $N_{ij}$, $2\times 10^5$ events with aTGC are generated at LO with one fixed reference set of coupling strengths using \sherpa{} and the CT10 PDF set. Based on the kinematic properties of each event, the coefficients $N_{ij}$ are extracted using a framework \cite{Bella:2008wc} based on the BHO program \cite{Baur:1997kz}. The yield for all other values of the coupling strengths can then be calculated using Eq.~\eqref{eq:atgc_yield_parametrization}.

The SM prediction $N_{\text{SM}}$ is constructed separately using the highest-order calculations available. The nominal \SHERPA{} setup is used, except that the $\Pquark\APquark$-initiated process is generated with \POWHEGpy{} and each event reweighted by NNLO QCD and NLO EW corrections binned in \ptz{}. The SM \ZZ{} predictions, estimated backgrounds, as well as observed yields are shown in Table~\ref{tab:atgc_yields} as a function of \ptz{}. These contributions are also shown in Figure~\ref{fig:atgc_input} together with two different aTGC predictions. The considered systematic uncertainties of the predictions are the same as in the integrated cross-section measurement. An additional uncertainty associated with the combination of NNLO QCD and NLO EW corrections for the SM \ZZllll{} process is assigned as described in Section~\ref{sec:uncertainties}, ranging from $\sim$1\% in the lowest to $\sim$10\% in the highest \ptz{} bin. The \ptz{} bins are optimized using the predictions to maximize the expected sensitivity.

\begin{table}[h!]
\centering
\begin{tabular}{llllll}
\toprule
\textbf{\ptz{} range [\GeV{}]} &  \textbf{0--295} & \textbf{295--415} &  \textbf{415--555} & \textbf{555--3000}\\
\midrule
Data & 998 & 16 & 3 & 0\\
\midrule
Total SM prediction & $950\phantom{.0}\pm 40$& $10.6\phantom{0}\pm 0.9$& $2.50\pm 0.33$& $1.18\phantom{0}\pm 0.21$\\
\midrule
SM \ZZllll{} & $930\phantom{.0} \pm 40$& $10.0\phantom{0}\pm 0.9$& $2.34\pm 0.33$& $1.10\phantom{0}\pm 0.21$\\ 
Triboson, $\Ptop\APtop\kern1pt\PZ$, $\PZ\PZ\to \tau^{+}\tau^{-}[\ell^{+}\ell^{-}, \tau^{+}\tau^{-}]$ & $\phantom{00}9.2 \pm \phantom{0}2.8$ & $\phantom{0}0.43\pm 0.13$ & $0.15\pm 0.05$ & $0.078 \pm 0.028$ \\
Misid.~lepton background & $\phantom{0}12\phantom{.0}\pm \phantom{0}8$ & $\phantom{0}0.17\pm 0.11$ & $< 0.1$ & $< 0.1$\\
\bottomrule 
\end{tabular} 
\caption{Observed and predicted yields in bins of the transverse momentum of the leading-\pt{} \PZ{} boson candidate. All statistical and systematic uncertainties are included in the prediction uncertainties, including the uncertainty associated with the combination of NNLO QCD and NLO EW corrections for the SM \ZZllll{} process.}
\label{tab:atgc_yields}
\end{table}

\begin{figure}[h!]
\centering
\includegraphics[width=0.5\columnwidth]{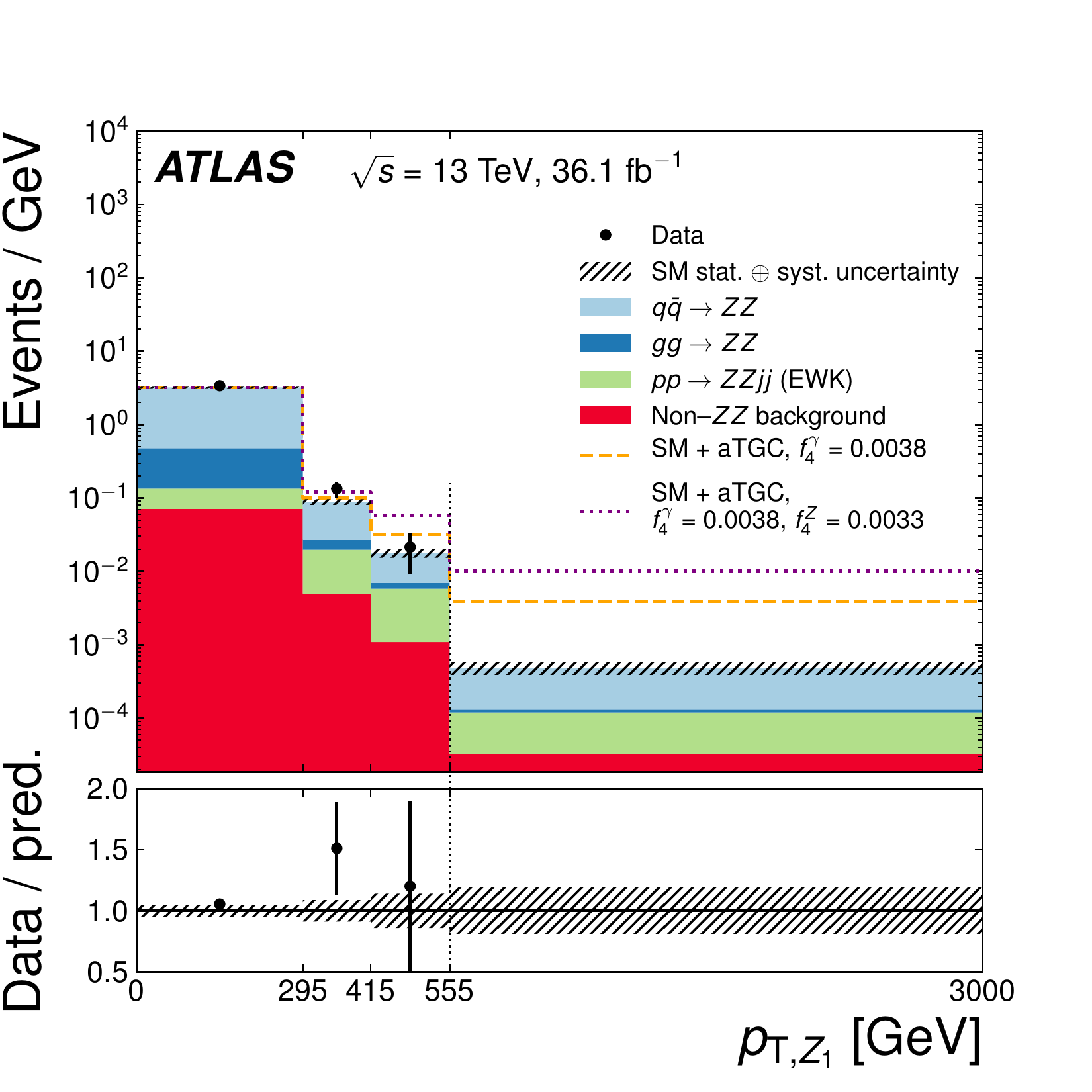}
\caption{Data and SM predictions as a function of the transverse momentum of the leading-$\pt$ $\PZ$ boson candidate. Also shown is the SM plus aTGC signal prediction with $f_4^{\gamma} = 3.8 \times 10^{-4}$ as well as with $f_4^{\gamma} = 3.8 \times 10^{-4}$ and $f_4^{\PZ} = 3.3 \times 10^{-4}$. In both cases all other aTGC coupling strengths are set to zero. The shaded band shows the total SM prediction uncertainty including the statistical and all systematic uncertainties. \brokenaxiscomment{}}
\label{fig:atgc_input}
\end{figure}

The data are found to be consistent with the SM predictions, and no indication of aTGCs is observed. Confidence intervals of aTGC parameters are determined using the expected and observed yields in bins of \ptz{} as reconstructed by the detector.

A frequentist method~\cite{Feldman:1997qc} is used to find the 95\% confidence level (CL) intervals for the aTGC parameters. The predicted and observed yields are assumed to follow Poissonian probability density functions, while the systematic uncertainties are treated as nuisance parameters constrained by Gaussian functions. The expected confidence intervals and their one- and two-standard-deviation confidence bands are established using many independent sets of randomly generated pseudodata following a Poisson distribution whose expectation value is the SM prediction in each bin. 

Confidence intervals are set for each coupling strength individually, setting all others to zero, using 2500 sets of pseudodata. The expected and observed 95\% CL intervals are listed in Table~\ref{tab:oneD_results}. The one-dimensional confidence intervals are more stringent than those derived in previous measurements by the ATLAS and CMS collaborations \cite{STDM-2014-16,CMS-SMP-13-005,CMS-SMP-12-016} and at the Tevatron and LEP colliders  \cite{Abazov:2007ad,Alcaraz:2006mx}. In addition, two-dimensional 95\% CL intervals are obtained by allowing pairs of aTGC parameters to vary simultaneously, while setting the others to zero, using 26000 sets of pseudodata. They are shown in Figure~\ref{fig:2D_results}. No significant deviations from the SM are observed.

\begin{table}[h!]
\centering
\begin{tabular}{lll} 
\toprule
Coupling strength & Expected 95\% CL $\big[\times 10^{-3}\big]$ & Observed 95\% CL $\big[\times 10^{-3}\big]$\\
\midrule
\fyfour{} & $-2.4$, 2.4 & $-1.8$, 1.8 \\ 
\fzfour{} & $-2.1$, 2.1 & $-1.5$, 1.5 \\ 
\fyfive{} & $-2.4$, 2.4 & $-1.8$, 1.8 \\ 
\fzfive{} & $-2.0$, 2.0 & $-1.5$, 1.5\\ 
\bottomrule
\end{tabular}
\caption{One-dimensional expected and observed 95\% CL intervals of the aTGC coupling strengths. Each limit is obtained setting all other aTGC coupling strengths to zero.}
\label{tab:oneD_results}
\end{table}

\begin{figure}[p]
 \centering
\subfigure[]{\includegraphics[width=0.48\textwidth]{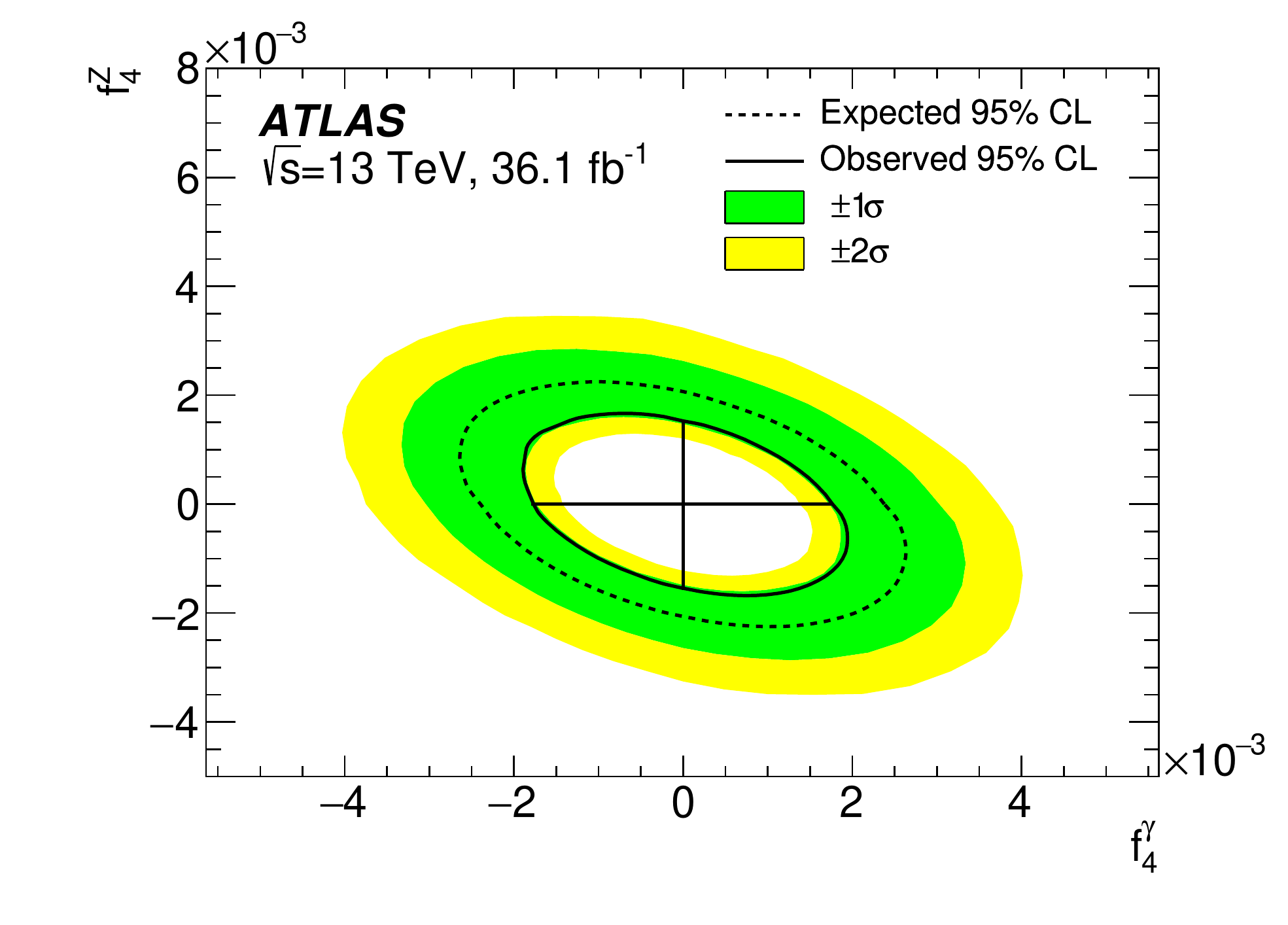}}
\subfigure[]{\includegraphics[width=0.48\textwidth]{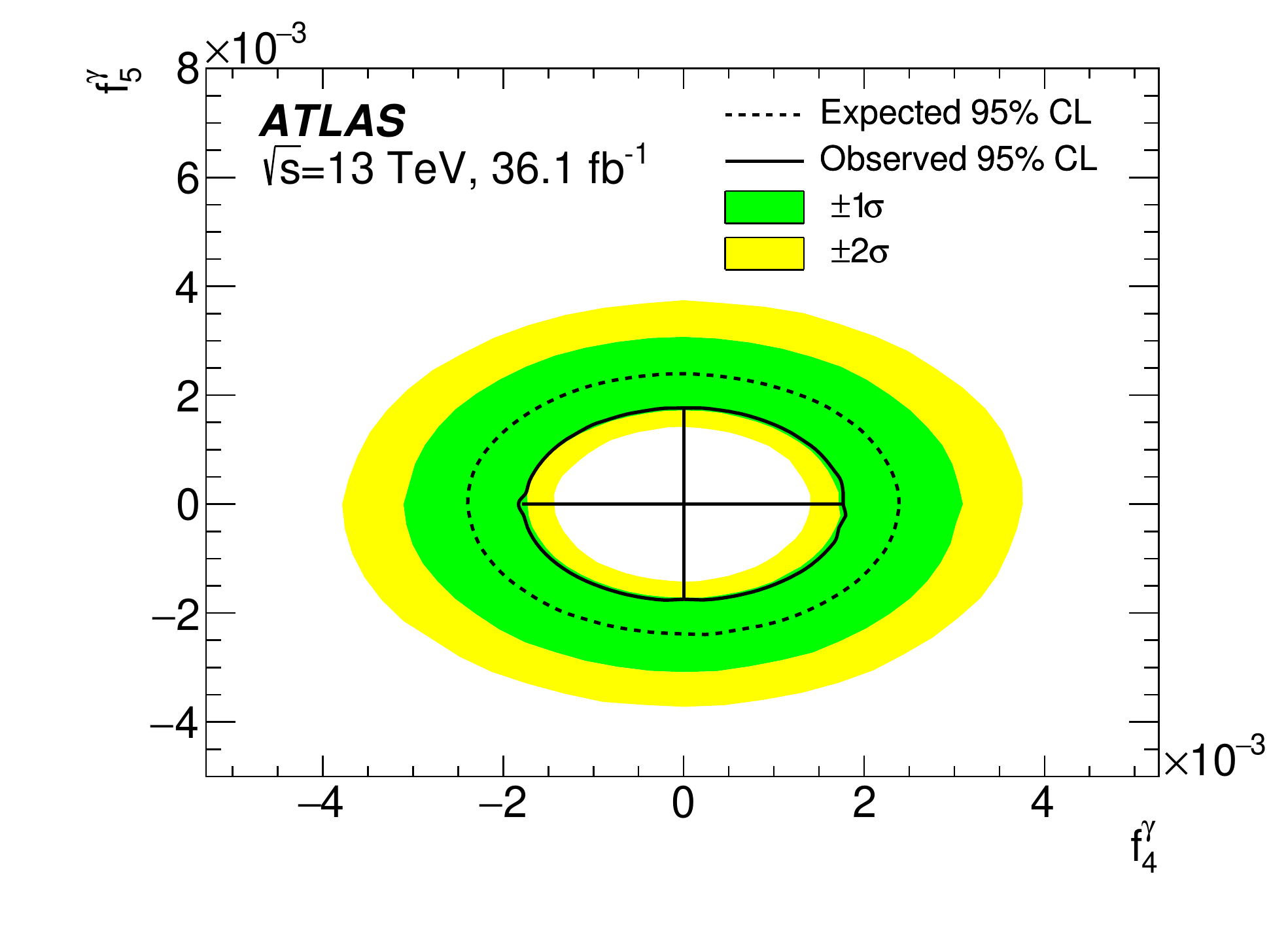}}
\subfigure[]{\includegraphics[width=0.48\textwidth]{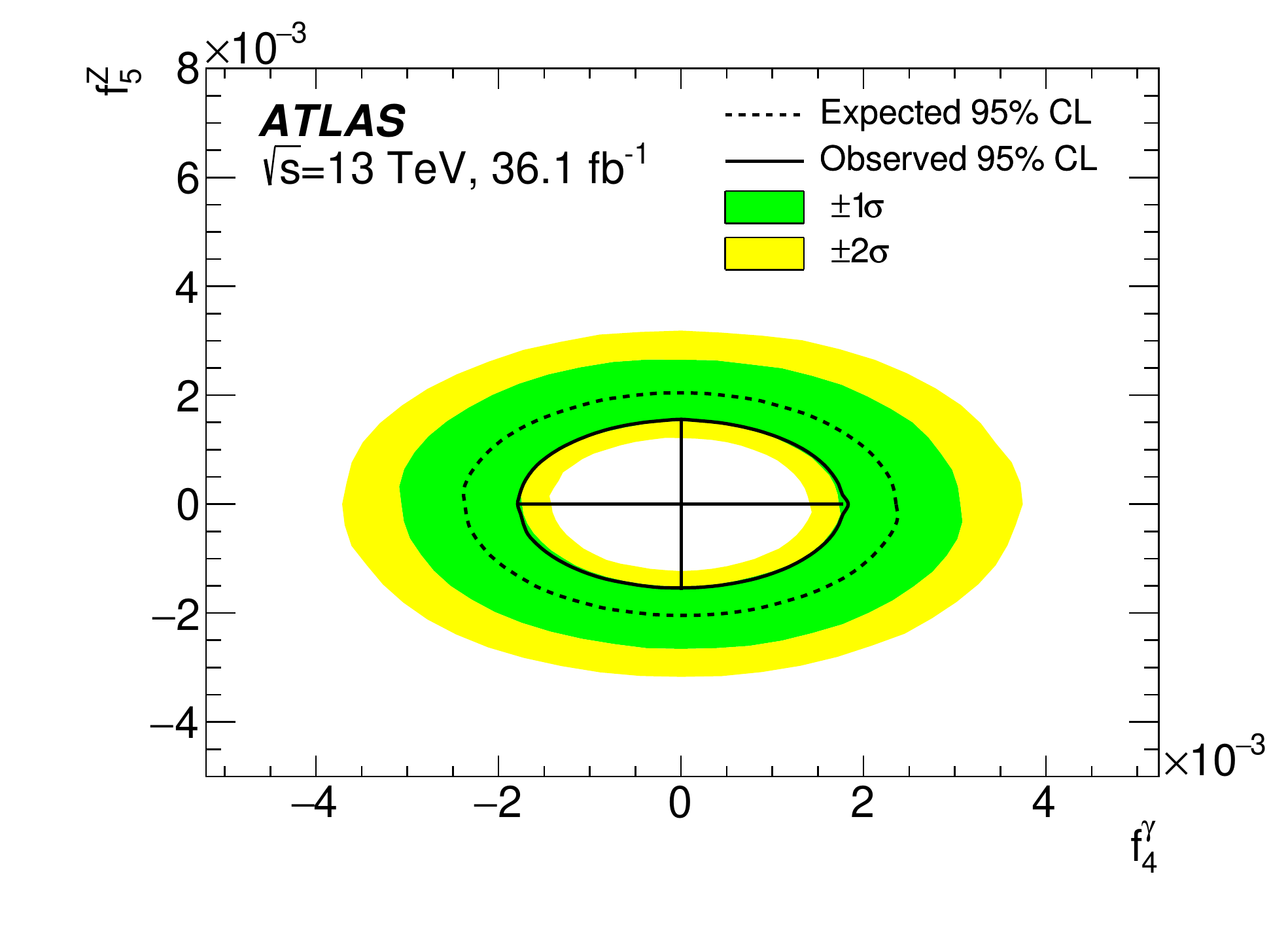}}
\subfigure[]{\includegraphics[width=0.48\textwidth]{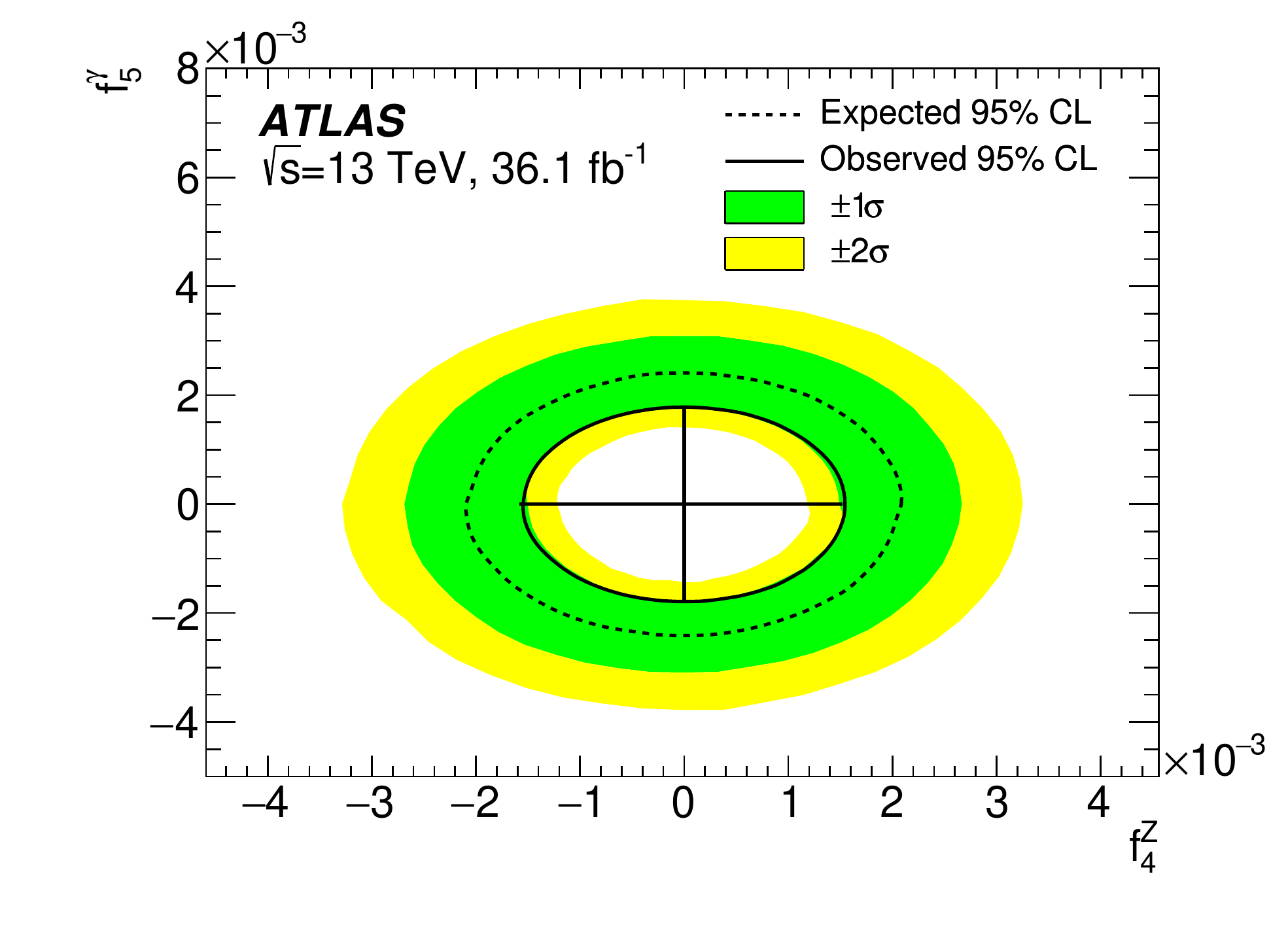}}
\subfigure[]{\includegraphics[width=0.48\textwidth]{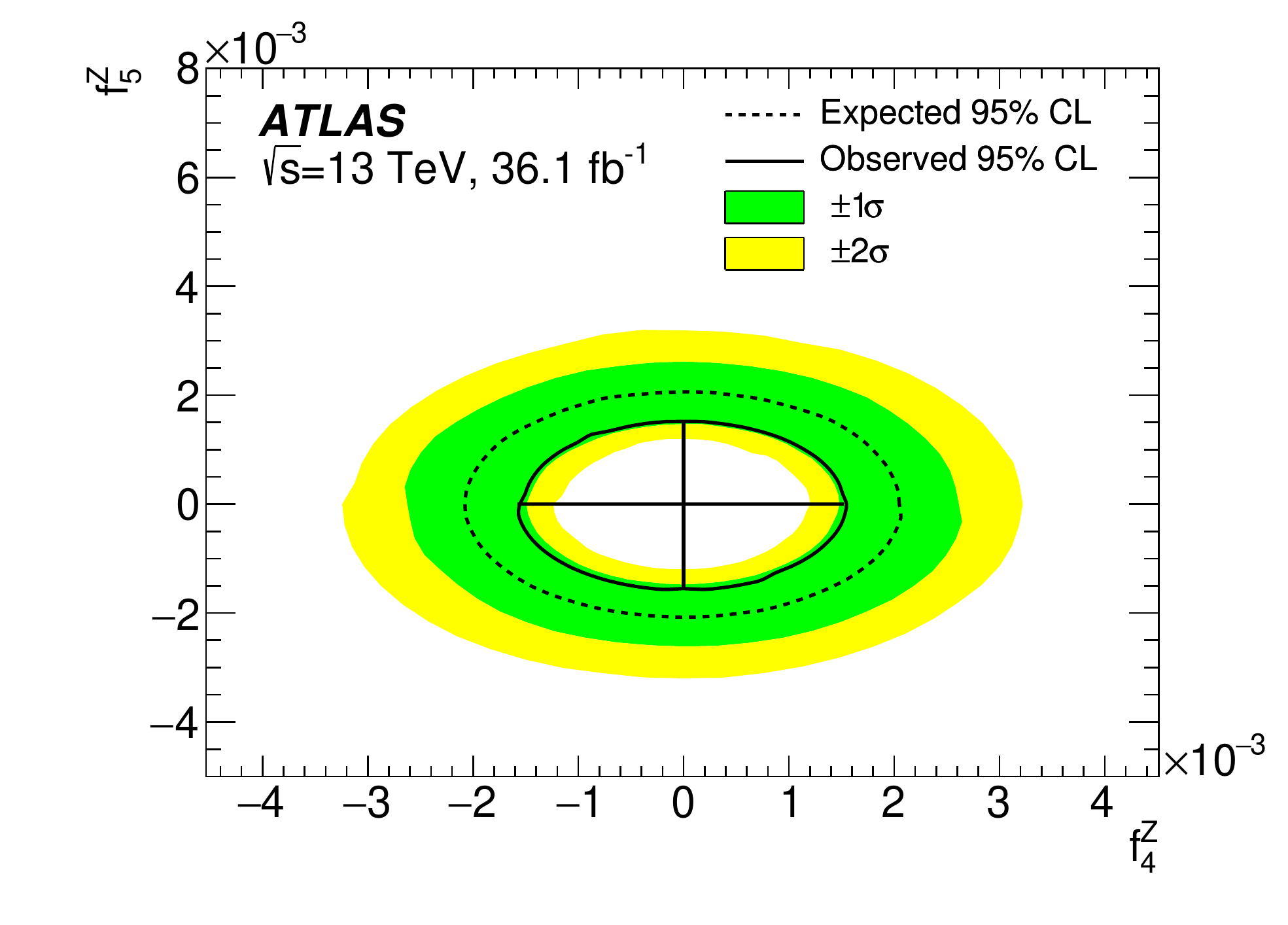}}
\subfigure[]{\includegraphics[width=0.48\textwidth]{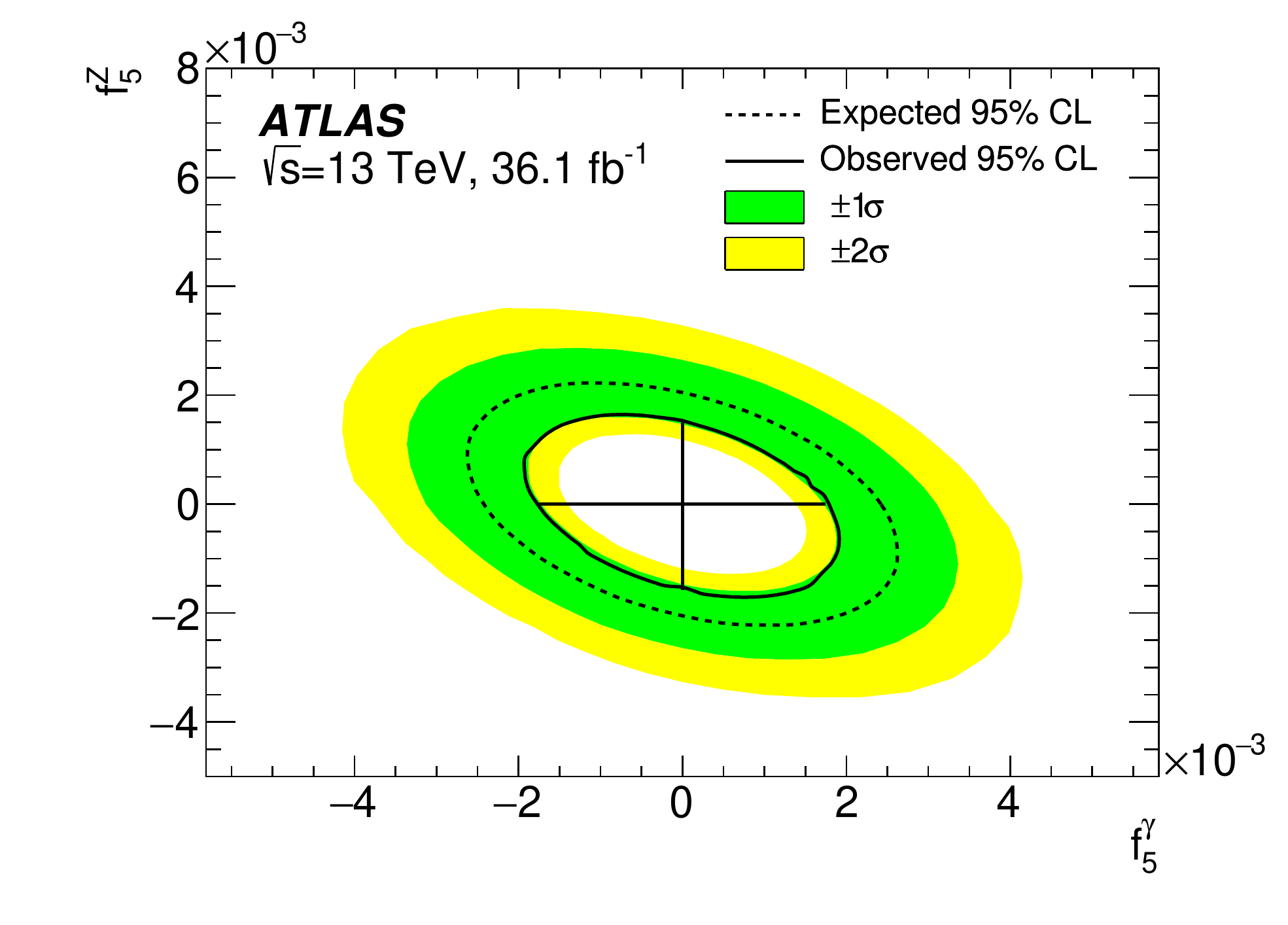}}
 \caption{Observed and expected two-dimensional 95\% CL intervals in planes of different pairs of aTGC coupling strengths. The aTGC coupling strengths other than those shown are set to zero. The black straight lines indicate the observed one-dimensional confidence intervals at 95\% CL.}
 \label{fig:2D_results}
\end{figure}

Confidence intervals are also provided for parameters of the effective field theory (EFT) in Ref.~\cite{Degrande:2012wf}, which includes four dimension-8 operators describing aTGC interactions of neutral gauge bosons. The coefficients of the operators are denoted $C_{\tilde{B}W}/\Lambda^{4}$, $C_{BW}/\Lambda^{4}$, $C_{WW}/\Lambda^{4}$, and $C_{BB}/\Lambda^{4}$, where $\Lambda$ is the energy scale of the new physics described by the EFT. They can be linearly related to the parameters \fyfour{}, \fzfour{}, \fyfive{}, and \fzfive{} as described in Ref.~\cite{Degrande:2013kka}. Thus Eq.~\eqref{eq:atgc_yield_parametrization} can be reformulated in terms of the EFT coefficients and confidence intervals set in the same way as for the coupling strengths. The resulting one-dimensional EFT confidence intervals are given in Table~\ref{tab:oneD_EFTLimits}.

Two-dimensional EFT confidence intervals are shown in Figure~\ref{fig:2D_results_EFT}.

\begin{table}[h!]
\centering
\begin{tabular}{lll} 
\toprule
EFT parameter & Expected 95\% CL $\big[\TeV^{-4}\big]$ & Observed 95\% CL $\big[\TeV^{-4}\big]$\\
\midrule
$C_{\tilde{B}W}/\Lambda^{4}$ & $-8.1$, 8.1 & $-5.9$ ,  5.9 \\ 
$C_{WW}/\Lambda^{4}$ & $-4.0$, 4.0 & $-3.0$ ,  3.0 \\ 
$C_{BW}/\Lambda^{4}$ & $-4.4$, 4.4 & $-3.3$ ,  3.3 \\ 
$C_{BB}/\Lambda^{4}$ & $-3.7$, 3.7 & $-2.7$ ,  2.8 \\ 
\bottomrule
\end{tabular}
\caption{One-dimensional expected and observed 95\% CL intervals of EFT parameters using the transformation from Ref.~\cite{Degrande:2013kka}. Each limit is obtained setting all other EFT parameters to zero.}
\label{tab:oneD_EFTLimits}
\end{table}

\begin{figure}[p]
 \centering
\subfigure[]{\includegraphics[width=0.48\textwidth]{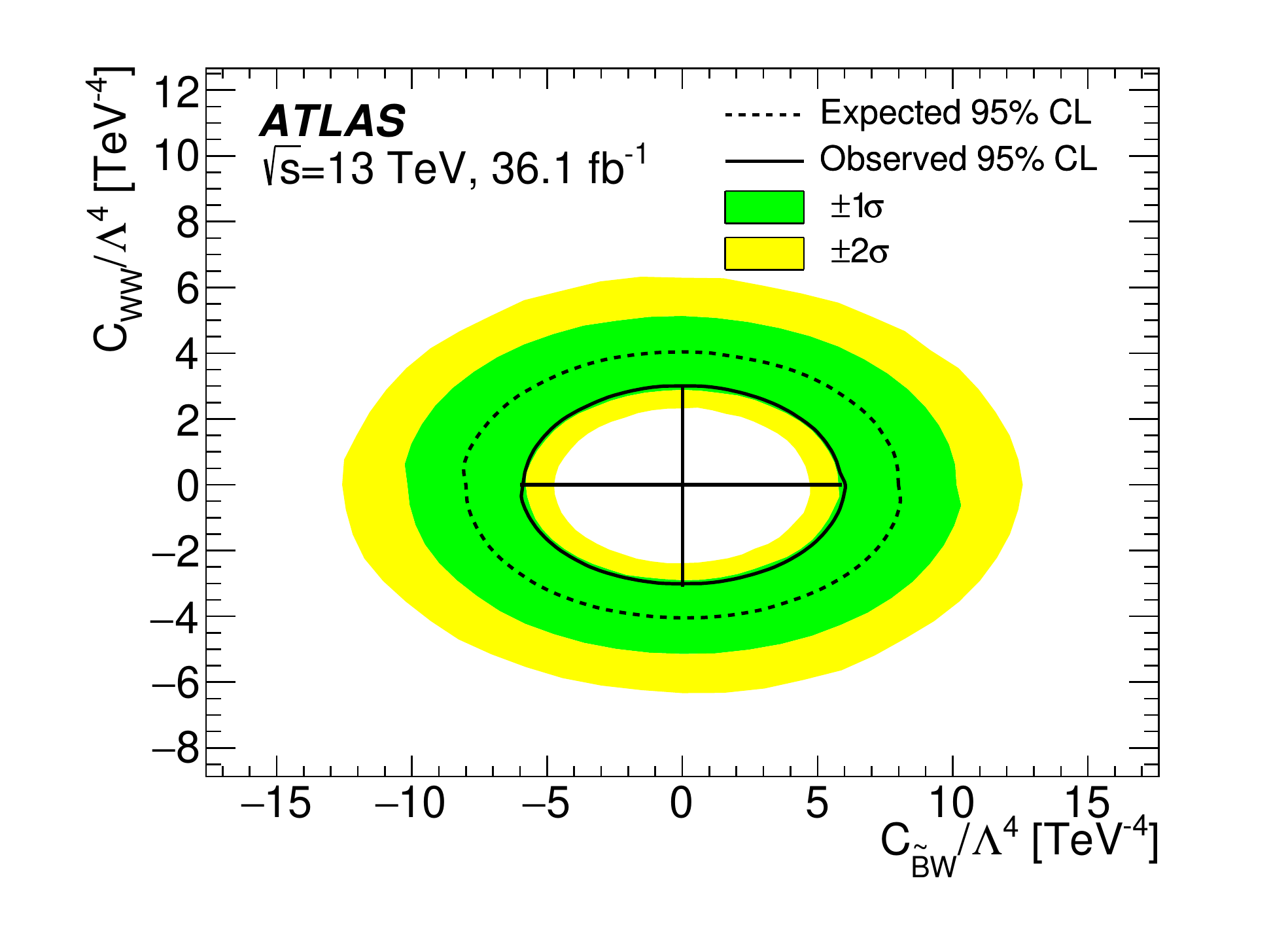}}
\subfigure[]{\includegraphics[width=0.48\textwidth]{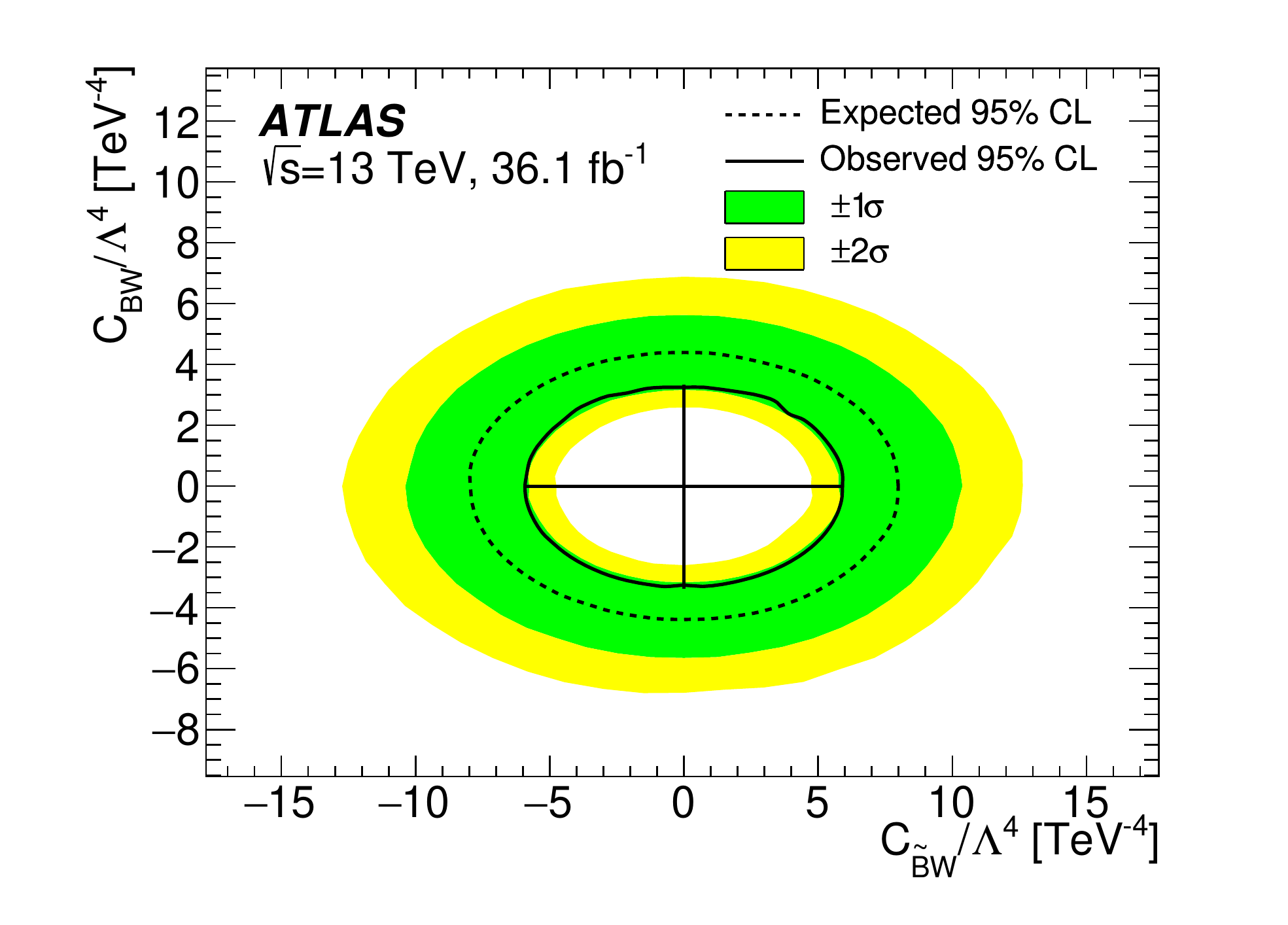}}
\subfigure[]{\includegraphics[width=0.48\textwidth]{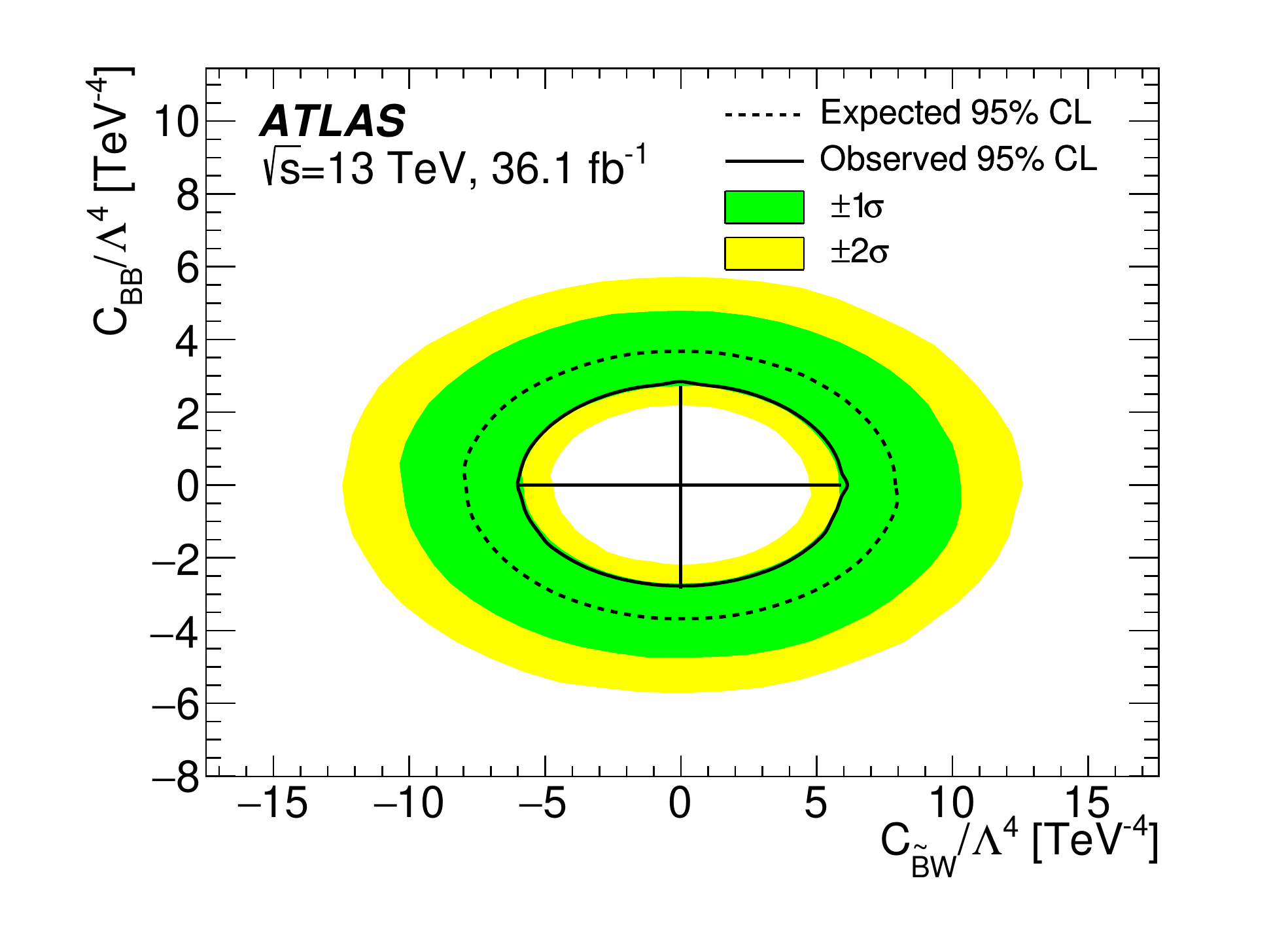}}
\subfigure[]{\includegraphics[width=0.48\textwidth]{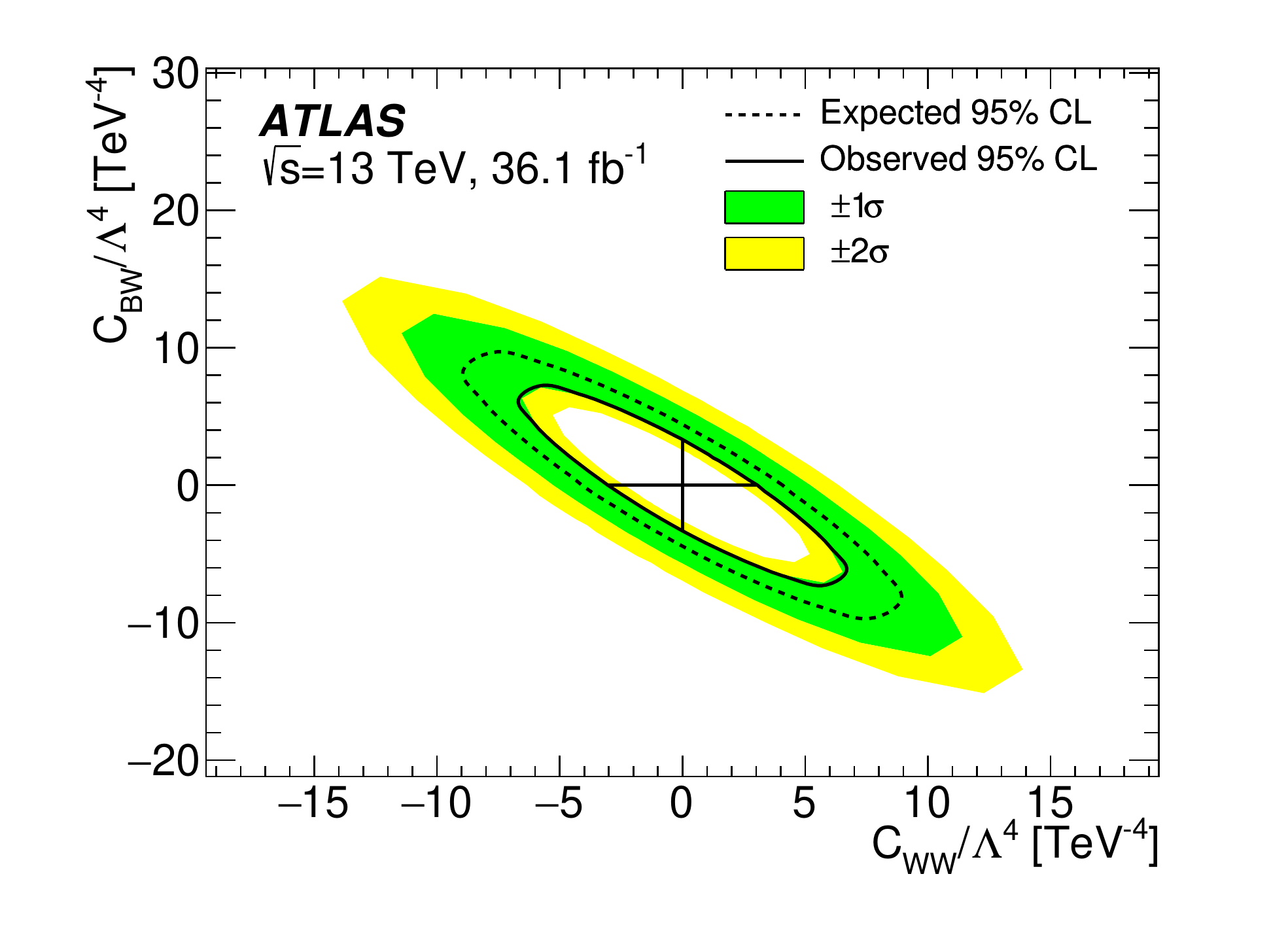}}
\subfigure[]{\includegraphics[width=0.48\textwidth]{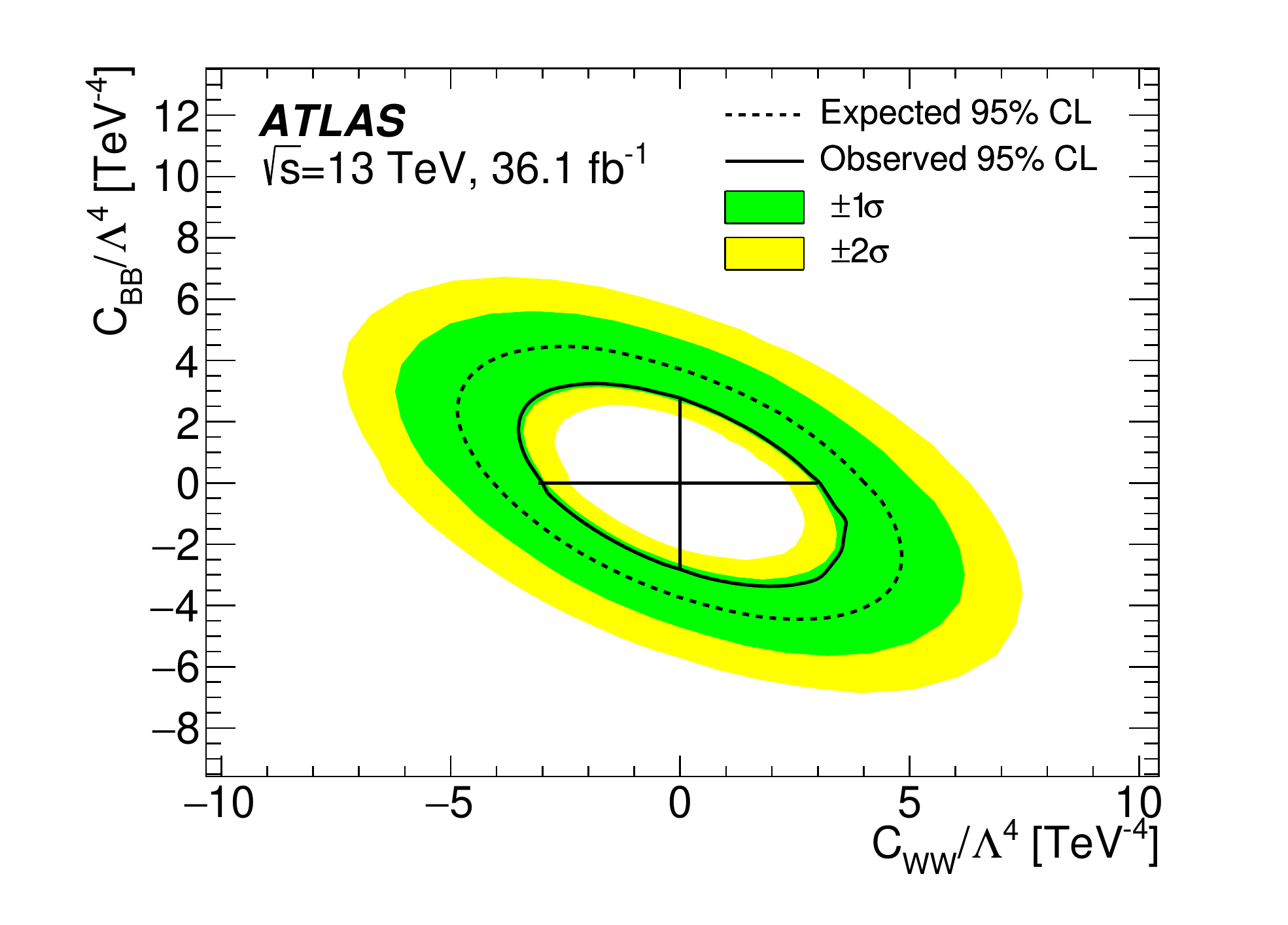}}
\subfigure[]{\includegraphics[width=0.48\textwidth]{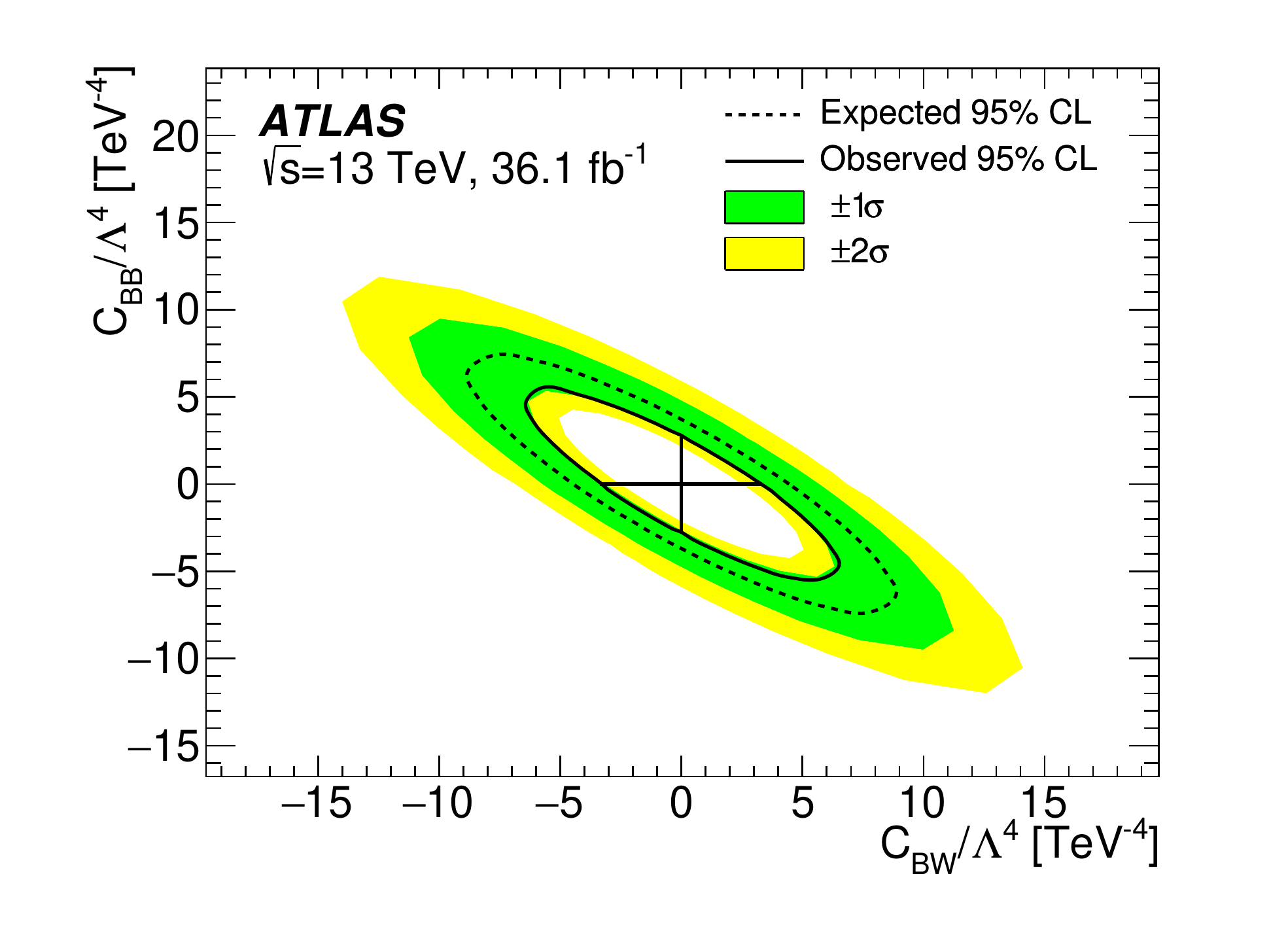}}
 \caption{Observed and expected two-dimensional 95\% CL intervals in planes of different pairs of EFT parameters using the transformation from Ref.~\cite{Degrande:2013kka}. The EFT parameters other than those shown are set to zero. The black straight lines indicate the observed one-dimensional confidence intervals at 95\% CL.}
 \label{fig:2D_results_EFT}
\end{figure}

\clearpage

\section{Conclusion}
The production of pairs of $\PZ$ bosons is studied in the \ZZllll{} channel in 13~\TeV{} proton--proton collisions produced at the LHC and recorded with the ATLAS detector, using data corresponding to an integrated luminosity of $36.1 \pm 1.1$~\ifb{}. Integrated fiducial cross sections are measured separately in the three decay channels \eeee{}, \eemm, and \mmmm{} as well as in their combination. They are found to agree well with NNLO SM predictions with NLO QCD corrections for the \gluongluon-initiated production process as well as NLO EW corrections applied.
The combined cross section is extrapolated to a total phase space and all SM \PZ{} boson decays. Differential cross sections are measured for 20 observables. They are compared to NLO predictions with parton shower, to fixed-order NNLO predictions, and to fixed-order predictions from calculations at the highest known orders for the different subprocesses (NNLO $\Pproton\Pproton \to \ZZ$, NLO $\gluongluon \to \ZZ$, NLO EW corrections, electroweak $\Pproton\Pproton \to \ZZ jj$). In general, the predictions describe the observables reasonably well. 
Using the transverse momentum of the leading-\pt{} \PZ{} boson candidate, confidence intervals are obtained for parameters of aTGCs forbidden at tree level in the SM, both parameterized as aTGC coupling strengths and in an effective field theory approach. No significant deviations from the SM are observed.

\section*{Acknowledgments}

We thank CERN for the very successful operation of the LHC, as well as the
support staff from our institutions without whom ATLAS could not be
operated efficiently.

We acknowledge the support of ANPCyT, Argentina; YerPhI, Armenia; ARC, Australia; BMWFW and FWF, Austria; ANAS, Azerbaijan; SSTC, Belarus; CNPq and FAPESP, Brazil; NSERC, NRC and CFI, Canada; CERN; CONICYT, Chile; CAS, MOST and NSFC, China; COLCIENCIAS, Colombia; MSMT CR, MPO CR and VSC CR, Czech Republic; DNRF and DNSRC, Denmark; IN2P3-CNRS, CEA-DRF/IRFU, France; SRNSF, Georgia; BMBF, HGF, and MPG, Germany; GSRT, Greece; RGC, Hong Kong SAR, China; ISF, I-CORE and Benoziyo Center, Israel; INFN, Italy; MEXT and JSPS, Japan; CNRST, Morocco; NWO, Netherlands; RCN, Norway; MNiSW and NCN, Poland; FCT, Portugal; MNE/IFA, Romania; MES of Russia and NRC KI, Russian Federation; JINR; MESTD, Serbia; MSSR, Slovakia; ARRS and MIZ\v{S}, Slovenia; DST/NRF, South Africa; MINECO, Spain; SRC and Wallenberg Foundation, Sweden; SERI, SNSF and Cantons of Bern and Geneva, Switzerland; MOST, Taiwan; TAEK, Turkey; STFC, United Kingdom; DOE and NSF, United States of America. In addition, individual groups and members have received support from BCKDF, the Canada Council, CANARIE, CRC, Compute Canada, FQRNT, and the Ontario Innovation Trust, Canada; EPLANET, ERC, ERDF, FP7, Horizon 2020 and Marie Sk{\l}odowska-Curie Actions, European Union; Investissements d'Avenir Labex and Idex, ANR, R{\'e}gion Auvergne and Fondation Partager le Savoir, France; DFG and AvH Foundation, Germany; Herakleitos, Thales and Aristeia programmes co-financed by EU-ESF and the Greek NSRF; BSF, GIF and Minerva, Israel; BRF, Norway; CERCA Programme Generalitat de Catalunya, Generalitat Valenciana, Spain; the Royal Society and Leverhulme Trust, United Kingdom.

The crucial computing support from all WLCG partners is acknowledged gratefully, in particular from CERN, the ATLAS Tier-1 facilities at TRIUMF (Canada), NDGF (Denmark, Norway, Sweden), CC-IN2P3 (France), KIT/GridKA (Germany), INFN-CNAF (Italy), NL-T1 (Netherlands), PIC (Spain), ASGC (Taiwan), RAL (UK) and BNL (USA), the Tier-2 facilities worldwide and large non-WLCG resource providers. Major contributors of computing resources are listed in Ref.~\cite{ATL-GEN-PUB-2016-002}.

\printbibliography

\newpage 
\begin{flushleft}
{\Large The ATLAS Collaboration}

\bigskip

M.~Aaboud$^\textrm{\scriptsize 137d}$,
G.~Aad$^\textrm{\scriptsize 88}$,
B.~Abbott$^\textrm{\scriptsize 115}$,
O.~Abdinov$^\textrm{\scriptsize 12}$$^{,*}$,
B.~Abeloos$^\textrm{\scriptsize 119}$,
S.H.~Abidi$^\textrm{\scriptsize 161}$,
O.S.~AbouZeid$^\textrm{\scriptsize 139}$,
N.L.~Abraham$^\textrm{\scriptsize 151}$,
H.~Abramowicz$^\textrm{\scriptsize 155}$,
H.~Abreu$^\textrm{\scriptsize 154}$
% ...
\bigskip
\\
% ...
$^{12}$ Institute of Physics, Azerbaijan Academy of Sciences, Baku, Azerbaijan\\
% ...
$^{88}$ CPPM, Aix-Marseille Universit{\'e} and CNRS/IN2P3, Marseille, France\\
% ...
$^{115}$ Homer L. Dodge Department of Physics and Astronomy, University of Oklahoma, Norman OK, United States of America\\
% ...
$^{119}$ LAL, Univ. Paris-Sud, CNRS/IN2P3, Universit{\'e} Paris-Saclay, Orsay, France\\
% ...
$^{137}$ $^{(a)}$ Facult{\'e} des Sciences Ain Chock, R{\'e}seau Universitaire de Physique des Hautes Energies - Universit{\'e} Hassan II, Casablanca; $^{(b)}$ Centre National de l'Energie des Sciences Techniques Nucleaires, Rabat; $^{(c)}$ Facult{\'e} des Sciences Semlalia, Universit{\'e} Cadi Ayyad, LPHEA-Marrakech; $^{(d)}$ Facult{\'e} des Sciences, Universit{\'e} Mohamed Premier and LPTPM, Oujda; $^{(e)}$ Facult{\'e} des sciences, Universit{\'e} Mohammed V, Rabat, Morocco\\
% ...
$^{139}$ Santa Cruz Institute for Particle Physics, University of California Santa Cruz, Santa Cruz CA, United States of America\\
% ...
$^{151}$ Department of Physics and Astronomy, University of Sussex, Brighton, United Kingdom\\
% ...
$^{154}$ Department of Physics, Technion: Israel Institute of Technology, Haifa, Israel\\
$^{155}$ Raymond and Beverly Sackler School of Physics and Astronomy, Tel Aviv University, Tel Aviv, Israel\\
% ...
$^{161}$ Department of Physics, University of Toronto, Toronto ON, Canada\\
% ...
$^{a}$ Also at Department of Physics, King's College London, London, United Kingdom\\
$^{b}$ Also at Institute of Physics, Azerbaijan Academy of Sciences, Baku, Azerbaijan\\
$^{c}$ Also at Novosibirsk State University, Novosibirsk, Russia\\
$^{d}$ Also at TRIUMF, Vancouver BC, Canada\\
$^{e}$ Also at Department of Physics {\&} Astronomy, University of Louisville, Louisville, KY, United States of America\\
% ...
$^{*}$ Deceased
\end{flushleft}


\end{document}